\documentclass[12pt]{iopart}

\usepackage{iopams}

\usepackage{amssymb}
\usepackage{gensymb}

\usepackage{color}
\usepackage{dcolumn}
\usepackage{float}
\usepackage{graphicx}
\usepackage{multirow}
\usepackage{rotating}
\usepackage{subfigure} 
\usepackage{psfrag}
\usepackage{tabularx}
\usepackage[hyphens]{url}
\usepackage{wrapfig}
\usepackage{multicol}
\usepackage{widetext}
\usepackage[utf8]{inputenc}

\usepackage[bookmarks, bookmarksopen, bookmarksnumbered]{hyperref}
\usepackage[all]{hypcap}

\usepackage[sort&compress,numbers]{natbib}

\graphicspath{{figures/}
{cactus-benchmarks/}
{examples/kerr/figs/}
{examples/bbh/figs/}
{examples/tov/}
{examples/collapse/}
{examples/cosmology/figs/}
}

\def\grb#1{gamma-ray burst#1 (GRB#1)\gdef\grb{GRB}}
\def\gw#1{gravitational wave#1 (GW#1)\gdef\gw{GW}}
\def\bh#1{black hole#1 (BH#1)\gdef\bh{BH}}
\def\ns#1{neutron star#1 (NS#1)\gdef\ns{NS}}
\def\gt#1{Georgia Tech#1 (GaTech#1)\gdef\gt{GaTech}}
\def\ahz#1{apparent horizon#1 (AH#1)\gdef\ahz{AH}}

\newcommand{\dfrac}[2]{{\frac{#1}{#2}}}

\newcommand{\codename}[1]{\texttt{#1}}
\newcommand{\todo}[1]{}

\usepackage[top=2.5cm, bottom=2.5cm, left=2.5cm, right=2.5cm]{geometry}

\sloppypar

\begin{document}

\title[The Einstein Toolkit]{The Einstein Toolkit: A Community Computational Infrastructure for Relativistic Astrophysics}

\author{
Frank L\"{o}ffler $^1$,
Joshua Faber $^2$,
Eloisa Bentivegna $^3$,
Tanja Bode $^4$,
Peter Diener $^1$,
Roland Haas $^{5,4}$,
Ian Hinder $^3$,
Bruno C. Mundim $^2$,
Christian D. Ott $^{5,1,6}$,
Erik Schnetter $^{7,8,1}$,
Gabrielle Allen $^{1,9,10}$,
Manuela Campanelli $^2$
and Pablo Laguna $^4$}

\address{$^1$ Center for Computation \& Technology, Louisiana State
  University, Baton Rouge, LA, USA}
\address{$^2$ Center for Computational Relativity and Gravitation, School of Mathematical Sciences, Rochester Institute of Technology, Rochester, NY, USA}
\address{$^3$ Max-Planck-Institut f\"{u}r Gravitationsphysik, Albert-Einstein-Institut, Golm, Germany}
\address{$^4$ Center for Relativistic Astrophysics, School of Physics, Georgia Institute of Technology, Atlanta, GA, USA}
\address{$^5$ TAPIR, California Institute of Technology, Pasadena, CA, USA}
\address{$^6$ Institute for the Physics and Mathematics of the Universe, University of Tokyo, Kashiwa, Japan}
\address{$^7$ Perimeter Institute for Theoretical Physics, Waterloo,
  ON, Canada}
\address{$^8$ Department of Physics, University of Guelph, Guelph, ON,
  Canada}
\address{$^9$ Department of Computer Science, Louisiana State University, Baton Rouge, LA, USA}
\address{$^{10}$ National Science Foundation, USA}
\ead{knarf@cct.lsu.edu}


    
\begin{abstract}
We describe the Einstein Toolkit, a community-driven, freely accessible
computational infrastructure intended for use in numerical relativity,
relativistic astrophysics, and other applications.  The Toolkit, developed by a
collaboration involving researchers from multiple institutions around the world,
combines a core set of components needed to simulate astrophysical objects such
as black holes, compact objects, and collapsing stars, as well as a full suite
of analysis tools.  The Einstein Toolkit is currently based on the Cactus Framework for
high-performance computing and the Carpet adaptive mesh refinement driver.  It
implements spacetime evolution via the BSSN evolution system and
general-relativistic hydrodynamics in a finite-volume discretization. The
toolkit is under continuous development and contains many new code components
that have been publicly released for the first time and are described in this
article.  We discuss the motivation behind the release of the toolkit, the
philosophy underlying its development, and the goals of the project.  A summary
of the implemented numerical techniques is included, as are results of
numerical test covering a variety of sample astrophysical problems. 
\end{abstract}

\maketitle


\section{Introduction}

Scientific progress in the field of numerical relativity has always
been closely tied to the availability and ease-of-use of enabling
software and computational infrastructure. This document describes
the Einstein Toolkit, which provides such an infrastructure,
developed openly and made available freely with grant support from 
the National Science Foundation.

Now is a particularly exciting time for numerical relativity and
relativistic astrophysics, with major advances having been achieved 
in the study of astrophysical systems containing black holes (BHs) 
and neutron stars (NSs).  The first fully general relativistic (GR) 
simulations of merging NS-NS binaries were reported in 1999, with further 
advances over the next few years~\cite{Shibata:1999wm,Shibata:2002jb,
Shibata:2003ga,Shibata:2005ss,Shibata:2006nm}. However, systems containing BHs proved 
much more difficult to evolve numerically until 2005.  That year, computational 
breakthroughs were made   following the development of a generalized harmonic formulation~\cite{Pretorius:2005gq} and then a ``moving puncture'' approach 
\cite{Campanelli:2005dd, Baker:2005vv} in the BSSN 
(Baumgarte-Shapiro-Shibata-Nakamura) formalism~\cite{Shibata:1995we,Baumgarte:1998te} 
that lead to the first stable long-term evolutions of moving single 
and multiple BH systems.  These results quickly transformed the field 
which was now able to effectively evolve the Einstein field equations 
for coalescing BH-BH binaries and other systems containing moving 
BHs, including merging BH-NS binaries.

These breakthroughs had direct relevance 
to astrophysics, and enabled exciting new results on recoil velocities 
from BH-BH mergers~(e.g,
\cite{Baker:2006vn,Campanelli:2007ew,HolleyBockelmann:2007eh,
  Pollney:2007ss,Lousto:2007db,Lousto:2008dn} and references therein),
post-Newtonian (PN) and numerical waveform comparisons and waveform
template generation~(e.g.,~\cite{Baker:2006ha,
  Husa:2007rh,Baumgarte:2006en,Buonanno:2006ui,
  Hannam:2007wf,Gopakumar:2007vh, Campanelli:2008nk, Buonanno:2007pf, 
  Ajith:2007kx} and references
therein), comparisons between numerical
waveforms~\cite{Baker:2006yw,Baker:2007fb}, determination of the spin of
the remnant BH formed in BH-BH mergers~(e.g,~\cite{Campanelli:2006uy,
  Campanelli:2006fg,Campanelli:2006fy,
Herrmann:2007ex,Rezzolla:2007rz,Berti:2007nw} and references therein), 
and studies of eccentric BH-BH binaries
\cite{Pretorius:2007jn,Sperhake:2007gu,Hinder:2007qu,Grigsby:2007fu,
Pfeiffer:2007yz,Stephens:2011as}. 

Meanwhile, general relativistic magneto-hydrodynamics (GRMHD) 
on fixed background spacetimes has been successful in multi-dimensional
settings since the mid-1990s, focusing on BH accretion processes and 
relativistic jet production and evolution
(see~\cite{Font:2008aa} for a review of the numerical formalism
and~\cite{Hawley2009apss} for a review of work on disk and jet models). 
GRMHD coupled with
curvature evolution, on the other hand, which is crucial for modeling large-scale bulk
dynamics in compact binary star coalescence or single-star collapse scenarios, has
started to produce astrophysically interesting results only in the
past $\sim 3-5$ years, enabled primarily by the availability of long-term
stable curvature evolution systems as well as improved GRMHD
algorithms~(see~\cite{Font:2008aa} for a review). 
In addition to these developments, substantial progress has been made 
in importing more physically motivated equations of state (EOS), 
including tabulated versions (e.g.,~\cite{Pandharipande:1989hn,
Douchin:2001sv,Akmal:1998cf}) and temperature-dependent models 
(e.g.,~\cite{Shen:1998by,Shen:1998gq,Lattimer:1991nc}).  Some codes
have also begun to incorporate microphysical effects of neutrino emission
and deleptonization~\cite{Sekiguchi:2011zd,O'Connor:2009vw}.

Many of the successful techniques used to
evolve BH-BH binaries have proven to be equally applicable to merging 
NS-NS \cite{Anderson:2007kz,Anderson:2008zp,Baiotti:2008ra,Baiotti:2009gk,Baiotti:2010xh,
Baiotti:2011am,Bernuzzi:2011aq,Giacomazzo:2009mp,
Giacomazzo:2010bx,Gold:2011df,Hotokezaka:2011dh,Kiuchi:2009jt, Kiuchi:2010ze,Liu:2008xy,Rezzolla:2010fd,Rezzolla:2011da,Sekiguchi:2011zd,Sekiguchi:2011mc,Thierfelder:2011yi,Yamamoto:2008js} and BH-NS \cite{Chawla:2010sw,Duez:2008rb,Duez:2009yy,Etienne:2007jg,Etienne:2008re,Foucart:2010eq,Foucart:2011mz,Kyutoku:2010zd,Kyutoku:2011vz,Lackey:2011vz,Loffler:2006nu,Shibata:2006bs,
Shibata:2006ks,Shibata:2007zm,Shibata:2009cn,
Shibata:2010zz,Stephens:2011as,Yamamoto:2008js} binaries (for reviews, see also~\cite{Faber:2009zz,Duez:2009yz}), allowing for further investigations into the former
and  the first full GR simulations of the latter.  All recent results use 
either the general harmonic formalism or the
BSSN formalism in the ``moving puncture'' gauge.  Nearly all include some form of adaptive mesh 
refinement, since unigrid models cannot produce accurate long-term evolutions 
without requiring exorbitant computational resources, though some BH-NS simulations have been performed with a pseudospectral code \cite{Duez:2008rb,Duez:2009yy,Foucart:2010eq,Foucart:2011mz}.  Many groups' codes 
now include GRMHD (used widely for NS-NS mergers, and for BH-NS mergers 
in~\cite{Chawla:2010sw}, and some include microphysical effects as
well~(e.g.,~\cite{Duez:2009yy,Sekiguchi:2011zd,Sekiguchi:2011mc}).

In addition to studying binary mergers, numerical relativity is a necessary 
element for understanding stellar collapse and dynamical instabilities 
in NSs.  GRHD has been used to study, among many other applications, 
massive stars collapsing to protoneutron stars  
\cite{Ott:2006eu,Ott:2006eh,Shibata:2004kb}, the collapse of rotating,
hypermassive NSs to BHs in 2D and 3D (see, e.g.,~\cite{Shibata:2006hr,
Shibata:1999yx,Duez:2005sf,Duez:2005cj,Baiotti:2004wn,Baiotti:2005vi,
Baiotti:2006wn}),  and non-axisymmetric instabilities in
rapidly rotating polytropic NS models~\cite{Shibata:1999yx,Baiotti:2006wn,
Manca:2007ca}.
 
Simultaneously with the advances in both our physical understanding of 
relativistic dynamics and the numerical techniques required to study them,
a set of general computational tools and libraries has been developed with the
aim of providing a computational core that can enable new science,
broaden the community, facilitate collaborative and interdisciplinary research,  promote software reuse and take
advantage of emerging petascale computers and advanced cyberinfrastructure:
the Cactus computational toolkit~\cite{Cactuscode:web}. 
Although the development of Cactus was driven directly from the numerical relativity community, it was developed in collaboration 
with computer scientists and other computational fields to facilitate the incorporation of innovations in computational 
theory and technology. 

This success prompted usage of the {\tt Cactus} computational toolkit in other
areas, such as ocean forecast models~\cite{Djikstra2005} and chemical reaction 
simulations~\cite{Camarda2001}. At the same time, the growing
number of results in numerical relativity increased the need for commonly
available utilities such as comparison and analysis tools, typically
those specifically designed for astrophysical problems. Including them within
the {\tt Cactus} computational toolkit was not felt to fit within its rapidly
expanding scope. This triggered the creation of the Einstein
Toolkit~\cite{EinsteinToolkit:web}. Large parts of the Einstein toolkit
presently do make use of the {\tt Cactus} toolkit, but this is not an
requirement, and other contributions are welcome, encouraged and have been
accepted in the past.

\section{Requirements}

\subsection{Scientific}

While the aforementioned studies collectively represent
breakthrough simulations that have significantly advanced the modeling of
relativistic astrophysical systems, all simulations are presently
missing one or more critical physical ingredients and are lacking the
numerical precision to accurately and realistically model the
large-scale and small-scale dynamics of their target systems simultaneously.

One of the aims of the Einstein Toolkit is to provide or extend
some of these missing ingredients in the course of its development.
Over the past three years, routines have been added to the code to allow for a wider range of initial data choices, 
to allow for multithreading in hydrodynamic evolutions, and to refine the {\tt Carpet} adaptive mesh refinement
driver.  Looking forward,
three possible additions to future releases are the inclusion of magnetic fields 
into the dynamics via an ideal MHD treatment, more physical nuclear matter 
equations of state (EOSs) including the ability to model finite-temperature effects, 
and higher-order numerical techniques.  All of these are under active development, 
with MHD and finite-temperature evolution code already available, though not completely
documented, within the public toolkit releases, and will be made available once they are
thoroughly tested and validated against known results.

\subsection{Academic and Social}

A primary concern for research groups is securing reliable funding
to support graduate students and postdoctoral researchers. This
is easier to achieve if it can be shown that scientific goals can be
attacked directly with fewer potential infrastructure problems, one
of the goals of the Einstein Toolkit.

While the Einstein Toolkit does have a large group of users, many of them do not directly collaborate on science problems, and some compete.
However, many groups agree that sharing the development
of the underlying infrastructure is mutually beneficial for every group and the wider community as well.
This is achieved by lifting  off the research groups' shoulders much of the
otherwise necessary burden of creating such an infrastructure, while at the same
time increasing the amount of code review and thus, code quality.
In addition, the Einstein Toolkit provides computer scientists an
ideal platform to perform state-of-the-art research, which directly
benefits research in other areas of science and provides an
immediate application of their research.

\section{Design and Strategy}

The mechanisms for the development and support of the Einstein Toolkit are
designed to be open, transparent and community-driven. The complete source code,
documentation and tools included in the Einstein Toolkit are distributed
under open-source licenses. The Einstein Toolkit maintains a version
control system ({\tt svn.einsteintoolkit.org}) with open access that contains software
supported by the Einstein Toolkit, the toolkit web pages, and
documentation. An open wiki for documentation
({\tt docs.einsteintoollkit.org}) has been established where the community
can contribute either anonymously or through personal authentication. Almost
all discussions about the toolkit take place on an open mail list
({\tt users@einsteintoolkit.org}). The regular weekly meetings for the
Einstein Toolkit are open and the community is invited to
participate. Meeting minutes are recorded and publicly available as well.
The Einstein Toolkit
blog requires users to first request a login, but then allows for 
posting at will. Any user can post comments to entries already on the blog. 
The community makes heavy use of an issue tracking system
({\tt trac.einsteintoolkit.org}), with submissions also open to the public.

Despite this open design, some actions naturally have to be restricted to a smaller
group of maintainers. This is true for administrative tasks like the
setup and maintenance of the services themselves, or to avoid large amounts
of spam. One of the most important tasks of an Einstein Toolkit
maintainer is to review and apply patches sent by users in order to ensure
a high software quality level. Every substantial change or addition to
the toolkit must be reviewed by another Einstein Toolkit maintainer,
and is generally open for discussion on the users mailing list. This convention,
though not being technically enforced, works well in practice and promotes
active development.

\section{Core Technologies}

The Einstein Toolkit modules center around a set of core modules that provide
basic functionality to create, deploy and manage a numerical simulation
starting with code generation all to way to archiving of simulation
results: (i) the {\tt Cactus} framework ``flesh'' provides the underlying
infrastructure to build complex simulation codes out of independently
developed modules and facilities communication between these modules. (ii) the
adaptive mesh refinement driver, {\tt Carpet}, is build on top of {\tt Cactus}
and provides problem independent adaptive mesh refinement support for
simulations that need to resolve physics on length scales differing by many
orders of magnitude, while relieving the scientist of the need to worry about
internal details of the mesh refinement driver. (iii) {\tt Kranc}, which generates
code in a computer language from a high-level description in Mathematica and
(iv) the Simulation Factory, which provides a uniform, high-level interface to
common operations, such as submission and restart of jobs, for a large number of
compute clusters.

\subsection{Cactus Framework}

The {\tt Cactus}
  Framework~\cite{Cactuscode:web,Goodale:2002a,CactusUsersGuide:web} is
an open source, modular, portable programming environment for
collaborative HPC computing primarily developed at Louisiana State University,
which originated at the Albert Einstein Institute and also has roots
at the National Center for Supercomputing Applications~(see, e.g.,~\cite{Anninos:1995am,Anninos:1996ai,Seidel:1999ey} for historical reviews).
The {\tt Cactus} computational toolkit consists of general modules which provide
parallel drivers, coordinates, boundary conditions, interpolators,
reduction operators, and efficient I/O in different data
formats. Generic interfaces make it possible to use
external packages and improved modules which are made immediately available
to users.

The structure of the {\tt Cactus} framework is completely modular, with
only a very small core (the ``flesh'') which provides the interfaces between
modules
both at compile- and run-time. The {\tt Cactus} modules (called ``thorns'')
may (and typically do) specify inter-module dependencies, e.g., to share or
extend configuration information, common variables, or runtime parameters.
Modules compiled into an executable can remain dormant at run-time.
This usage of modules and a common interface between them enables researchers
to 1) easily use modules written by others without the need to understand
all details of their implementation and 2) write their own modules
without the need to change the source code of other parts of a simulation
in the (supported) programming language of their choice.
The number of active modules within a typical {\tt Cactus} simulation ranges
from tens to hundreds and often has an extensive set of inter-module
dependencies.

The {\tt Cactus} Framework was developed originally by the
numerical relativity community, and although it is now a general component
framework that supports different application domains, its core user
group continues to be comprised of numerical relativists. 
It is not surprising therefore, that one of the science modules provided in
the Einstein Toolkit is a set of state of the art modules to simulate binary
black hole mergers. All modules to simulate and analyze the data are provided
out of the box. This set of modules also provides a way of testing the
Einstein Toolkit modules in a production type simulation rather than synthetic
test cases. Some of these modules have been developed specifically for the
Einstein Toolkit while others are modules used in previous publications and
have been contributed to the toolkit. In these cases the Einstein Toolkit
provides documentation and best practice guidelines for the contributed modules.

\subsection{Adaptive Mesh Refinement}

In {\tt Cactus}, infrastructure capabilities such as memory management,
parallelization, time evolution, mesh refinement, and I/O are
delegated to a set of special \emph{driver} components. This helps
separate physics code from infrastructure code; in fact, a typical
physics component (implementing, e.g., the Einstein or relativistic MHD
equations) does not contain any code or subroutine calls having to do
with parallelization, time evolution, or mesh refinement. The
information provided in the interface declarations of the individual
components allows a highly efficient execution of the combined
program.

The Einstein Toolkit offers two drivers, \codename{PUGH} and
{\tt Carpet}. \codename{PUGH} provides domains consisting of a uniform 
grid with Cartesian topology, and is highly scalable (up to more than
130,000~\cite{Cactuscode:BlueGene:web}.)
{\tt Carpet}~\cite{Schnetter:2003rb, Schnetter:2006pg,
  CarpetCode:web} provides multi-block methods and adaptive mesh
refinement (AMR\@). Multi-block methods cover the domain with a set of
(possibly distorted) blocks that exchange boundary information via techniques such as
interpolation or penalty methods.\footnote{Although multi-block
  methods are supported by {\tt Carpet}, the Einstein Toolkit itself
  does not currently
  contain any multi-block coordinate systems.} The AMR capabilities
employ the standard Berger-Oliger algorithm~\cite{Berger:1984zza} with
subcycling in time.

AMR implies that resolution in the simulation
domain is dynamically adapted to the current state of the simulation,
i.e., regions that require a higher resolution are covered with blocks
with a finer grid (typically by a factor of two); these are called 
\emph{refined levels}. Finer grids can be also recursively refined.
At regular intervals, the resolution requirements in the
simulation are re-evaluated, and the grid hierarchy is updated; this
step is called \emph{regridding}.

\begin{figure}
  \centering
  \includegraphics[width=0.3\textwidth]{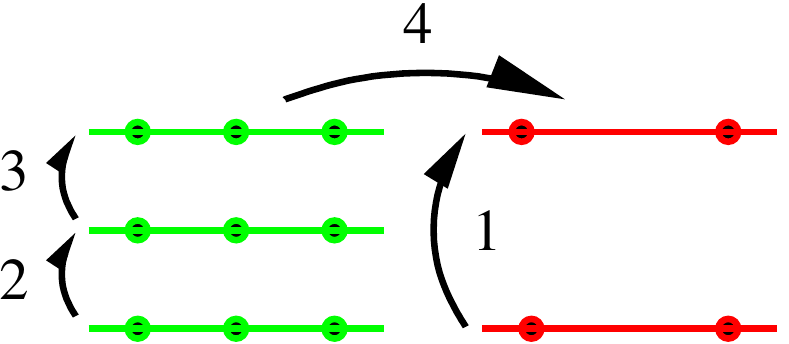}
  \hspace{3em}
  \includegraphics[width=0.3\textwidth]{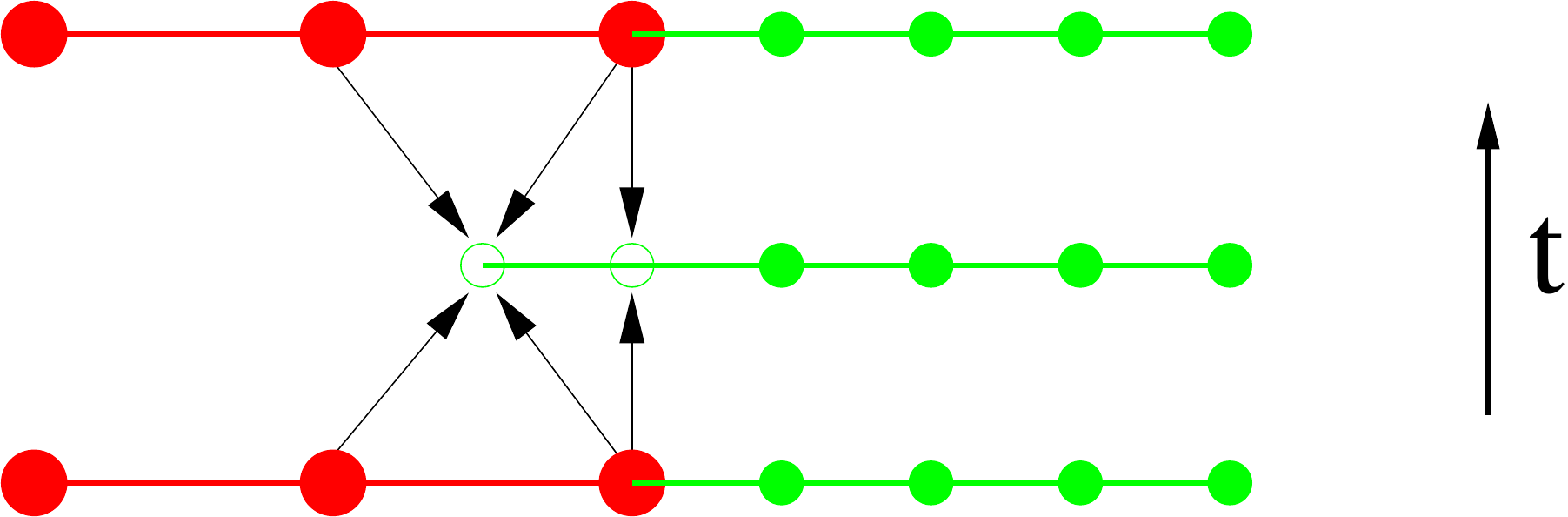}
  \caption{Berger-Oliger time stepping details, showing a coarse and a
    fine grid, with time advancing upwards. \textbf{Left:} Time stepping
    algorithm. First the coarse grid takes a large time step, then the
    refined grid takes two smaller steps. The fine grid solution
    is then injected into the coarse grid where the grids overlap.
    \textbf{Right:} Fine grid boundary conditions. The boundary points
    of the refined grids are filled via interpolation. This may
    require interpolation in space and in time.}
  \label{fig:carpet-details}
\end{figure}

Since a finer grid spacing also requires smaller time steps for
hyperbolic problems, the finer grids perform multiple time steps for
each coarse grid time step, leading to a recursive time evolution
pattern that is typical for Berger-Oliger AMR\@. If a simulation uses
11 levels, then the resolutions (both in space and time) of the
the coarsest and finest levels differ by a factor of $2^{11-1}=1024$. This
non-uniform time stepping leads to a certain complexity that is also
handled by the {\tt Carpet} driver; for example, applying boundary
conditions to a fine level requires interpolation in space and time
from a coarser level. Outputting the solution at a time in between
coarse grid time steps also requires interpolation. These parallel
interpolation operations are implemented efficiently in {\tt Carpet} and
are applied automatically as specified in the execution schedule,
i.e.\ without requiring function calls in user code.
Figure~\ref{fig:carpet-details} describes some details of the
Berger-Oliger time stepping algorithm; more details are described in
\cite{Schnetter:2003rb}.

{\tt Carpet} is the main driver used today for {\tt Cactus}-based astrophysical
simulations. {\tt Carpet} offers hybrid MPI/OpenMP parallelization and is
used in production on up to several thousand cores~\cite{Reisswig:2010cd,Lousto:2010ut}.
\begin{figure}
  \centering
  \includegraphics[width=0.85\textwidth]{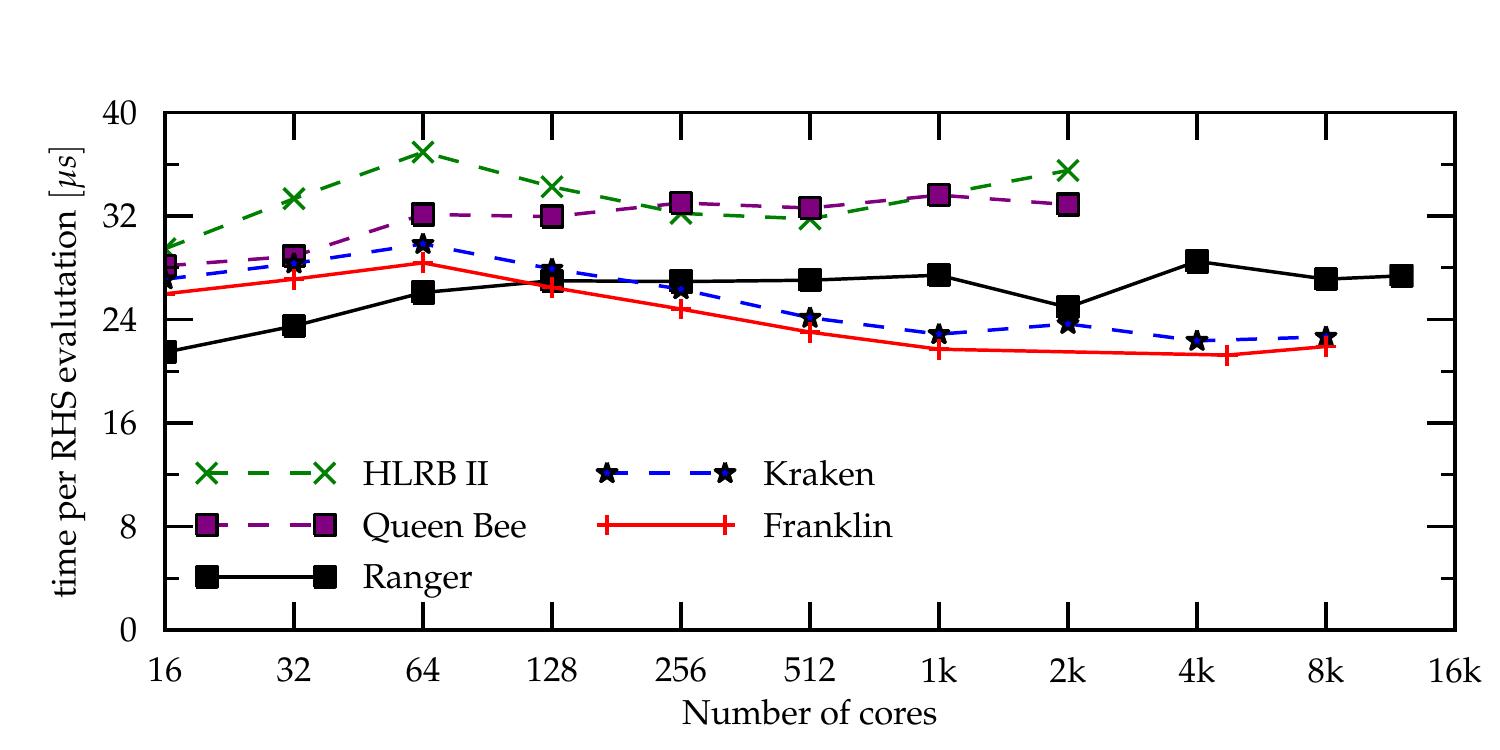}
  \caption{Results from weak scaling tests evolving the Einstein
    equations on a mesh refinement grid structure with nine levels.
    This shows the time required per grid point,
    where smaller numbers are
    better (the ideal scaling is a horizontal line). This
    demonstrates excellent scalability to up to more than 10,000
    cores. Including hydrodynamics approximately doubles
    calculation times without negatively influencing scalability.}
  \label{fig:weak-scaling}
\end{figure}
Figure \ref{fig:weak-scaling} shows a weak scaling test of \texttt{Carpet},
where \texttt{McLachlan} (see section \ref{sec:Kevol} below) solves
the Einstein equations on a grid structure with
nine levels of mesh refinement. This demonstrates excellent
scalability up to more than ten thousand cores. In production
simulations, smaller and more complex grid structures, serial
tasks related to online data analysis and other
necessary tasks reduce scalability by up to a factor of ten.

We estimate that
in 2010, about 7,000 core years of computing time (45 million core
hours) will have been used via {\tt Carpet} by more than a dozen research groups
world-wide. To date, more than 90 peer-reviewed publications and more
than 15 student theses have been based on {\tt Carpet}~\cite{CarpetCode:web}.

\subsection{Simulation Factory}

Today's supercomputers differ significantly in
their hardware configuration, available software, directory structure,
queuing system, queuing policy, and many other user-visible
properties. In addition, the system architectures and user interfaces
offered by supercomputers are very different from those offered by
laptops or workstations. This makes performing large,
three-dimensional time-dependent simulations a complex, time-consuming
and difficult task. However, most of these
differences are only superficial, and the basic capabilities of
supercomputers are very similar; most of the complexity of managing
simulations lies in menial tasks that require no physical or numerical
insight.

The Simulation Factory~\cite{Thomas:2010aa, SimFactory:web} offers a
set of abstractions for the tasks necessary to set up and successfully
complete numerical simulations based on the {\tt Cactus} framework. These
abstractions hide tedious low-level management operations,
capture ``best practices'' of experienced users, and create a log
trail ensuring repeatable and well-documented scientific results.
Using these abstractions, most operations are simplified and many
types of potentially disastrous user errors are avoided, allowing different
supercomputers to be used in a uniform manner.

Using the Simulation Factory, we offer a
tutorial for the Einstein Toolkit~\cite{EinsteinToolkit:web} that teaches
new users how to download, configure, build, and run full simulations of the
coupled Einstein/relativistic hydrodynamics equations on a
supercomputer with a few simple commands. Users need no prior
knowledge about either the details of the architecture of a
supercomputer nor its particular software configuration. 
The same exact set of SimFactory commands can be used
on all other supported supercomputers to run the same simulation
there.

The Simulation Factory supports and simplifies three kinds of
operations:
\begin{description}
\item[1. Remote Access.] The actual access commands and authentication
  methods differ between systems, as do the user names that a person
  has on different systems. Some systems are not directly
  accessible, and one must log in to a particular ``trampoline''
  server first. The Simulation Factory hides this complexity.
\item[2. Configuring and Building.] Building {\tt Cactus} requires certain
  software on the system, such as compilers, libraries, and build
  tools. Many systems offer different versions of these, which may
  be installed in non-default locations. Finding a working combination
  that results in efficient code is extremely tedious and requires
  low-level system experience. The Simulation Factory provides a
  \emph{machine database} that enables users to store and exchange
  this information. In many cases, this allows people to begin to use
  a new machine in a very short time with just a few, simple
  commands.
\item[3. Submitting and Managing Simulations.] Many simulations run for
  days or weeks, requiring frequent checkpointing and job
  re-submission because of short queue run-time limits. Simple user
  errors in these menial tasks can potentially destroy weeks of
  information. The Simulation Factory offers safe commands that
  encapsulate best practices that prevent many common errors and leave
  a log trail.
\end{description}
The above features make running simulations on supercomputers much
safer and simpler.

\subsection{Kranc}
\label{sec:kranc}

\codename{Kranc}\cite{Husa:2004ip,Lechner:2004cs,Kranc:web} is a Mathematica application which converts a high-level
continuum description of a PDE into a highly optimized module for
{\tt Cactus}, suitable for running on anything from a laptop to the world's
largest HPC systems. Many codes contain a large amount of complexity,
including expanded tensorial expressions, numerical methods, and the
large amount of ``glue'' code needed for interfacing a modern HPC
application with the underlying framework.  \codename{Kranc} absorbs this
complexity, allowing the scientist to concentrate on writing only the
\codename{Kranc} script which describes the continuum equations.

This approach brings with it many advantages.  With these complicated
elements factored out, a scientist can write many different \codename{Kranc}
codes, all taking advantage of the features of \codename{Kranc} and avoiding
unnecessary or trivial but painstaking duplication.  The codes might
be variations on a theme, perhaps versions which use different
sets of variables or formulations of the equations, or they could
represent completely different physical systems.  The use of a
symbolic algebra package, Mathematica, enables high-level
optimizations which are not performed by
the compiler to be implemented in \codename{Kranc}.

Any enhancements to \codename{Kranc} can be automatically applied to all codes
which are generated using \codename{Kranc}.  Existing codes have easily benefited
from the following features added to \codename{Kranc} after the codes themselves
were written: (i) OpenMP parallelization support, necessary for
efficient use of modern multi-core processors; (ii) support for
multipatch domains with the Llama~\cite{Pollney:2009yz} code; (iii) automatic generation of
vectorized code, where the equations are evaluated simultaneously by
the processor for two grid points at the same time; and (iv) common
sub-expression elimination, and various other optimization strategies.

Within the Einstein Toolkit, the Einstein evolution thorn \codename{McLachlan},
as well as the wave extraction thorn \codename{WeylScal4}, are both generated
using \codename{Kranc}, and hence support all the above features.

\section{Components}
The Einstein Toolkit uses the modular {\tt Cactus} framework as its underlying infrastructure.
A simulation within {\tt Cactus} could just
use one module, but in practice simulations are often composed from hundreds of components. 
Some of these modules provide common definitions
and conventions (see section~\ref{sec:base_modules}). Others provide
initial data (see section~\ref{sec:initial_data}), which may
be evolved using the different evolution methods for vacuum and matter configurations described in
sections~\ref{sec:Kevol} and~\ref{sec:GRHydro}, respectively.
The thermodynamic properties of fluids are encoded in equations
of state (see section~\ref{sec:eoss}).
Finally, additional quantities which are not directly evolved are often
interesting for a detailed analysis of the simulation's results.
Modules providing commonly used analysis methods are described in
section~\ref{sec:analysis}.

\subsection{Base Modules}
\label{sec:base_modules}
Modular designs have proven to be essential for distributed development
of complex software systems and require the use of well-defined interfaces.
Low-level interoperability within the Einstein Toolkit is provided by
the {\tt Cactus} infrastructure. One example of this is the usage of one module
from within another, e.g., by calling a function within another thorn
independent of programming language used for both the calling and
called function. Solutions for technical challenges like this can be and are
provided by the underlying framework, in this case {\tt Cactus}.

However, certain other standards are very hard or impossible to
enforce on a technical level. Examples for these include the exact
definitions of physical variables, their units, and,  on a more
technical level, the variable names used for the physical quantities.
Even distinct simulation codes typically use very similar
scheduling schemes, so conventions describing the behavior of the scheduler can help
coordinate the order in which functions in different modules
are called.

The Einstein Toolkit provides modules whose sole purpose is to
declare commonly used variables and define their meaning and units.
These conditions are not strictly enforced, but instead documented for the 
convenience of the user community. Three of these base modules, \codename{ADMBase}, \codename{HydroBase}, and \codename{TmunuBase}, 
are described in more detail below.

In the following, we assume that the reader is familiar with the
basics of numerical relativity and GR hydrodynamics, including the underlying differential geometry and tensor analysis. Detailed
introductions to numerical relativity have recently been given by
Alcubierre~\cite{Alcubierre:2008it}, Baumgarte \&
Shapiro~\cite{Baumgarte:2010nu}, and Centrella et al.~\cite{Centrella:2010mx}.
GR hydrodynamics has been reviewed by Font~\cite{Font:2008aa}.  
 We set $G = c
=1$ throughout this paper, and $M_\odot = 1$ where appropriate.

\subsubsection{ADMBase}

The Einstein Toolkit provides code to evolve the Einstein equations
\begin{equation}
G^{\mu\nu} = 8 \pi T^{\mu\nu}\,,
\label{eq:einstein}
\end{equation}
where $G^{\mu\nu}$ is the Einstein tensor, describing the curvature of 4-dimensional spacetime, and $T^{\mu\nu}$  is the stress-energy
tensor.
Relativistic spacetime evolution methods used within the {\tt Cactus} framework employ
different formalisms to accomplish this goal, but essentially all are based on the $3+1$ ADM
construction~\cite{Arnowitt:1962hi}, which makes it the
natural choice of a common foundation for exchange data between
modules using different formalisms. In the $3+1$ approach, 4-dimensional spacetime is foliated into sequences of spacelike
3-dimensional hypersurfaces (slices) connected by timelike normal vectors. The $3+1$ split introduces 4
gauge degrees of freedom: the lapse function $\alpha$ that describes
the advance of proper time with respect to coordinate time for a normal
observer\footnote{A normal observer follows a worldline tangent to the
  unit normal on the 3-hypersurface.} and the shift vector $\beta^i$
that describes how spatial coordinates change from one slice to the
next. 

 According to the ADM formulation, the spacetime metric is assumed to take the form
\begin{equation}
ds^2=g_{\mu\nu}dx^\mu dx^\nu\equiv (-\alpha^2+\beta_i\beta^i)dt^2+2\beta_i dt~dx^i+\gamma_{ij} dx^idx^j,\label{eq:adm}
\end{equation}
where $g_{\mu\nu},~\alpha,~\beta^i$, and $\gamma_{ij}$ are the spacetime 4-metric, lapse function, shift vector, and spatial 3-metric, respectively, and we follow the standard relativistic convention where  Latin letters are used to index 3-dimensional spatial quantities and
Greek letters to index 4-dimensional spacetime quantities, with the index running from 0 to 3.
The remaining dynamical component of the spacetime is contained in the definition of the extrinsic curvature $K_{ij}$, which is defined in terms of the time derivative of the metric after incorporating a Lie derivative with respect to the shift vector:
\begin{equation}
K_{ij}\equiv -\frac{1}{2\alpha}(\partial_t-\mathcal{L}_\beta)\gamma_{ij}.
\end{equation}
The three-metric, extrinsic curvature, lapse function, and shift vector are all
declared as variables in the \codename{ADMBase} module,
the latter two together with their first time derivatives.
The variables provided by {\tt ADMBase} are:
\begin{itemize}
 \item
  The 3-metric tensor, $\gamma_{ij}$:
  {\tt gxx}, {\tt gxy}, {\tt gxz},{\tt gyy}, {\tt gyz}, {\tt gzz}
 \item The extrinsic curvature tensor, $K_{ij}$:
  {\tt kxx}, {\tt kxy}, {\tt kxz}, {\tt kyy}, {\tt kyz}, {\tt kzz}
 \item The lapse function, $\alpha$:
  {\tt alp}
 \item The shift vector $\beta^i$:
  {\tt betax}, {\tt betay}, {\tt betaz}
\end{itemize}

This base module also defines common parameters to manage interaction
between different modules. Examples are the type of requested initial data
or the used evolution method.

The type of initial data chosen for a simulation is specified by the 
parameters {\tt initial\_data} (3-metric and extrinsic curvature),
{\tt initial\_lapse}, {\tt initial\_shift}.  The time derivatives of the gauge variables 
(the lapse and shift) are set by the parameters
{\tt initial\_dtlapse} and
{\tt initial\_dtshift}, respectively.
By default, {\tt ADMBase} initializes the 3-metric and extrinsic
curvature to Minkowski (i.e., $\gamma_{ij}=\delta_{ij}$, the Kronecker delta, and $K_{ij}=0$), the shift to zero,  and the lapse to unity. Initial data thorns
override these defaults by extending the parameters.

Analogous to specifying initial data, evolution methods are chosen by
the parameters {\tt evolution\_method} (3-metric and extrinsic curvature),
{\tt lapse\_evolution\_method}, {\tt shift\_evolution\_method},
{\tt dtlapse\_evolution\_method} and {\tt dtshift\_evolution\_method}.
{\tt ADMBase} does not evolve the
3-metric or extrinsic curvature, and holds the lapse and shift static.
Evolution thorns extend the ranges of these parameters and
contain the evolution code.

The variables defined in ADMBase typically are not used for the actual
evolution of the curvature. Instead,
it is expected that every evolution module converts its internal representation
to the form defined in ADMBase after each evolution step. This procedure
enables modules which perform analysis on the spacetime variables to use
the ADMBase variables without direct dependency on any of the existing
curvature evolution methods.

\subsubsection{HydroBase}
Similar to \codename{ADMBase}, the module \codename{HydroBase} defines a common
basis for interactions between modules of a given evolution problem, in this
case relativistic hydrodynamics.  \codename{HydroBase} extends the Einstein Toolkit to
include an interface within which  magnetohydrodynamics may work. \codename{HydroBase}'s main
function is to store variables which are common to most if not all
hydrodynamics codes solving the Euler equations, the so-called primitive
variables.  These are also the variables which are needed to couple to a
spacetime solver, and often by analysis thorns as well. As with
ADMBase, the usage of a common set of variables by different hydrodynamics
codes creates the possibility of sharing parts of the code, e.g., initial data
solvers or analysis routines.
\codename{HydroBase} also defines commonly needed parameters and schedule
groups for the main functions of a hydrodynamics code.

\codename{HydroBase} uses a set of conventions known as the Valencia
formulation~\cite{Marti:1991wi,Banyuls:1997zz,Ibanez:2001:godunov}.
In particular, \codename{HydroBase} defines the primitive variables (see~\cite{Font:2008aa} for
details):
\begin{itemize}
 \item \verb|rho|: rest mass density $\rho$
 \item \verb|press|: pressure $P$
 \item \verb|eps|: internal energy density $\epsilon$
 \item \verb|vel[3]|: contravariant fluid three velocity $v^i$ defined as
  \begin{equation}
      v^i = \frac{u^i}{\alpha u^0} + \frac{\beta^i}{\alpha}
  \end{equation}
  in terms of the four-velocity $u^\mu$, lapse, and shift vector
.
 \item \verb|Y_e|: electron fraction $Y_e$
 \item \verb|temperature|: temperature $T$
 \item \verb|entropy|: specific entropy per particle $s$
 \item \verb|Bvec[3]|: contravariant magnetic field vector defined as
  \begin{equation}
      B^i = \frac{1}{\sqrt{4\pi}} n_{\nu} F^{*\nu i}
  \end{equation}
  in terms of the dual
  $F^{*\mu\nu} = \frac{1}{2}\varepsilon^{\mu\nu\alpha\beta}F_{\alpha\beta}$
  to the Faraday tensor and the unit normal of the foliation of spacetime
  $n^\mu\equiv \alpha^{-1}[1,-\beta^i]^{\rm T}$.
\end{itemize}

\codename{HydroBase} also sets up scheduling blocks that organize the main functions which modules of a
hydrodynamics code may need. All of those scheduling blocks are optional, but when used
they simplify existing codes and make them more interoperable. \codename{HydroBase} itself does
not schedule routines inside most of the groups that it provides.
Currently the scheduling blocks are:
\begin{itemize}
 \item Initializing the primitive variables
 \item Converting primitive variables to conservative variables
 \item Calculating the right hand side (RHS) in the method of lines (MoL)
 \item Setting and updating an excision mask
 \item Applying boundary conditions
\end{itemize}

Through these, the initiation of the primitive variables, methods to recover the conservative
variables, and basic atmosphere handling can be implemented in different thorns while allowing
a central access point for analysis thorns.

\subsubsection{TmunuBase}
In the Einstein Toolkit, we typically  choose the stress energy tensor $T^{\mu\nu}$ to be that of an ideal relativistic fluid,
\begin{equation}
T^{\mu\nu} = \rho h u^\mu u^\nu - g^{\mu\nu} P\,\,,
\end{equation}
where $\rho$, $u^\mu$, and
$g^{\mu\nu}$ are defined above, and $h = 1 + \epsilon + P/\rho$ is
the relativistic specific enthalpy.

The thorn \codename{TmunuBase} provides grid functions for the stress-energy
tensor $T_{\mu\nu}$ as well as schedule groups to manage when $T_{\mu\nu}$ is
calculated.  In a simulation, many different thorns may contribute to the
stress-energy tensor and this thorn allows them to do so without explicit
interdependence.  The resulting stress-energy tensor can then be used
by the spacetime evolution thorn (again without explicit dependence on any
matter thorns).  When thorn \codename{MoL} is used for time evolution this
provides a high-order spacetime-matter coupling.

The grid functions provided by \codename{TmunuBase} are:
\begin{itemize}
 \item The time component $T_{00}$: {\tt eTtt}
 \item The mixed components $T_{0i}$: {\tt eTtx}, {\tt eTty}, {\tt eTtz}
 \item The spatial components $T_{ij}$: {\tt eTxx}, {\tt eTxy}, {\tt eTxz}, {\tt eTyy}, {\tt eTyz}, {\tt eTzz}
\end{itemize}
In addition, the grid scalar {\tt stress\_energy\_state} has the value 1 if storage is provided for the stress-energy tensor and 0 if not.

Thorn \codename{ADMCoupling} provides a similar (but older) interface between
space-time and matter thorns. However, since it is based on an include file
mechanism it is more complicated to use. We recommend all new matter thorns
to use \codename{TmunuBase} instead.

\subsection{Initial Data}
\label{sec:initial_data}

The Einstein Toolkit contains many modules used to generate initial data for 
GR simulations, including both vacuum and hydrodynamic
configurations.
These include modules used primarily for testing of various components, as
well as physically motivated configurations that 
describe, for example, single or binary BHs and/or NSs.  
Many of the modules
are self-contained, consisting of either all the code to generate exact 
initial solutions or the numerical tools required to construct solutions 
known semi-analytically. Others, though, require the installation of 
numerical software packages that are included in the toolkit as external 
libraries. One example is the \codename{TwoPunctures}
module~\cite{Ansorg:2004ds} --- commonly 
used in numerical relativity to generate BH-BH binary initial data --- which makes
use of the GNU Scientific Library [GSL;~\cite{GSL:web,Galassi:2009}].
Several modules have also been implemented to read in data files generated by
the {\tt Lorene} code~\cite{Lorene:web,Gourgoulhon:2000nn}. 

Initial data setup is in most cases clearly separated from the
evolution that follows. Typically, initial data routines provide the data in terms of the
quantities defined in the Base modules (see section~\ref{sec:base_modules}),
and the evolution modules will later convert these quantities to forms
used for the evolution. For example, an initial data module must supply 
$g_{ij}$, the spatial 3-metric, and $K_{ij}$, the extrinsic curvature.  
The conversion between the physical  metric and extrinsic 
curvature and conformal versions of these is handled solely within evolution modules, which are responsible 
for calculating the conformally related three metric $\tilde{\gamma}_{ij}$, 
the conformal factor $\psi$, the conformal traceless extrinsic curvature 
$\tilde{A}_{ij}$ and the trace of the extrinsic curvature $K$, as well as 
initializing the BSSN variable $\tilde{\Gamma}^i$ should that be the evolution 
formalism chosen (see section~\ref{sec:Kevol} for definitions of these).  Optionally, many initial data modules also supply values 
for the lapse and shift vector and in some cases their time derivatives.
It is important to note, though, that many dynamical calculations 
run better from initial gauge choices set by ansatz rather than those derived 
from stationary approximations that are incompatible with the gauge evolution 
equations.  In particular, conformal thin-sandwich initial data for binaries 
include solutions for the lapse and shift that are frequently replaced by 
simpler analytical values that lead to more circular orbits under standard 
``moving puncture'' gauge conditions (see, e.g., 
\cite{York:1998hy,Etienne:2007jg} and other works).

We turn our attention next to a brief discussion of the capabilities of
the aforementioned modules as well as their implementation.

\subsubsection{Simple Vacuum initial data}

Vacuum spacetime tests in which the constraint equations are explicitly violated  are provided by \codename{IDConstraintViolate} and
\codename{Exact}, a set of exact spacetimes in
various coordinates including Lorentz-boosted versions of traditional solutions.
Vacuum gravitational wave configurations can be obtained by using either
\codename{IDBrillData}, providing a Brill wave spacetime~\cite{Brill:1959zz};
or \codename{IDLinearWaves}, for a spacetime containing a linear gravitational
wave.
Single BH configurations include \codename{IDAnalyticBH} which generates
various analytically known BH configurations; as well as
\codename{IDAxibrillBH}, \codename{IDAxiOddBrillBH}, \codename{DistortedBHIVP}
and \codename{RotatingDBHIVP}, which introduce perturbations to isolated BHs.

\subsubsection{Hydrodynamics Tests}

Initial data to test different parts of hydrodynamics evolution systems are provided 
by \codename{GRHydro\_InitData}.  This module includes several shock tube problems 
that may be evolved in any of the Cartesian directions or diagonally. All of these
have been widely used throughout the field to evaluate a diverse set of solvers 
\cite{Marti:1999wi}.  
Conservative to primitive variable conversion and vice versa are also supported, as are 
tests to check on the reconstruction of hydrodynamical variables at cell faces 
(see Sec.~\ref{sec:GRHydro} for more on this).  Along similar lines, the 
\codename{Hydro\_InitExcision} module sets up masks for different kinds of excised 
regions, which is convenient for hydrodynamics tests.

\subsubsection{TwoPunctures: Binary Black Holes and extensions}\label{sec:twopunctures}
\begin{figure}
 \centering\includegraphics[width=0.5\textwidth]{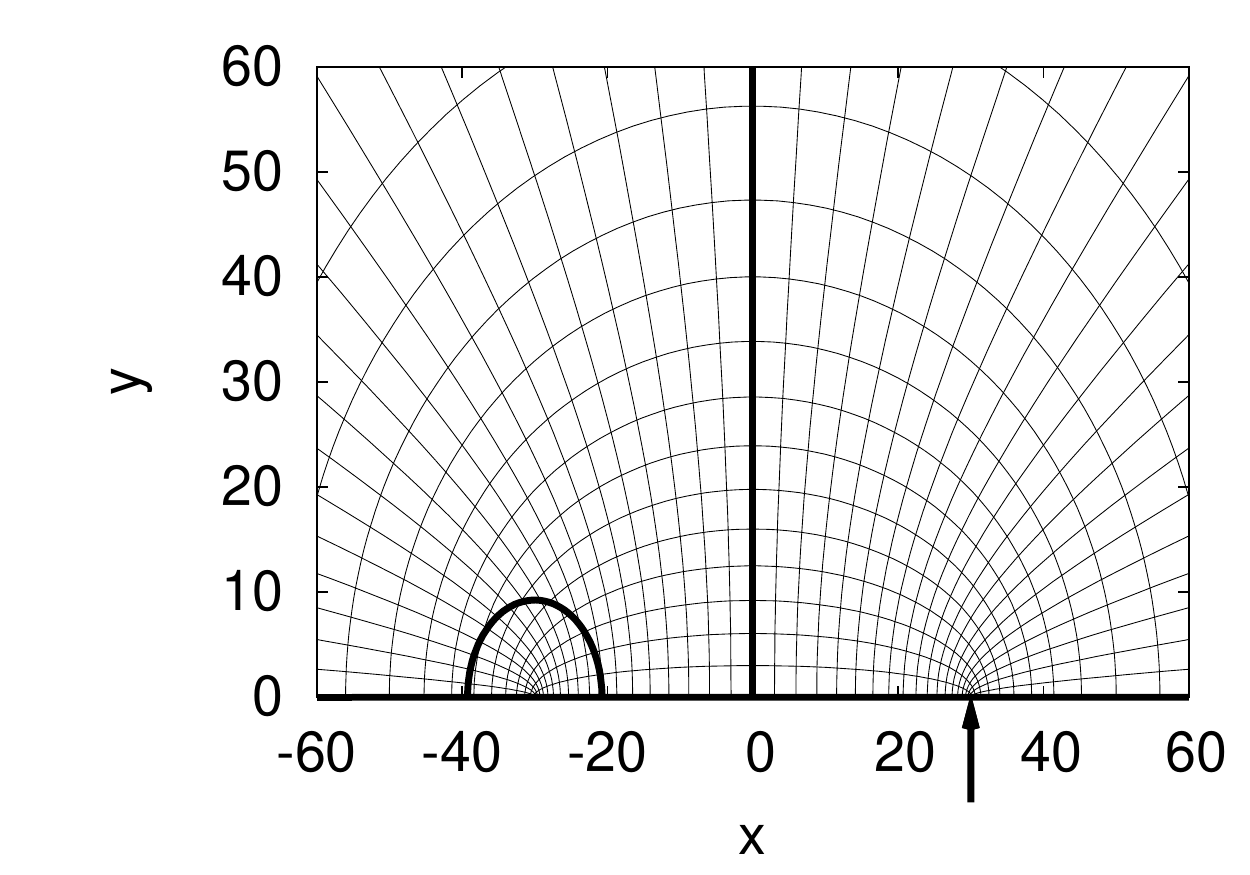}\\
 \caption{Example of a TwoPunctures coordinate system for BH-NS binary initial data}
 \label{fig:TP_BHNS_coordinates}
\end{figure}

A substantial fraction of the published work on the components of the Einstein toolkit 
involves the evolution of BH-BH binary systems.
The most widely used routine to generate initial data for these is the 
\codename{TwoPunctures} code (described originally in~\cite{Ansorg:2004ds}) which solves 
the binary puncture equations for a pair of BHs~\cite{Brandt:1997tf}.
To do so, one assumes the extrinsic curvature for each BH corresponds to 
the Bowen-York form~\cite{Bowen:1980yu}:
\begin{eqnarray}
K_{(m)}^{ij}&=&\frac{3}{2r^2}(p_{(m)}^i\hat{N}^j+p_{(m)}^j\hat{N}^i-(\gamma^{ij}-\hat{N}^i\hat{N}^j)p_{(m)}^k\hat{N}_k))\nonumber\\
&&+\frac{3}{r^3}(\varepsilon^{ikl}S^{(m)}_k\hat{N}_l\hat{N}^j+\varepsilon^{jkl}S^{(m)}_k\hat{N}_l\hat{N}^i),
\end{eqnarray}
where the sub/superscript $(m)$ refers to the contribution from BH $m=1,2$; the 
3-momentum is $p^i$; the BH spin angular momentum is $S_i$; the conformal 3-metric 
$\gamma^{ij}$ is assumed to be flat, i.e.\ $\gamma_{ij}=\eta_{ij}$, and $\hat{N}^i=x^i/r$ 
is the Cartesian normal vector relative to the position of each BH in turn.  This 
solution automatically satisfies the momentum constraint, and the Hamiltonian constraint 
may be written as an elliptic equation for the conformal factor, defined by the condition $g_{ij}=\psi^4\gamma_{ij}$ or equivalently $\psi\equiv (\det|g_{ij}|)^{1/12}$:
\begin{equation}
\Delta \psi+\frac{1}{8}K^{ij}K_{ij}\psi^{-7}=0
\end{equation}
Decomposing the conformal factor into a singular analytical term and a regular term $u$, 
such that
\begin{equation}
\psi = \frac{m_1}{2r_1}+\frac{m_2}{2r_2}+u\equiv \frac{1}{\Psi}+u
\end{equation}
where $m_1,~m_2$ and $r_1,~r_2$ are the mass of and distance to each BH, respectively, and $\Psi$ is defined by the equation itself, the Hamiltonian 
constraint may be written as
\begin{equation}
\Delta u +\left[\frac{1}{8}\Psi^7K^{ij}K_{ij}\right](1+\Psi u)^{-7}\label{eq:twopunc_u}
\end{equation}
subject to the boundary condition $u\rightarrow 1$ as $r\rightarrow\infty$.  In Cartesian 
coordinates, the function $u$ is infinitely differentiable everywhere except the 
two puncture locations.  \codename{TwoPunctures} resolves this problem by performing 
a coordinate transformation modeled on confocal elliptical/hyperbolic coordinates. This
transforms the spatial domain into a finite cube with the puncture locations mapped 
to two parallel edges, as can be seen in figure~\ref{fig:TP_BHNS_coordinates}.  
The coordinate transformation is:
\begin{eqnarray}
x&=&b\frac{A^2+1}{A^2-1}\frac{2B}{1+B^2}\nonumber\\
y&=&b\frac{2A}{1-A^2}\frac{1-B^2}{1+B^2}\cos\phi\nonumber\\
z&=&b\frac{2A}{1-A^2}\frac{1-B^2}{1+B^2}\sin\phi
\end{eqnarray}
which maps $\mathcal{R}^3$ onto $0\le A\le 1$ (the elliptical quasi-radial coordinate), 
$-1\le B\le 1$ (the hyperbolic quasi-latitudinal coordinate), and $0\le\phi<2\pi$ 
(the longitudinal angle).  Since $u$ is smooth everywhere in the interior of the remapped 
domain, expansions into modes in these coordinates are {\em spectrally convergent} and 
thus capable of extremely high accuracy.  In practice, the field is expanded into Chebyshev 
modes in the quasi-radial and quasi-latitudinal coordinates, and into Fourier modes around 
the axis connecting the two BHs.  The elliptic solver uses a stabilized biconjugate gradient 
method to achieve rapid solutions and to avoid ill-conditioning of the spectral matrix.

\subsubsection{Lorene-based binary data}

The ET contains three routines that can read in publicly available data generated 
by the {\tt Lorene} code~\cite{Lorene:web,Gourgoulhon:2000nn}, though it does not 
currently include the capability of generating such data from scratch.  For a 
number of reasons, such functionality is not truly required; in particular, 
{\tt Lorene} is a serial code and to call it as 
an ET initial data generator saves no time.  Also, it is not guaranteed to be convergent for 
an arbitrary set of parameters; thus the initial data routine itself may never 
finish its iterative steps.  Instead, recommended practice is to let Lorene output 
data into files, and then read those into ET at the beginning of a run.

\begin{figure}
 \centering\includegraphics[width=0.5\textwidth]{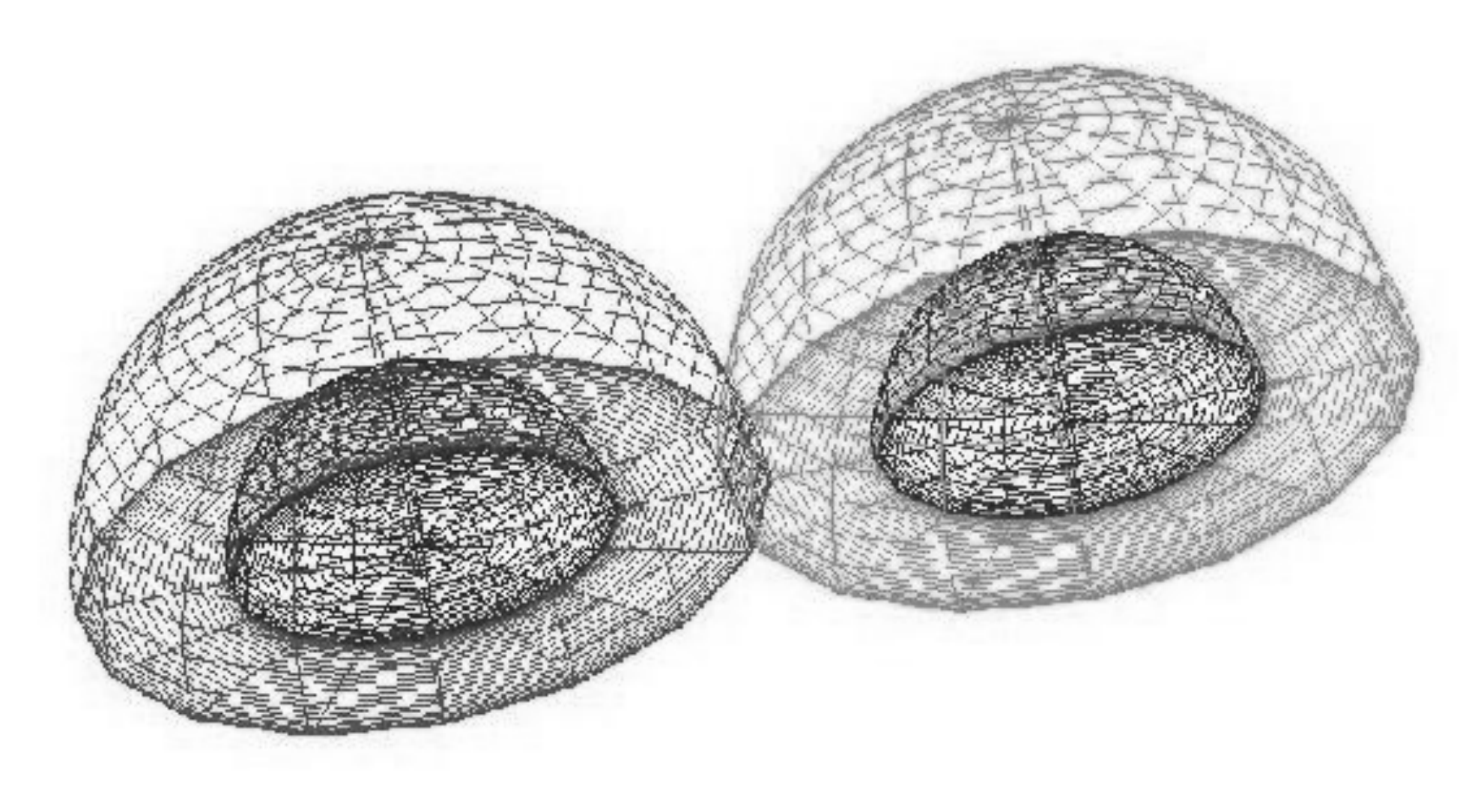}\\
 \caption{Example of a Lorene multi-domain coordinate system for binary initial data.  
The outermost, compactified domain extending to spatial infinity is not shown.}
 \label{fig:Lorene_coordinates}
\end{figure}

Lorene uses a multigrid spectral approach to solve the conformal thin-sandwich 
equations for binary initial configurations~\cite{York:1998hy} and a single-grid 
spectral method for rotating stars.  For binaries, five elliptic equations for 
the shift, lapse, and conformal factor are written down and the source terms 
are divided into pieces that are attributed to each of the two objects.  
Matter source terms are ideal for this split, since they are compactly supported, 
while extrinsic curvature source terms are spatially extended but with sufficiently 
rapid falloff at large radii to yield convergent solutions.  Around each object, 
a set of nested spheroidal sub-domains (see figure~\ref{fig:Lorene_coordinates}) is constructed to extending through all 
of space, with the outermost domain incorporating a compactification to allow 
it to extend to spatial infinity.  Within each of the nested sub-domains, 
fields are decomposed into Chebyshev modes radially and into spherical harmonics 
in the angular directions, with elliptic equation solving reduced to a matrix 
problem.  The nested sub-domains need not be perfectly spherical, and 
indeed one may modify the outer boundaries of each to cover any convex shape.  
For NSs, this allows one to map the surface of a particular sub-domain 
to the NS surface, minimizing Gibbs error there.  For BHs, excision 
boundary conditions are imposed at the horizon.  To read a field solution 
describing a given quantity onto a {\tt Cactus}-based grid, one must incorporate the data
 from each star's  domains at that point.

 \codename{Meudon\_Bin\_BH} can read in BH-BH binary initial data described 
in~\cite{Grandclement:2001ed}, while  \codename{Meudon\_Bin\_NS} 
handles binary NS data from~\cite{Gourgoulhon:2000nn}.  \codename{Meudon\_Mag\_NS} 
may be used to read in magnetized isolated NS data~\cite{Lorene:web}.

\subsubsection{TOVSolver}
\label{sec:TOVSolver}
The \codename{TOVSolver} routine in the ET solves the standard TOV equations 
\cite{Tolman:1939jz,Oppenheimer:1939ne} expressed using the Schwarzschild (areal) 
radius $r$ in the interior of a spherically symmetric star in hydrostatic equilibrium:
\begin{eqnarray}
  \label{eq:TOViso}
  \frac{d P}{d r} & = & -(e + P) \frac{m + 4\pi r^3 P}{r(r - 2m)}\nonumber\\
  \frac{d m}{d r} & = & 4 \pi r^2 e\nonumber\\
  \frac{d \Phi}{d r} & = & \frac{m + 4\pi r^3 P}{r(r -
    2m)},
\end{eqnarray}
where $e\equiv \rho(1+\epsilon)$ is the energy density of the fluid, including the internal energy contribution\footnote[1]{We note that since different application thorns may define their own local variables, the energy density is referred to as {\tt rho} within \codename{TOVSolver}, as the projected energy density $E$, defined in Sec.~\protect\ref{sec:Kevol}, is within \codename{McLachlan} and a few other thorns.  Similar ambiguities exist for other commonly used variable names, particularly $\phi$ and $\alpha$.},  $m$ is the gravitational mass inside a sphere of radius $r$, and
$\Phi$ the logarithm of the lapse.  The routine also supplies the analytically known 
solution in the exterior,
\begin{eqnarray}
     P & = & {\tt TOV\_atmosphere},\nonumber \\
     m & = & M, \nonumber\\
  \Phi & = &\dfrac{1}{2} \log(1-2M / r)
  \label{eq:TOVexterior}
\end{eqnarray}
where {\tt TOV\_atmosphere} is a parameter used to define the density of the 
ambient atmosphere.  Since the isotropic radius $\bar{r}$ is the more 
commonly preferred choice to initiate dynamical calculations, the code then 
transforms all variables into isotropic coordinates, integrating the radius 
conversion formula
\begin{equation}
\label{eq:rbar}
\frac{d (\log(\bar{r} / r))}{\partial r} =  \frac{r^{1/2} - (r-2m)^{1/2}}{r(r-2m)^{1/2}} \ .
\end{equation}
subject to the boundary condition that in the exterior,
\begin{eqnarray}
\bar{r} &=& \dfrac{1}{2}\left(\sqrt{r^2-2Mr}+r -M\right)\nonumber \\
r&=&\bar{r}\left(1+\dfrac{M}{2\bar{r}}\right)^2 \ .
\end{eqnarray}
handling with some care the potentially singular terms that appear at the origin.

To facilitate the construction of stars in more complicated dynamical configurations, 
one may also apply a uniform velocity to the NS, though this does not affect 
the ODE solution nor the resulting density profile.

\subsection{Spacetime Curvature Evolution}
\label{sec:Kevol}
The Einstein Toolkit
curvature evolution code \codename{McLachlan}~\cite{Brown:2008sb,Reisswig:2010cd} is
auto-generated from tensor equations via \codename{Kranc}
(Sec.~\ref{sec:kranc}).  It implements the Einstein equations in a
$3+1$ split as a Cauchy initial boundary value problem
\cite{York:1979sg}. For this, the Baumgarte-Shapiro-Shibata-Nakamura
(BSSN) conformal-tracefree reformulation
\cite{Shibata:1995we,Baumgarte:1998te,Alcubierre:2000xu} of the original
Arnowitt-Deser-Misner (ADM) formalism~\cite{Arnowitt:1962hi} is employed.
\codename{McLachlan} uses fourth-order accurate finite differencing
for the spacetime variables and adds a fifth-order Kreiss-Oliger
dissipation term to remove high frequency noise.
The evolved variables are the conformal factor $\Phi$, the conformal
3-metric $\tilde{\gamma}_{ij}$, the trace $K$ of the extrinsic curvature,
the trace free extrinsic curvature $A_{ij}$ and the conformal connection
functions $\tilde{\Gamma}^i$. These are defined in terms of the
standard ADM 4-metric $g_{ij}$, 3-metric $\gamma_{ij}$, and extrinsic
curvature $K_{ij}$ by:

\begin{eqnarray}
  \phi & \equiv & \log \left[ \frac{1}{12} \det \gamma_{ij} \right]\,,
  \\
  \tilde\gamma_{ij} & \equiv & e^{-4\phi}\; \gamma_{ij}\,,
  \\
  K & \equiv & g^{ij} K_{ij}\,,
  \\
  \tilde A_{ij} & \equiv & e^{-4\phi} \left[ K_{ij} - \frac{1}{3} g_{ij} K\,,
    \right]
  \\
  \tilde\Gamma^i & \equiv & \tilde\gamma^{jk} \tilde\Gamma^i_{jk} .
\end{eqnarray}

The evolution equations are then:
\begin{widetext}
\begin{eqnarray}
  \partial_0 \alpha & = & -\alpha^2 f(\alpha, \phi, x^\mu) (K -
  K_0(x^\mu))
  \\
  \partial_0 K & = & -e^{-4\phi} \left[ \tilde{D}^i \tilde{D}_i \alpha
    + 2 \partial_i \phi \cdot \tilde{D}^i \alpha \right] + \alpha
  \left( \tilde{A}^{ij} \tilde{A}_{ij} + \frac{1}{3} K^2 \right) - \alpha S
  \\
  \partial_0 \beta^i & = & \alpha^2 G(\alpha,\phi,x^\mu) B^i
  \\
  \partial_0 B^i & = & e^{-4\phi} H(\alpha,\phi,x^\mu)
  \partial_0\tilde{\Gamma}^i - \eta^i(B^i,\alpha,x^\mu)
  \\
  \partial_0 \phi & = & -\frac{\alpha}{6}\, K +
  \frac{1}{6}\partial_k\beta^k
  \\
  \partial_0 \tilde{\gamma}_{ij} & = & -2\alpha\tilde{A}_{ij} 
  + 2\tilde{\gamma}_{k(i}\partial_{j)}\beta^k 
  - \frac{2}{3}\tilde{\gamma}_{ij}\partial_k\beta^k
  \\
  \partial_0 \tilde{A}_{ij} & = & e^{-4\phi}\left[ 
    \alpha\tilde{R}_{ij} + \alpha R^\phi_{ij} - \tilde{D}_i\tilde{D}_j\alpha 
    + 4\partial_{(i}\phi\cdot\tilde{D}_{j)}\alpha\right]^{TF}
  \nonumber\\
  & & {} + \alpha K\tilde{A}_{ij} - 2\alpha\tilde{A}_{ik}\tilde{A}^k_{\; j}
  + 2\tilde{A}_{k(i}\partial_{j)}\beta^k 
  - \frac{2}{3}\tilde{A}_{ij}\partial_k\beta^k
  - \alpha e^{-4\phi} \hat{S}_{ij}
  \\
  \partial_0\tilde{\Gamma}^i & = & 
  \tilde{\gamma}^{kl}\partial_k\partial_l\beta^i
  + \frac{1}{3} \tilde{\gamma}^{ij}\partial_j\partial_k\beta^k 
  + \partial_k\tilde{\gamma}^{kj} \cdot \partial_j\beta^i
  - \frac{2}{3}\partial_k\tilde{\gamma}^{ki} \cdot \partial_j\beta^j
  \nonumber\\
  & & {} - 2\tilde{A}^{ij}\partial_j\alpha 
  + 2\alpha\left[ (m-1)\partial_k\tilde{A}^{ki} - \frac{2m}{3}\tilde{D}^i K
  \right. \nonumber \\
  & & {} + m(\tilde{\Gamma}^i_{\; kl}\tilde{A}^{kl} +
    6\tilde{A}^{ij}\partial_j\phi) \Bigg] - S^i,
\end{eqnarray}
\end{widetext}
with the momentum constraint damping constant set to $m=1$. The stress
energy tensor $T_{\mu\nu}$ is incorporated via the projections
\begin{eqnarray}
  E & \equiv & \frac{1}{\alpha^2} \left( T_{00} - 2 \beta^i T_{0i} +
  \beta^i \beta^j T^{ij} \right)
  \\
  S & \equiv & g^{ij} T_{ij}
  \\
  S_i & \equiv & - \frac{1}{\alpha} \left( T_{0i} - \beta^j T_{ij} \right) .
\end{eqnarray}
We have introduced the notation $\partial_0 = \partial_t -
\beta^j\partial_j$. All quantities with a tilde involve
the conformal 3-metric $\tilde{\gamma}_{ij}$, which is used to
raise and lower indices. In particular, $\tilde{D}_i$ and
$\tilde{\Gamma}^k_{ij}$ refer to the covariant derivative and the
Christoffel symbols with respect to $\tilde{\gamma}_{ij}$. The
expression $[ \cdots ]^{TF}$ denotes the trace-free part of the
expression inside the parentheses, and we define the Ricci tensor
contributions as:
\begin{widetext}
\begin{eqnarray}
\tilde{R}_{ij} 
 &=& -\frac{1}{2} \tilde{\gamma}^{kl}\partial_k\partial_l\tilde{\gamma}_{ij} 
  + \tilde{\gamma}_{k(i}\partial_{j)}\tilde{\Gamma}^k
  - \tilde{\Gamma}_{(ij)k}\partial_l\tilde{\gamma}^{lk} 
  + \tilde{\gamma}^{ls}\left( 2\tilde{\Gamma}^k_{\; l(i}\tilde{\Gamma}_{j)ks} 
  + \tilde{\Gamma}^k_{\; is}\tilde{\Gamma}_{klj} \right)
\\
R^\phi_{ij} &=& -2\tilde{D}_i\tilde{D}_j\phi 
  - 2\tilde{\gamma}_{ij} \tilde{D}^k\tilde{D}_k\phi
  + 4\tilde{D}_i\phi\, \tilde{D}_j\phi 
  - 4\tilde{\gamma}_{ij}\tilde{D}^k\phi\, \tilde{D}_k\phi .
\end{eqnarray}
\end{widetext}
This is a so-called $\phi$-variant of BSSN. The evolved gauge
variables are lapse $\alpha$, shift $\beta^i$, and a quantity $B^i$
related to the time derivative of the shift. The gauge parameters $f$,
$G$, $H$, and $\eta$ are determined by our choice of a $1+\log$
\cite{Alcubierre:2002kk} slicing:
\begin{eqnarray}
  f(\alpha,\phi,x^\mu) & \equiv & 2/\alpha
  \\
  K_0(x^\mu) & \equiv & 0
\end{eqnarray}
and $\Gamma$-driver shift condition~\cite{Alcubierre:2002kk}:
\begin{eqnarray}
  G(\alpha,\phi,x^\mu) & \equiv & (3/4)\, \alpha^{-2}
  \\
  H(\alpha,\phi,x^\mu) & \equiv & \exp\{4\phi\}
  \\\label{eq:eta}
  \eta(B^i,\alpha,x^\mu) & \equiv & (1/2)\, B^i q(r) .
\end{eqnarray}
The expression $q(r)$ attenuates the $\Gamma$-driver depending on the
radius as described below.

The $\Gamma$-driver shift condition is symmetry-seeking,
driving the shift $\beta^i$ to a state that renders the conformal 
connection functions $\tilde\Gamma^i$
stationary. Of course, such a stationary state cannot be achieved
while the metric is evolving, but in a stationary spacetime the time
evolution of the shift
$\beta^i$ and thus that of the spatial coordinates $x^i$ will be exponentially
damped. This damping time scale is set by the gauge parameter $\eta$
(see~\eref{eq:eta}) which has dimension $1/T$ (inverse time).
As described in~\cite{Muller:2009jx, Schnetter:2010cz}, this
time scale may need to be adapted in different regions of the domain
to avoid spurious high-frequency behavior in regions that otherwise
evolve only very slowly, e.g., far away from the source.

Here we use the simple damping mechanism described in (12) of
\cite{Schnetter:2010cz}, which is defined as:
\begin{eqnarray}
  \label{eq:varying-simple}
  q(r) & \equiv & \left\{
    \begin{array}{llll}
      1 & \mathrm{for} & r \le R & \textrm{(near the origin)}
      \\
      R/r & \mathrm{for} & r \ge R & \textrm{(far away)}
    \end{array}
    \right.
\end{eqnarray}
with a constant $R$ defining the transition radius between the
interior, where $q\approx1$, and the exterior, where $q$ falls off as
$1/r$.  A description of how $q$ appears in the gauge
parameters may be found in~\eref{eq:eta}.

\subsubsection{Initial Conditions}

Initial conditions for the ADM variables $g_{ij}$, $K_{ij}$, lapse
$\alpha$, and shift $\beta^i$ are provided by the initial data routines
discussed in Sec.~\ref{sec:initial_data}. From these the BSSN
quantities are calculated via their definitions, setting $B^i=0$, and
using cubic extrapolation for $\tilde\Gamma^i$ at the outer
boundary. This extrapolation is necessary since the $\tilde\Gamma^i$
are calculated from derivatives of the metric, and one cannot use
centered finite differencing stencils near the outer boundary. 

The extrapolation stencils distinguish between points on the faces,
edges, and corners of the grid. Points on the faces are extrapolated
via stencils perpendicular to that face, while points on the edges and
corners are extrapolated with stencils aligned with the average of the
normals of the adjoining faces. For example, points on the $(+x,+y)$
edge are extrapolated in the $(1,1,0)$ direction, while points in the
$(+x,+y+z)$ corner are extrapolated in the $(1,1,1)$ direction. Since
several layers of boundary points have to be filled for higher order
schemes (e.g., three layers for a fourth order scheme), one proceeds 
outwards starting from the innermost layer. Each subsequent layer is
then defined via the points in the interior and the previously
calculated layers.

\subsubsection{Boundary Conditions}
\label{sec:curv_boundaries}

During time evolution, a Sommerfeld-type radiative boundary condition
is applied to all components of the evolved BSSN variables as
described in~\cite{Alcubierre:2000xu}. The main feature of this boundary
condition is that it assumes approximate spherical symmetry of the
solution, while applying the actual boundary condition on the boundary
of a cubic grid where the face normals are not aligned with the radial
direction. This boundary condition defines the right hand side of the
BSSN state vector on the outer boundary, which is then integrated in
time as well so that the boundary and interior are calculated with
the same order of accuracy.

The main part of the boundary condition assumes that one has an
outgoing radial wave with some speed $v_0$:
\begin{eqnarray}
  X & = & X_0 + \frac{u(r - v_0 t)}{r},
\end{eqnarray}
where $X$ is any of the tensor components of evolved variables, $X_0$
the value at infinity, and $u$ a spherically symmetric perturbation.
Both $X_0$ and $v_0$ depend on the particular variable and have to be
specified. This implies the following differential equation:
\begin{eqnarray}
  \partial_t X & = & - v^i \partial_i X - v_0\, \frac{X - X_0}{r}\,,
\end{eqnarray}
where $v^i = v_0\, x^i/r$. The spatial derivatives $\partial_i$ are
evaluated using centered finite differencing where possible, and
one-sided finite differencing elsewhere.  Second order stencils
are used in the current implementation.

In addition to this main part, it is also necessary to account for
those parts of the solution that do not behave as a pure wave, e.g.,
Coulomb type terms caused by infall of the coordinate lines. It is
assumed that these parts decay with a certain power $p$ of
the radius. This is implemented by considering the radial derivative of
the source term above, and extrapolating according to this power-law
decay.

Given a source term $(\partial_t X)$, one defines the corrected source
term $(\partial_t X)^*$ via
\begin{eqnarray}
  (\partial_t X)^* & = & (\partial_t X) + \left( \frac{r}{r - n^i
    \partial_i r} \right)^p\; n^i \partial_i (\partial_t X)\,,
\end{eqnarray}
where $n^i$ is the normal vector of the corresponding boundary face.
The spatial derivatives $\partial_i$ are evaluated by comparing
neighboring grid points, corresponding to a second-order stencil
evaluated in the middle between the two neighboring grid points. 
Second-order decay is assumed, hence $p=2$.

As with the initial conditions above, this boundary condition is
evaluated on several layers of grid points, starting from the
innermost layer. Both the extrapolation and radiative boundary
condition algorithms are implemented in the publicly available
\texttt{NewRad} component of the Einstein Toolkit.

This boundary condition is only a coarse approximation of the actual
decay behavior of the BSSN state vector, and it does not capture the
correct behavior of the evolved variables. However, one finds that
this boundary condition leads to stable evolutions if applied
sufficiently far from the source. Errors introduced at the boundary
(both errors in the geometry and constraint violations) propagate
inwards with the speed of light~\cite{Brown:2008sb}. Gauge changes
introduced by the boundary condition, which are physically not
observable, propagate faster, with a speed up to $\sqrt{2}$ for the
 gauge conditions used in \codename{McLachlan}.

\subsection{Hydrodynamics Evolution}
\label{sec:GRHydro}

Hydrodynamic evolution in the Einstein Toolkit is designed so that it
interacts with the metric curvature evolution through a small set of
variables, allowing for maximum modularity in implementing, editing,
or replacing either evolution scheme.

The primary hydrodynamics evolution routine in the Einstein Toolkit is
\codename{GRHydro}, a code derived from the public \codename{Whisky}
code~\cite{Baiotti:2004wn,Hawke:2005zw,Baiotti:2010zf,Whisky:web} 
designed primarily by researchers at AEI and their collaborators.  It
includes a high resolution shock capturing (HRSC) scheme to evolve
hydrodynamic quantities, with several different reconstruction methods
and Riemann solvers, as we discuss below.  In such a scheme, we define
a set of ``conserved'' hydrodynamic variables, defined in terms of the
``primitive'' physical variables such as mass and internal energy
density, pressure, and velocity.  Wherever derivatives of hydrodynamic
terms appear in the evolution equations for the conserved variables,
they are restricted to appear only inside divergence terms
(referred to as fluxes) and never in the source terms.  By calculating fluxes at cell faces, we may
obtain a consistent description of the inter-cell values using
reconstruction techniques that account for the fact that hydrodynamic
variables are not smooth and may not be finite differenced accurately.
All other source terms in the evolution equations may contain only the
hydrodynamic variables themselves and the metric variables and derivatives of the latter, since the metric must formally be smooth and thus
differentiable using finite differencing techniques.  Summarizing
these methods briefly, the following stages occur every timestep:
\begin{itemize}
\item The primitive variables are ``reconstructed'' at cell faces
  using shock-capturing techniques, with total variation diminishing
  (TVD), piecewise parabolic (PPM), and essentially non-oscillatory
  (ENO) methods currently implemented.
\item A Riemann problem is solved at each cell face using an
  approximate solver.  Currently implemented versions include HLLE
  (Harten-Lax-van Leer-Einfeldt), Roe, and Marquina solvers.
\item The conserved variables are advanced one timestep, and used to
  recalculate the new values of the primitive variables.
\end{itemize}
We discuss the GRHD formalism,  the stages within a timestep, and the 
other aspects of the code below, noting that the documentation included 
in the released version is quite extensive and covers many of these topics 
in substantially more detail.

\subsubsection{Ideal general relativistic hydrodynamics (GRHD)}
The equations of ideal GR hydrodynamics evolved by \codename{GRHydro} are
derived from the local GR conservation laws of mass and
energy-momentum:
\begin{equation}
  \nabla_{\!\mu} J^\mu = 0, \qquad \nabla_{\!\mu} T^{\mu \nu} = 0\,\,,
  \label{eq:equations_of_motion_gr}
\end{equation}
where $ \nabla_{\!\mu} $ denotes the covariant derivative with respect
to the 4-metric, and $ J^{\,\mu} = \rho u^{\,\mu} $ is the mass current.

The 3-velocity $v^i$ may be calculated in the form
\begin{equation}
v^i = \frac{u^i}{W} + \frac{\beta^i}{\alpha}\,\,,
\label{eq:vel}
\end{equation}
where $W = (1-v^i v_i)^{-1/2}$ is the Lorentz factor.  The contravariant 4-velocity is then given by:
\begin{equation}
u^0  = \frac{W}{\alpha}\,,\qquad
u^i = W \left( v^i - \frac{\beta^i}{\alpha}\right)\,\,,
\end{equation}
and the covariant 4-velocity is:
\begin{equation}
u_0  = W(v^i \beta_i - \alpha)\,,\qquad
u_i = W v_i\,\,.
\end{equation}

The {\tt GRHydro} evolution scheme is a first-order hyperbolic
flux-conservative system for the conserved variables
$D$, $S^i$, and $\tau$, which may be defined in terms of the primitive
variables $\rho, \epsilon, v^i$, such that:
\begin{eqnarray}
  D &=& \sqrt{\gamma} \rho W,\label{eq:p2c1}\\
  S^i &=& \sqrt{\gamma} \rho h W^{\,2} v^i,\label{eq:p2c2}\\
  \tau &=& \sqrt{\gamma} \left(\rho h W^{\,2} - P\right) - D\label{eq:p2c3}\,,
\end{eqnarray}
where $ \gamma $ is the determinant of $\gamma_{ij} $.
The evolution system then becomes
\begin{equation}
  \frac{\partial \mathbf{U}}{\partial t} +
  \frac{\partial \mathbf{F}^{\,i}}{\partial x^{\,i}} =
  \mathbf{S}\,\,,
  \label{eq:conservation_equations_gr}
\end{equation}
with
\begin{eqnarray}
  \mathbf{U} & = & [D, S_j, \tau], \nonumber\\
  \mathbf{F}^{\,i} & = & \alpha
  \left[ D \tilde{v}^{\,i}, S_j \tilde{v}^{\,i} + \delta^{\,i}_j P,
  \tau \tilde{v}^{\,i} + P v^{\,i} \right]\!, \nonumber \\
  \mathbf{S} & = & \alpha
  \bigg[ 0, T^{\mu \nu} \left( \frac{\partial g_{\nu j}}{\partial x^{\,\mu}} - 
  \Gamma^{\,\lambda}_{\mu \nu} g_{\lambda j} \right), \nonumber\\
  & &\qquad\alpha \left( T^{\mu 0}
  \frac{\partial \ln \alpha}{\partial x^{\,\mu}} -
  T^{\mu \nu} \Gamma^{\,0}_{\mu \nu} \right) \bigg]\,.
\end{eqnarray}%
Here, $ \tilde{v}^{\,i} = v^{\,i} - \beta^i / \alpha $ and $
\Gamma^{\,\lambda}_{\mu \nu} $ are the 4-Christoffel symbols.  
 The time integration and coupling with curvature
are carried out with the Method of
Lines~\cite{Hyman:1976cm}.  The expressions for 
$\mathbf{S}$ are calculated in \codename{GRHydro} by using the 
definition of the extrinsic curvature to avoid any time derivatives 
whatsoever, as discussed in detail in the code's documentation, 
following a suggestion by Mark Miller based on experience with the 
\codename{GR3D} code.

\subsubsection{Reconstruction techniques}

In order to calculate fluxes at cell faces, we first must calculate 
function values on either side of the face.  In practice, reconstructing 
the primitive variables yields more stable and accurate evolutions 
than reconstructing the conservatives.  In what follows, we assume 
a Cartesian grid and loop over faces along each direction in turn.  
We define $q^L_{i+1/2}$ to be the value of a quantity $q$ on the 
left side of the face between $q_i\equiv q(x_i,y,z)$ and 
$q_{i+1}\equiv q(x_{i+1},y,z)$, where $x_i$ is the $i$th point 
in the $x$-direction, and $q^R_{i+1/2}$ the right side of the same face.  

For total variation diminishing (TVD) methods, we let:
\begin{equation}
q^L_{i+1/2} = q_i+\frac{f(q_i)\Delta x}{2};~~q^R_{i+1/2} = q_{i+1}-\frac{f(q_{i+1})\Delta x}{2}
\end{equation}
where $f(q_i)$ is a slope-limited gradient function, typically determined 
by the values of $q_{i+1}-q_i$ and $q_i-q_{i-1}$, with a variety of 
different forms of the slope limiter available.  In practice, all 
try to accomplish the same task of preserving monotonicity and removing 
the possibility of spuriously creating local extrema.  Implemented methods 
include minmod, superbee~\cite{Roe:1986cb}, and monotonized central~\cite{vanLeer:1977aa}.

The piecewise parabolic method (PPM) is a multi-step method based around 
a quadratic fit to nearby points interpolated to cell faces 
\cite{Colella:1982ee}, for which $q^L$ and $q^R$ are generally equivalent 
except near shocks and local extrema.  The version implemented in 
\codename{GRHydro} includes the steepening and flattening routines described 
in the original PPM papers, with a simplified dissipation procedure.
Essentially non-oscillatory (ENO) methods use a divided differences 
approach to achieve third-order accuracy via polynomial interpolation 
\cite{Harten:1987un,Shu:1999ho}.  

Both ENO and PPM yield third-order accuracy for smooth monotonic functions, 
whereas TVD methods typically yield second-order accurate values.  Regardless 
of the reconstruction scheme chosen, all of these methods reduce 
to first order near local extrema and shocks.

\subsubsection{Riemann solvers}

The Riemann problem involves the solution of the equation 
\begin{equation}
\partial_t q+\partial_i f^i(q)=0\label{eq:Riemann}
\end{equation}
at some point $X$ representing a discontinuity between constant states.  
The exact solution can be quite complicated, involving five different waves 
with different characteristic speeds for a hydrodynamic problem (8 for GRMHD), 
so \codename{GRHydro} implements three different approximate solvers 
to promote computational efficiency. In each case, the solution takes a 
self-similar form $q(\xi)$, where $\xi\equiv (x-X)/t$ represents 
the characteristic speed from the original shock location to the point 
in question in space and time.

The simplest method implemented is the Harten-Lax-van Leer-Einfeldt  
solver~\cite{Harten:1983on,Einfeldt:1988og} (HLL or HLLE, depending on the reference), 
which uses a two wave approximation to calculate the evolution along 
the shock front.  With $\xi_-$ and $\xi_+$ the most negative and 
most positive wave speeds present on either side of the interface, 
the solution $q(\xi)$ is assumed to take the form
\begin{equation}
  \label{hlle1}
  q = \left\{ \begin{array}[c]{r c l} q^L & {\rm if} & \xi
        < \xi_- \\  q_* & {\rm if} & \xi_- < \xi < \xi_+ \\
        q^R & {\rm if} & \xi   > \xi_+, \end{array}\right. 
\end{equation}
\noindent with the intermediate state $q_*$ given by
\begin{equation}
  \label{hlle2}
  q_* = \frac{\xi_+ q^R - \xi_- q^L - f(
  q^R) + f(q^L)}{\xi_+ - \xi_-}.
\end{equation}
\noindent The numerical flux along the interface
takes the form
\begin{equation}
  \label{hlleflux}
  f(q) = \frac{\widehat{\xi}_+f(q^L) -
  \widehat{\xi}_-f(q^R) + \widehat{\xi}_+ \widehat{\xi}_-
  (q^R - q^L)}{\widehat{\xi}_+ - \widehat{\xi}_-},
\end{equation}
\noindent where
\begin{equation}
  \label{hlle3}
  \widehat{\xi}_- = {\rm min}(0, \xi_-), \quad \widehat{\xi}_+ =
  {\rm max}(0, \xi_+). 
\end{equation}
It is these flux terms that are then used to evolve the hydrodynamic quantities.

The Roe solver~\cite{Roe:1981ar} involves linearizing the evolution system 
for the hydrodynamic evolution, defining 
the Jacobian matrix $A\equiv \frac{\partial f}{\partial q}$ (see~\eref{eq:Riemann}), and 
working out the eigenvalues $\lambda^i$ and left and right eigenvectors,  
$l_i$ and $r^j$, assumed to be orthonormalized so that 
$l_i\cdot r^j=\delta_i^j$.  Defining the characteristic variables 
$w_i=l_i\cdot q$, the characteristic equation becomes 
\begin{equation}
\partial_t w+\Lambda \partial_x w=0
\end{equation}
with $\Lambda$ the diagonal matrix of eigenvalues.  Letting 
$\Delta w_i\equiv w_i^L-w_i^R=l_i\cdot (q^L-q^R)$ represent the differences 
in the characteristic variables across the interface, the Roe flux 
is calculated as 
\begin{equation}
f(q)=\frac{1}{2}\left(f(q^L)+f(q^R)-\sum |\lambda_i| \Delta w_i r^i\right)
\end{equation}
where the eigenvector appearing in the summed term are evaluated for 
the approximate Roe average flux $q_{\rm Roe}=\frac{1}{2}(q^L+q^R)$.  
The Marquina flux routines use a similar approach to the Roe method, 
but provide a more accurate treatment for supersonic flows (i.e., those 
for which the characteristic wave with $\xi=0$ is within a rarefaction zone) 
\cite{Donat:1996cs,Aloy:1999ne}.

\subsubsection{Conservative to primitive conversion}

In order to invert eqs.~\eref{eq:p2c1}~--~\eref{eq:p2c3}, solving for 
the primitive variables based on the values of the conservative ones, 
\codename{GRHydro} uses a 1-dimensional Newton-Raphson approach that 
solves for a consistent value of the pressure.   Defining the (known) 
undensitized conservative variables $\hat{D}\equiv D/\sqrt{\gamma} = 
\rho W$, $\hat{S}^i=S^i/\sqrt{\gamma} = \rho h W^2 v^i$ and 
$\hat{\tau}\equiv \tau/\sqrt{\gamma} = \rho h W^2-P-\hat{D}$, 
as well as the auxiliary quantities $Q\equiv \rho h W^2 = \hat{\tau}+\hat{D}+P$
and $\hat{S}^2 = \gamma_{ij}\hat{S}^i\hat{S}^j = (\rho h W)^2(W^2-1)$, 
the former of which depends on $P$ and the latter of which is known, 
we find that $\sqrt{Q^2-\hat{S}^2} = \rho h W$ and thus
\begin{eqnarray}
\rho&=&\frac{\hat{D}\sqrt{Q^2-\hat{S}^2}}{Q}\\
W&=&\frac{Q}{\sqrt{Q^2-\hat{S}^2}}\\
\epsilon&=&\frac{\sqrt{Q^2-\hat{S}^2}-PW-\hat{D}}{\hat{D}}.
\end{eqnarray}
Given the new values of $\rho$ and $\epsilon$, one may then find 
the residual between the pressure and $P(\rho,\epsilon)$ and perform 
the Newton-Raphson step, so long as the values of 
$\frac{\partial P}{\partial \rho}$ and 
$\frac{\partial P}{\partial\epsilon}$ are known.

\subsubsection{Atmospheres, boundaries, and other code details}

\codename{GRHydro} uses an atmosphere, or extremely-low density floor, 
to avoid problems involving sound speeds and conservative-to-primitive 
variable conversion near the edges of matter distributions.  
The floor density value may be chosen in either absolute 
(\codename{rho\_abs\_min}) or relative (\codename{rho\_rel\_min}) terms.  
The atmosphere is generally assumed to have a specified polytropic EOS, 
regardless of the EOS describing the rest of the matter within the simulation.  
Whenever the numerical evolution results in a grid cell where conservative 
to primitive variable conversion yields negative values of either $\rho$ 
or $\epsilon$, the cell is reassigned to the atmosphere, with zero velocity.

At present, only flat boundary conditions are supported for hydrodynamic 
variables, since it is generally recommended that the outer boundaries 
of the simulation be placed far enough away so that all cells near 
the edge of the computational domain represent the atmosphere.

\codename{GRHydro} has the ability to advect a set of passive scalars, 
referred to as ``tracers'', as well as the electron fraction of a fluid, 
under the assumption that each tracer $X$ follows the conservation law 
\begin{equation}
\partial_t (DX)+\partial_i(\alpha \tilde{v}^iDX)=0.
\end{equation}

\subsection{Equations of State}
\label{sec:eoss}

An equation of state connecting the primitive state variables is
needed to close the system of GR hydrodynamics equations.  The module
\codename{EOS\_Omni} provides a unified general equation of state
(EOS) interface and back-end for simple analytic and complex
microphysical EOSs. 

The polytropic EOS
\begin{equation}
P = K\rho^\Gamma\,\,,
\end{equation}
where $K$ is the polytropic constant
and $\Gamma$ is the adiabatic
index, is appropriate for adiabatic (= isentropic) evolution without
shocks. When using the polytropic EOS, one does not need to evolve the
total fluid energy equation, since the specific internal energy
$\epsilon$ is fixed to
\begin{equation}
\epsilon = \frac{K\rho^\Gamma}{(\Gamma - 1)\rho}\,.
\end{equation} 
Note that the adiabatic index $\Gamma = d\ln{P}/
d\ln{\rho}$ is related to the frequently used polytropic index $n$ via
$n = 1 / (\Gamma - 1)$.

The gamma-law EOS\footnote{For historic reasons, this EOS is referred to
as the ``ideal fluid'' EOS in \codename{GRHydro}.},
\begin{equation}
P = (\Gamma - 1) \rho \epsilon\,\,,
\end{equation}
allows for non-adiabatic flow but still assumes fixed microphysics, which
is encapsulated in the constant adiabatic index $\Gamma$. This EOS has been
used extensively in simulations of NS-NS and BH-NS mergers.

The hybrid EOS, first introduced by~\cite{Janka:1993da}, is a
2-piecewise polytropic with a thermal component designed for the
application in simple models of stellar collapse. At densities below
nuclear density, a polytropic EOS with $\Gamma = \Gamma_1 \approx 4/3$ is
used.  To mimic the stiffening of the nuclear EOS at nuclear density,
the low-density polytrope is fitted to a second polytrope with 
$\Gamma = \Gamma_2 \gtrsim 2$. To allow for thermal contributions to the
pressure due to shock heating, a gamma-law with $\Gamma = \Gamma_\mathrm{th}$
is used. The full EOS then reads
\begin{eqnarray}
  P & = & \frac{\Gamma - \Gamma_{\rm th}}{\Gamma - 1}
  K \rho_{\rm nuc}^{\Gamma_1 - \Gamma}
  \rho^{\Gamma} - \frac{(\Gamma_{\rm th} - 1) (\Gamma - \Gamma_1)}
  {(\Gamma_1 - 1) (\Gamma_2 - 1)}
  K \rho_{\rm nuc}^{\Gamma_1 - 1} \rho \nonumber \\
  & & + (\Gamma_{\rm th} - 1) \rho \epsilon\,.
  \label{eq:hybrid_eos}
\end{eqnarray}%
In this,  the total specific internal energy $\epsilon$
consists of a polytropic and a thermal contribution. In iron core
collapse, the pressure below nuclear density is dominated by the
pressure of relativistically degenerate electrons. For this, one sets
$K = 4.897 \times 10^{14}$ [cgs] in the above. The thermal index
$\Gamma_{\rm th}$ is usually set to $1.5$, corresponding to a mixture
of relativistic ($\Gamma=4/3$) and non-relativistic ($\Gamma=5/3$)
gas. Provided appropriate choices of EOS parameters (e.g.,
\cite{Dimmelmeier:2007ui}), the hybrid EOS leads to qualitatively correct
collapse and bounce dynamics in stellar collapse.

\codename{EOS\_Omni} also integrates the \codename{nuc\_eos} driver
routine, which was first developed for the \codename{GR1D} code
\cite{O'Connor:2009vw} for tabulated microphysical finite-temperature EOS
which assume nuclear statistical equilibrium (NSE). \codename{nuc\_eos}
handles EOS tables in \codename{HDF5} format which contain entries for
thermodynamic variables $X = X(\rho,T,Y_e)$, where $T$ is the matter
temperature and $Y_e$ is the electron fraction.  \codename{nuc\_eos}
also supports calls for $X = X(\rho,\epsilon,Y_e)$ and carries out a
Newton iteration to find $T(\rho,\epsilon,Y_e)$.  For performance
reasons, \codename{nuc\_eos} employs simple tri-linear interpolation
in thermodynamic quantities and thus requires finely spaced tables to
maintain thermodynamic consistency at an acceptable level. EOS tables
in the format required by \codename{nuc\_eos} are freely available
from {\tt http://stellarcollapse.org}.

\subsection{Analysis}
\label{sec:analysis}
It is often beneficial and sometimes necessary to evaluate analysis quantities
during the simulation rather than post-processing variable output. Beyond
extracting physics, these quantities are often used as measures of how
accurately the simulation is progressing. In the following, we describe the 
common quantities available through Einstein Toolkit modules, and how different
modules approach these quantities with differing assumptions and algorithms.
The most common analysis quantities provided fall broadly into several different
categories, including horizons, masses and momenta, and gravitational waves. Note that several
modules bridge these categories and some fall outside them,
including routines to perform constraint monitoring and to compute commonly
used derived spacetime quantities. The following discussion is meant as an
overview of the most common tools rather than an exhaustive list of the
functionality provided by the Einstein Toolkit. In most cases, the analysis modules 
work on the variables stored in the base modules discussed in 
Sec.~\ref{sec:base_modules}, \codename{ADMBase}, \codename{TmunuBase}, 
and \codename{HydroBase}, to be as portable as possible.

\subsubsection{Horizons} 
When spacetimes contain a BH, localizing its horizon is necessary
for describing time-dependent quasi-local measures  such as its
mass and spin.  The Einstein Toolkit provides two modules --- \codename{AHFinder}
and \codename{AHFinderDirect} --- for locating the \ahz{s} defined locally on a
hypersurface.  The module \codename{EHFinder} is also available to search an
evolved spacetime for the globally defined event horizons.

\codename{EHFinder}~\cite{Diener:2003jc} 
evolves a null surface backwards in time given an initial guess (e.g.,
the last apparent horizon) which will, in the vicinity of an event 
horizon, converge exponentially to its location. This must be 
done after a simulation has already evolved the initial data forward 
in time with enough 3D data written out that the full 4-metric can be 
recovered at each timestep.

In \codename{EHFinder}, the null surface is represented by a function
$f(t,x^i)=0$ which is required to satisfy the null condition
$g^{\alpha\beta} \partial_\alpha f \partial_\beta f = 0$. In the 
standard numerical 3+1 form of the metric, this null condition can be 
expanded out into an evolution equation for $f$ as
\begin{equation}
   \partial_t f = \beta^i \partial_i f - \sqrt{\alpha^2 \gamma^{ij} 
	\partial_i f \partial_j f}
\end{equation}
where the roots are chosen to describe outgoing null geodesics.  The 
function $f$ is chosen such that it is negative within the initial 
guess of the horizon and positive outside, initially set to a distance
measure from the initial surface guess 
$f(t_0,x^i)=\sqrt{(x^i-x^i_0)(x_i-x_{i(0)})}-r_0$.
There is a numerical problem with the steepening of $\nabla f$ during
the evolution, so the function $f$ is regularly re-initialized during 
the evolution to satisfy $|\nabla f|\simeq1$. This is done by evolving 
\begin{equation}
   \frac{df}{d\lambda} = -\frac{f}{\sqrt{f^2+1}}\left(|\nabla f|-1\right)
\end{equation}
for some unphysical parameter $\lambda$ until a steady state has been
reached.  As the isosurface $f=0$ converges exponentially to the event
horizon, it is useful to evolve two such null surfaces which bracket
the approximate position of the anticipated event horizon to further 
narrow the region containing the event horizon.

However, event horizons can only be found after the full spacetime
has been evolved.  It is often useful to know the positions
and shapes of any BH on a given hypersurface for purposes such as 
excision, accretion, and local measures of its mass and spin.  The
Einstein Toolkit provides several algorithms of varying speed and 
accuracy to find marginally trapped surfaces, of which the outermost are 
\ahz{s}. All finders make use of the fact that null geodesics have 
vanishing expansion on an \ahz{} which, in the usual 3+1 quantities,
can be written
\begin{equation} \label{eq:ah_theta}
  \Theta \equiv \nabla_i n^i + K_{ij} n^i n^j - K = 0
\end{equation}
where $n^i$ is the unit outgoing normal to the 2-surface.

The module \codename{AHFinder} provides two algorithms for locating
\ahz{s}.  The minimization algorithm~\cite{Anninos:1996ez} finds the local
minimum of $\oint_S (\Theta - \Theta_o )^2 d^2S$ corresponding to a
surface of constant expansion $\Theta_o$, with $\Theta_o=0$
corresponding to the \ahz{.} For time-symmetric data, the option exists
to find instead the minimum of the surface area, which in this case 
corresponds to an \ahz{.} An alternative algorithm provided by
\codename{AHFinder}, the flow algorithm~\cite{Gundlach:1997us}, on 
which  \codename{EHFinder} is also based.  
Defining a surface as a level set $f(x^i)=r-h(\theta,\phi)=0$,
and introducing an unphysical timelike parameter $\lambda$ to
parametrize the flow of $h$ towards a solution, \eref{eq:ah_theta}
can be rewritten 
\begin{equation}
  \partial_\lambda h = - \left( \frac{\alpha}{\ell_\mathrm{max}
     (\ell_\mathrm{max}+1)} + \beta \right) \left( 1 -
     \frac{\beta}{\alpha} L^2\right)^{-1} \rho \Theta
\end{equation}
where $\rho$ is a strictly positive weight, $L^2$ is the Laplacian
of the 2D metric, and $\alpha$, $\beta$, and $\ell_\mathrm{max}$ are
free parameters.  Decomposing $h(\theta,\phi)$ onto a basis of spherical
harmonics, the coefficients $a_{\ell m}$ evolve iteratively towards a
solution as
\begin{equation}
   a_{\ell m}^{(n+1)} = a_{\ell m}^{(n)} - 
     \frac{\alpha + \beta \ell_\mathrm{max}
           \left(\ell_\mathrm{max}+1\right) }
          {\ell_\mathrm{max}\left(\ell_\mathrm{max}+1\right)
           \left(1+\beta\ell(\ell+1)/\alpha\right)} 
     \left(\rho\Theta\right)_{\ell m}^{(n)}
\end{equation}

The \codename{AHFinderDirect} module~\cite{Thornburg:2003sf} is a 
faster alternative to \codename{AHFinder}.  Its approach is to 
view~\eref{eq:ah_theta} as an elliptic PDE for $h(\theta,\phi)$ on $S^2$
using standard finite differencing methods. Rewriting~\eref{eq:ah_theta}
in the form
\begin{equation}
  \Theta \equiv \Theta\left(h,\partial_u h,\partial_{uv}h;
\gamma_{ij},K_{ij},\partial_k \gamma_{ij}\right) = 0\,,
\end{equation}
the expansion $\Theta$ is evaluated on a trial surface, then iterated
using a Newton-Raphson method to solve $\bf{J}\cdot\delta h=-\Theta$,
where $\bf{J}$ is the Jacobian matrix.
The drawback of this method is that it is not guaranteed to give the
outermost marginally trapped surface. In practice however, this limitation
can be easily overcome by either a single good initial guess, or multiple less accurate initial guesses.

\subsubsection{Masses and Momenta}
Two distinct measures of mass and momenta are available in relativistic
spacetimes.  First, ADM mass and angular momentum evaluated as either 
surface integrals at infinity or volume integrals over entire 
hypersurfaces give a measure of the total energy and angular momentum 
in the spacetime. The module \codename{ML\_ADMQuantities} of the 
McLachlan code~\cite{McLachlan:web} uses the latter method, creating 
gridfunctions containing the integrand of the volume 
integrals~\cite{Yo:2002bm}:
\begin{eqnarray}
   M &=& \frac{1}{16\pi} \int_\Omega d^3 x \left[ e^{5 \phi} 
	\left( 16 \pi E + \tilde{A}_{ij} \tilde{A}^{ij} - \frac23 K^2 
        \right) - e^\phi \tilde{R} \right] \\
   J_i &=& \frac{1}{8 \pi} \varepsilon_{ij}{}^k \int_\Omega d^3 x \left[ 
	e^{6\phi}\left( \tilde{A}^j{}_k + \frac23 x^j \tilde{D}_k K 
	- \frac12 x^j \tilde{A}_{\ell n} \partial_k \tilde{\gamma}^{\ell n}
	+ 8 \pi x^j S_k \right) \right]
\end{eqnarray}
on which the user can use the reduction functions provided by 
{\tt Carpet}  to perform the volume integral.  We note that
\codename{ML\_ADMQuantities} inherits directly from the BSSN variables
stored in \codename{McLachlan} rather than strictly from the base modules.
As the surface terms required when converting a surface integral to a 
volume integral are neglected, this procedure assumes
that the integrals of  $\tilde{D}^i e^\phi$ and
$e^{6\phi} \varepsilon_{ij}{}^k x^j \tilde{A}^\ell{}_k$ over the boundaries of
the computational domain vanish.  The ADM mass and angular momentum
can also be calculated from the 
variables stored in the base modules using the \codename{Extract} module, 
as surface integrals~\cite{Bowen:1980yu}
\begin{eqnarray}
  M &=& - \frac{1}{2\pi} \oint \tilde{D}^i \psi d^2 S_i \\
  J_i &=& \frac{1}{16\pi} \varepsilon_{ijk} \oint \left( x^j K^{km} 
	- x^k K^{jm} \right) d^2 S_m 
\end{eqnarray}
on a specified spherical surface, preferably one far from the center of 
the domain since these quantities are only properly defined when
calculated at infinity.

There are also the quasi-local measures of mass and angular momentum, from 
any \ahz{s} found during the spacetime.  Both 
\codename{AHFinderDirect} and \codename{AHFinder} output the 
corresponding mass derived from the area of the horizon $m_H = 
\sqrt{A/(16\pi)}$.

The module \codename{QuasiLocalMeasures} implements
the calculation of mass and spin multipoles from the isolated 
and dynamical horizon formalism~\cite{Dreyer:2002mx, Schnetter:2006yt},
as well as a number of other proposed formul\ae\@ for quasilocal mass, linear 
momentum and angular momentum that have been advanced over the
years~\cite{Szabados:2004ql}. Even though there are only a few rigorous proofs
that establish the properties of these latter quantities, they have 
been demonstrated to be surprisingly helpful in numerical simulations 
(see, e.g.,~\cite{Lovelace:2009dg}), and are therefore 
an indispensable tool in numerical relativity.
\codename{QuasiLocalMeasures} takes as input a horizon surface, or any
other surface that the user specifies (like a large coordinate
sphere) and
can calculate useful quantities such as the Weyl or Ricci scalars
or the three-volume element of the horizon world tube
in addition to physical observables such as mass and momenta. 

Finally, the module \codename{HydroAnalysis} additionally locates the 
(coordinate) center of mass as well as the point of maximum rest mass density of a 
matter field.

\subsubsection{Gravitational Waves} 
One of the main goals of numerical relativity to date is 
modeling  gravitational waveforms that may 
be used in template generation to help analyze 
data from the various gravitational wave detectors around the 
globe.  The Einstein Toolkit includes modules for extracting 
gravitational waves via either the Moncrief formalism of a perturbation 
on a Schwarzschild background or the calculation of the Weyl scalar $\Psi_4$.

The module \codename{Extract} uses the Moncrief formalism~\cite{
Moncrief:1974am} to extract gauge-invariant wave functions $Q_{\ell m}^\times$ and $Q_{\ell
m}^+$ given spherical surfaces of constant coordinate
radius. The spatial metric is expressed as a perturbation on
Schwarzschild and expanded into a tensor basis of
the Regge-Wheeler harmonics~\cite{Regge:1957td} described by six standard
Regge-Wheeler functions $\lbrace c_1^{\times\ell m}, c_2^{\times\ell m},
h_1^{+\ell m}, H_2^{+\ell m},K^{+\ell m}, G^{+ \ell m} \rbrace$. From
these basis functions the gauge-invariant quantities:
\begin{eqnarray}
  Q_{\ell m}^\times &=& \sqrt{\frac{2(\ell+2)!}{(\ell-2)!}} 
    \left[c_1^{\times\ell m} + \frac12\left( \partial_r - \frac{2}{r} 
    \right) c_2^{\times\ell m} \right] \frac{S}{r} \\
  Q_{\ell m}^+ &=& \frac{1}{\Lambda} \sqrt{ \frac{2(\ell-1)(\ell+2)}
    {\ell(\ell+1)}} \Bigg( \ell(\ell+1)S(r^2 \partial_r G^{+ \ell m}
    - 2 h_1^{+\ell m} ) \nonumber \\
   & & + 2rS(H_2^{+\ell m}-r\partial_rK^{+\ell m})
    + \Lambda r K^{+\ell m} \Bigg)
\end{eqnarray}
are calculated, where $S=1-2M/r$ and $\Lambda=(\ell-1)(\ell+2)+6M/r$.  
These functions then satisfy the wave equations:
\begin{eqnarray}
   (\partial_t^2-\partial_{r^*}^2)Q_{\ell m}^\times &=& 
      - S \left[ \frac{\ell(\ell+1)}{r^2}-\frac{6M}{r^3} \right] 
      Q_{\ell m}^\times \\
   (\partial_t^2-\partial_{r^*}^2)Q_{\ell m}^+ &=&
      - S \Bigg[ \frac{1}{\Lambda^2} \left( \frac{72M^3}{r^5}-\frac{12M
      (\ell-1)(\ell+2)}{r^3}\left(1-\frac{3M}{r}\right) \right)
      \nonumber \\ 
   & & + 
      \frac{\ell(\ell^2-1)(\ell+2)}{r^2\Lambda} \Bigg] Q_{\ell m}^+
\end{eqnarray}
where $r^*=r+2M \ln(r/2M-1)$.
Since these functions describe the 4-metric as a perturbation on 
Schwarzschild, the spacetime must be approximately spherically 
symmetric for the output to be interpreted as first-order 
gauge-invariant waveforms.

For more general spacetimes, the module \codename{WeylScal4} calculates
the complex Weyl scalar $\Psi_4=C_{\alpha\beta\gamma\delta}\,n^\alpha
\bar{m}^\beta n^\gamma \bar{m}^\delta$, which is a projection
of the Weyl tensor onto components of a null tetrad.
\codename{WeylScal4} uses the fiducial tetrad~\cite{Baker:2001sf},
written in 3+1 decomposed form as:
\begin{eqnarray}
   \ell^\mu &=& \frac{1}{\sqrt{2}}\left(u^\mu+\tilde{r}^\mu\right) \\
   n^\mu &=& \frac{1}{\sqrt{2}}\left(u^\mu-\tilde{r}^\mu\right) \\
   m^\mu &=&\frac{1}{\sqrt{2}}\left(\tilde{\theta}^\mu+i\tilde{\phi}^\mu\right)
\end{eqnarray}
where $u^\mu$ is the unit normal to the hypersurface.  The spatial
vectors $\lbrace \tilde{r}^\mu, \tilde{\theta}^\mu, \tilde{\phi}^\mu
\rbrace$ are created by initializing $\tilde{r}^\mu =
\lbrace0,x^i\rbrace$, $\tilde{\phi}^\mu = \lbrace0,-y,x,0\rbrace$, and
$\tilde{\theta}^\mu=\lbrace0,\sqrt{\gamma} \gamma^{ik} \varepsilon_{k\ell
m} \phi^\ell r^m\rbrace$, then orthonormalizing starting with
$\tilde{\phi}^i$ and invoking a Gram-Schmidt procedure at each step to
ensure the continued orthonormality of this spatial triad.

The Weyl scalar $\Psi_4$ is calculated
explicitly in terms of projections of the 3-Riemann tensor onto a null
tetrad, such that
\begin{eqnarray}
  \Psi_4 &=& \mathcal{R}_{ijk\ell} n^i \bar{m}^j n^k \bar{m}^\ell
      + 2 \mathcal{R}_{0jk\ell} \left( n^0 \bar{m}^j n^k \bar{m}^\ell
      - \bar{m}^0 n^j n^k \bar{m}^\ell \right) \nonumber \\
    &+& \mathcal{R}_{0j0\ell} \left( n^0 \bar{m}^j n^0 \bar{m}^\ell
      + \bar{m}^0 n^j \bar{m}^0 n^\ell - 2n^0 \bar{m}^j \bar{m}^0
      n^\ell \right)\,.
\end{eqnarray}
For a suitably chosen tetrad, this scalar in the radiation zone is
related to the strain of the gravitational waves since
\begin{equation}
   h = h_+ - i h_\times = - \int_{-\infty}^t dt^\prime
     \int_{-\infty}^{t^\prime} \Psi_4 dt^{\prime\prime}\,.
\end{equation}

While the waveforms generated by \codename{Extract} are 
already decomposed on a convenient basis to separate modes, the 
complex quantity $\Psi_4$ is provided by \codename{WeylScal4} as 
a complex grid function.  For this quantity, and any other real or
complex grid function, the module \codename{Multipole} interpolates 
the field $u(t,r,\theta,\phi)$ onto coordinate spheres of given radii
and calculates the coefficients
\begin{equation}
  C^{\ell m} \left(t,r\right) = \int {}_s Y_{\ell m}^* u(t,r,\theta,\phi)
    r^2 d\Omega
\end{equation}
of a projection onto spin-weighted spherical harmonics ${}_s Y_{\ell m}$.

\subsubsection{Object tracking}
\label{sec:object-tracking}
We provide a module (\codename{PunctureTracker}) for tracking BH
positions evolved with moving puncture techniques.  It can be used
with (\codename{CarpetTracker}) to have the mesh refinement regions follow the
BHs as they move across the grid.  The BH position is
stored as the centroid of a spherical surface (even though there is no surface)
provided by \codename{SphericalSurface}.

Since the punctures only move due to the shift advection terms in
the BSSN equations, the puncture location is evolved very simply as
\begin{equation}
  \frac{d x^i}{d t} = -\beta^i, \label{eq:puncturetracking}
\end{equation}
where $x^i$ is the puncture location and $\beta^i$ is the shift. Since the
puncture location usually does not coincide with grid points, the shift is
interpolated to the location of the puncture.  
Equation~(\eref{eq:puncturetracking}) is implemented with a simple first-order
Euler scheme, accurate enough for controlling the location
of the mesh refinement hierarchy.

Another class of objects which often needs to be tracked are neutron stars.
Here is it usually sufficient to locate the position of the maximum density
and adapt AMR resolution in these regions accordingly, coupled with the
condition that this location can only move at a specifiable maximum speed.

\subsubsection{Other analysis modules} 
The remaining analysis capabilities of the Einstein Toolkit span a 
variety of primarily vacuum-based functions.  
First, modules are provided to calculate the Hamiltonian and momentum 
constraints which are used to monitor how well the evolved spacetime 
satisfies the Einstein field equations.  Two modules, 
\codename{ADMConstraints} and \codename{ML\_ADMConstraints} provide these 
quantities.  Both calculate these directly from variables stored in the 
base modules described in Sec.~\ref{sec:base_modules}, explicitly written 
as:
\begin{eqnarray}
  H &=& R - K^i{}_j K^j{}_i + K^2 - 16 \pi E \label{eqn:analysis_hamiltonian_constraint}\\
  M_i &=& \nabla_j K_i{}^j - \nabla_i K - 8 \pi S_i
\end{eqnarray}
where $S_i=-\frac{1}{\alpha} \left( T_{i0} - \beta^j T_{ij} \right)$.  
The difference between these modules lies in how they access the stress 
energy tensor $T_{\mu\nu}$, as the module \codename{ADMConstraints}
uses a deprecated functionality which does not require storage
for $T_{\mu\nu}$.

Finally, \codename{ADMAnalysis} calculates a variety of derived spacetime 
quantities that are often useful in post-processing such as the determinant
of the 3-metric $\det{\gamma}$, the trace of the extrinsic curvature $K$, 
the 3-Ricci tensor in Cartesian coordinates $\mathcal{R}_{ij}$ and its trace 
$\mathcal{R}$, as well as the 3-metric and extrinsic curvature converted to
spherical coordinates.

\subsection{Simulation Domain, Symmetries, Boundaries}
\subsubsection{Domains and Coordinates.}
{\tt Cactus} distinguishes between
the \emph{physical} domain, which lives in the continuum, and
\emph{discrete} domain, which consists of a discrete set of grid
points. The physical domain is defined by its coordinate extent and
is independent of the numerical resolution; in particular, the
boundary of the physical domain has a width of zero (and is thus a set
of measure zero). The discrete domain is defined indirectly via a
discretization procedure that specifies the number of
boundary points, their location with respect to the physical boundary,
and either the grid spacing or the number of grid points spanning the
domain. This defines the number and
location of the grid points in the discrete domain. The discrete
domain may have grid points outside of the physical domain, and may
have a non-zero boundary width. This mechanism ensures that changes in
the numerical resolution do not affect the extent of the physical
domain, i.e., that the discrete domains converge to the physical
domain in the limit of infinite resolution.
The Einstein Toolkit provides the \codename{CoordBase} thorn that facilitates
the definition of the simulation domain independent of the actual
evolution thorn used, allowing it to be specified at run time via  parameters 
in the same way that parameters describing the physical system are specified.
\codename{CoordBase} exposes a public runtime interface that allows 
other
thorns to query the domain description in a uniform way. This is used
by 
\codename{Carpet} to query \codename{CoordBase} for the discrete
grid when creating the hierarchy of grids, automatically ensuring a
consistent grid description between the two thorns. 
Evolution thorns such as \codename{McLachlan} use the domain 
information to decide which points are evolved and therefore require the
evaluation of the right-hand-side expression, and which ones are set via
boundary or symmetry conditions.

\subsubsection{Symmetries and Boundary Conditions.}
\begin{figure}[htbp]
    \begin{center}
        \includegraphics[width=0.2\textwidth]{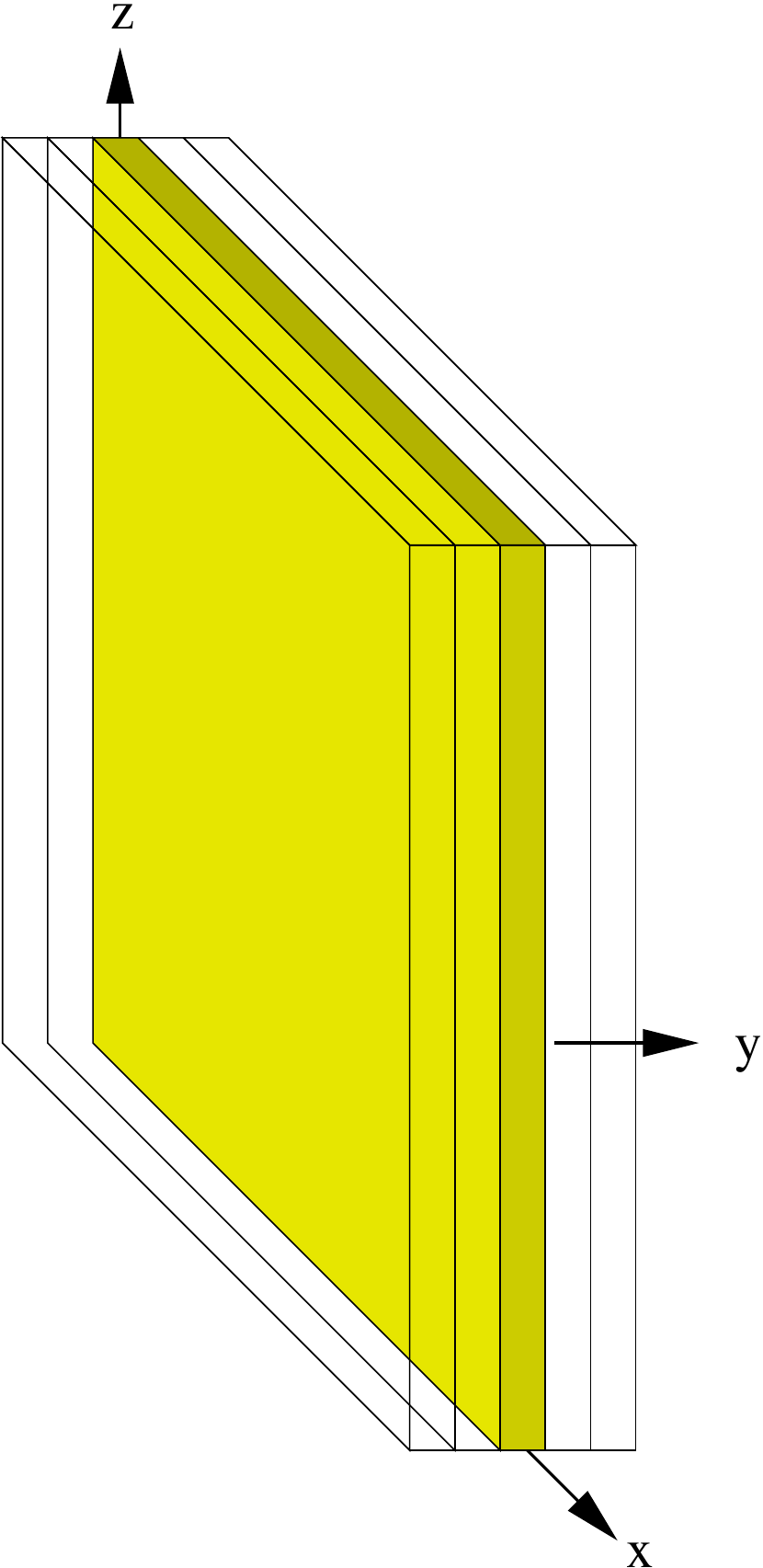}
    \end{center}
    \caption{Grid layout of a simulation using \codename{Cartoon2D}. The
    $z$-axis is the axis of rotational symmetry. Image courtesy of
    Denis Pollney.}
    \label{fig:cartoon-plane}
\end{figure}

The Einstein Toolkit includes two thorns, \codename{Boundary}
and \codename{SymBase}, to provide a generic interface to specify and
implement boundary and
symmetry conditions.
The toolkit includes built-in support for a 
set of reflecting or 
rotating symmetry
conditions that can be used to reduce the size of the simulation
domain. These symmetries include periodicity in any of the coordinate
directions
(via the \codename{Periodic} module), reflections across the
coordinate planes (via the \codename{Reflection} module), 
$90^{\circ}$ and $180^{\circ}$ rotational symmetries
(via the \codename{RotatingSymmetry90} and
\codename{RotatingSymmetry180} modules respectively)
about the $z$ axis, and a continuous rotational symmetry (via
the \codename{Cartoon2D} thorn)~\cite{Alcubierre:1999ab}. 
\codename{Cartoon2D} allows fully
three dimensional codes to be used in axisymmetric problems by evolving
a slice in the $y=0$ plane and using the rotational symmetry to populate
ghost points off the plane (see Figure~\ref{fig:cartoon-plane}). 

\begin{figure}[htbp]
    \begin{center}
        \includegraphics{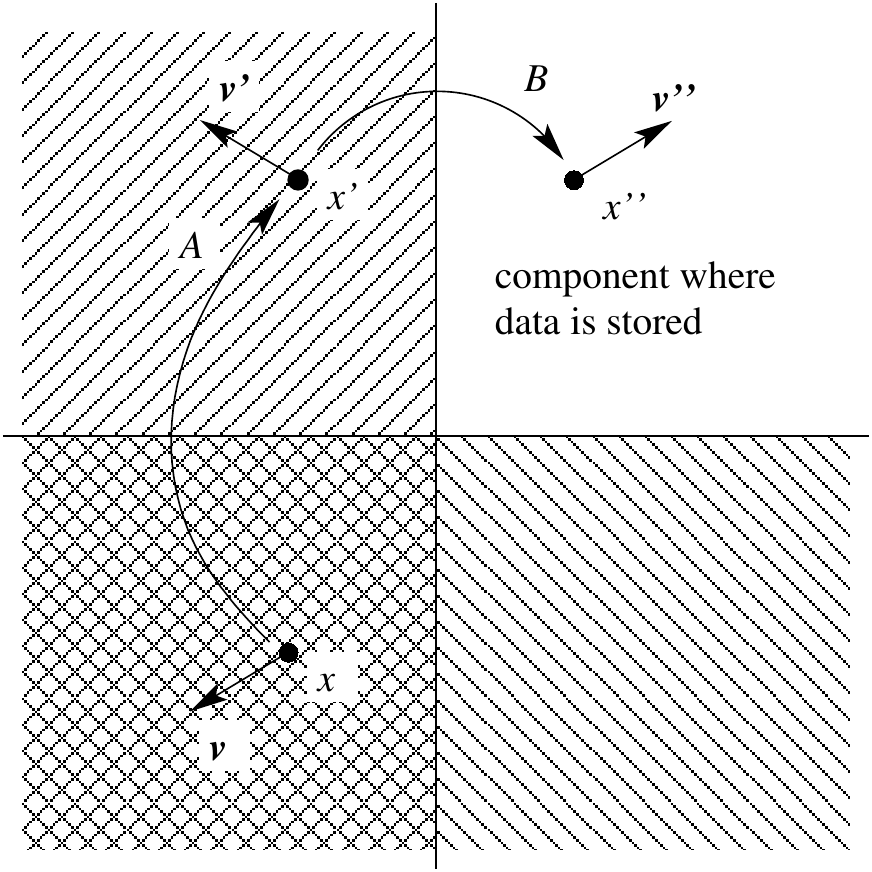}
    \end{center}
    \caption{Iterative transformation of a point $x$ in quadrant 3 to the
      corresponding
    point $x''$ for which there is actual data stored. In this
    example, two reflection symmetries along the horizontal and vertical axis
    are present. Notice how the vector components change in
    transformations $A$ and $B$.} 
    \label{fig:faces}
\end{figure}

In applying symmetries to populate ghost zones, the
transformation properties of tensorial quantities (including tensor
densities and non-tensors such as Christoffel symbols) are correctly
taken into account, just as they are in the interpolation routines present in {\tt Cactus}.
Thus, symmetries are handled transparently
from the point of view of user modules (see Figure~\ref{fig:faces} for an
illustration).

The \codename{Boundary} thorn serves as a registry for available boundary
conditions and provides basic scheduling to enforce all requested
boundary conditions at the proper times. It also provides a basic set of
boundary conditions to be used by user thorns. The ``flat'' boundary
conditions often used for hydrodynamic variables that approach an
atmosphere value fall in this category. More complicated boundary
conditions are often implemented as modifications to the evolution
equations and are not handled directly by \codename{Boundary}. Examples
are the radiative (Sommerfeld) and extrapolation boundary conditions
provided by thorn \codename{NewRad}.

\subsubsection{Adaptive Mesh Refinement}
\begin{figure}[htbp]
    \begin{center}
        \includegraphics{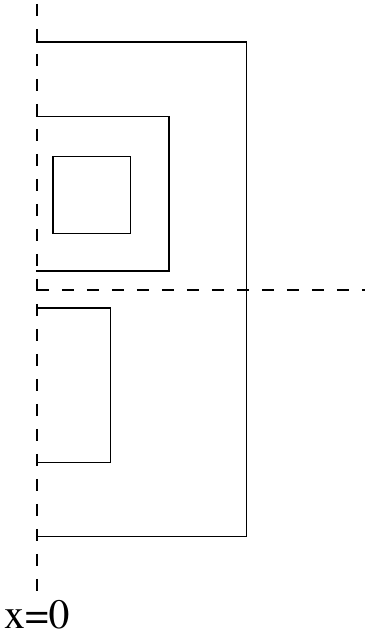}
    \end{center}
    \caption{Nested boxes following the individual BHs in binary
    BH merger simulation (see Section~\ref{sec:bbh-example}),
    with the location of the individual BHs  found by
    \codename{PunctureTracker}. The innermost three of the nine
    levels of mesh refinement used in this simulation are shown. Notice the use of
    \codename{RotatingSymmetry180} to reduce the computational domain.}
    \label{fig:bbh-boxes}
\end{figure}

The Einstein toolkit currently supports feature-based mesh refinement,
which is
based on
extracting quantities such as the locations of BHs or NSs and
then constructing a mesh hierarchy (stacks of refined regions) based
on the locations, sizes, and speeds of these objects. This allows
tracking objects as they move through the domain. One can also add or
remove stacks if, for instance,  the number of objects changes. Full AMR based on
a local error estimate is supported by \codename{Carpet}, but the
Einstein Toolkit does not presently provide a suitable regridding thorn
to create
such a grid. If initial
conditions are constructed outside of {\tt Carpet} (which is often the
case), then the initial mesh hierarchy has to be defined manually.
In order to facilitate the description of the mesh hierarchy the
Einstein toolkit provides two regridding modules in
the \codename{CarpetRegrid} and \codename{CarpetRegrid2} thorns.
Both thorns primarily support box-in-box type refined meshes, which are
well suited to current binary BH simulations in which the
high-resolution regions are centered on the individual BHs.
Figure~\ref{fig:bbh-boxes} shows a typical set of nested boxes during
the inspiral phase of a binary BH merger simulation.

\begin{figure}[htbp]
    \begin{center}
        \includegraphics[width=0.3\textwidth]{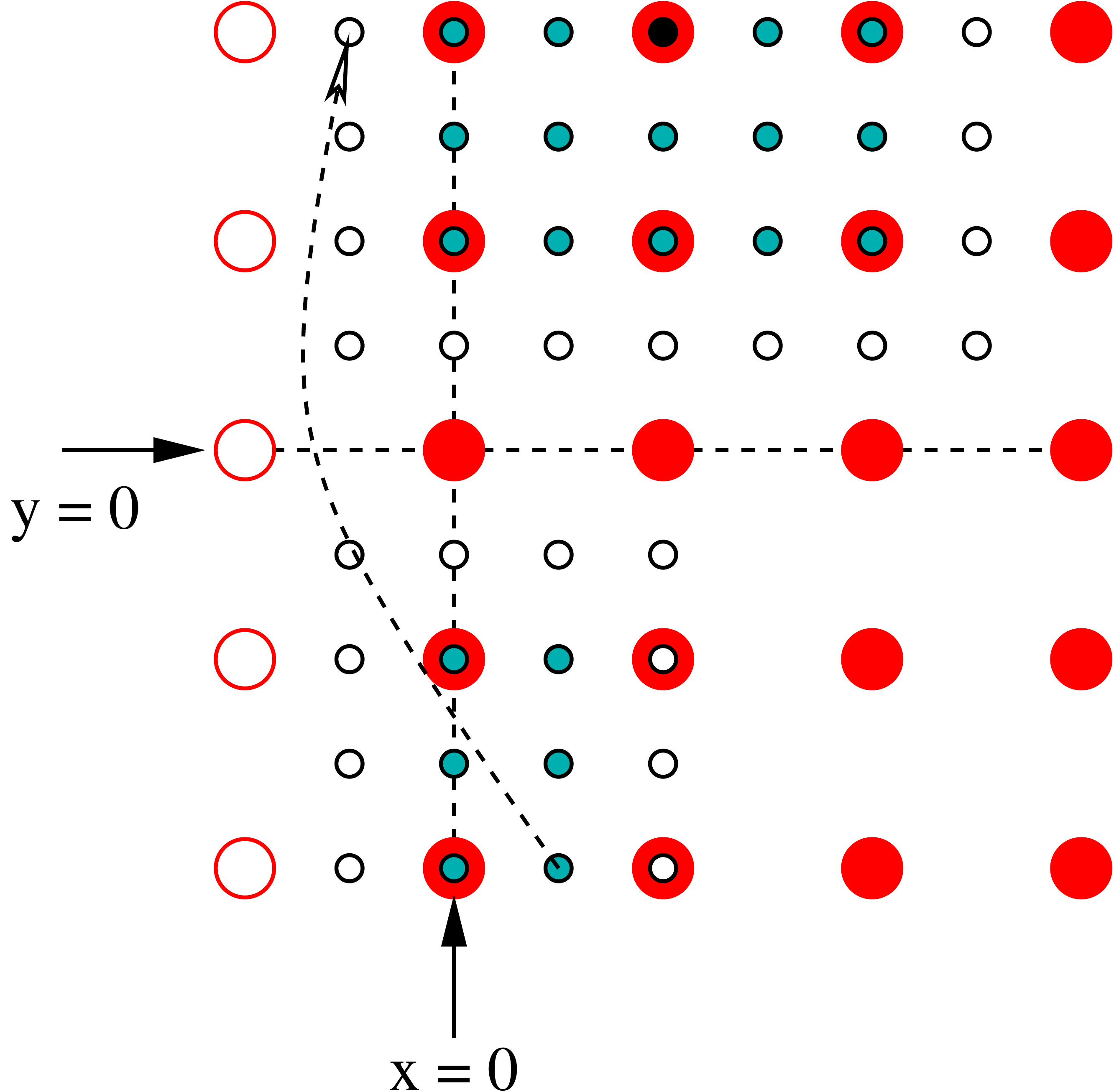}
    \end{center}
    \caption{Grid layout created by \codename{CarpetRegrid2}. In this
    example we use one ghost point, one boundary point, and two buffer 
    points
    as well as \codename{RotatingSymmetry180}. There are two refinement
    levels present, a coarse one represented by big red circles and a
    fine one represented by small black circles. The
    filled black circle is the single point specified by the user.
    \codename{CarpetRegrid2} surrounded it with a layer of buffer
    points, indicated by the cyan filled circles. The open circles are
    ghost and boundary points which are maintained by \codename{Carpet}.
    The presence of the $\pi$-symmetry forces \codename{CarpetRegrid2}
    to create the tiny region to the bottom left of the grid. It serves
    only as a source for the boundary condition.}
    \label{fig:rot180-grid}
\end{figure}

\codename{CarpetRegrid} provides a number of different ways to specify
the refined regions, e.g., as a set of boxes centered around the origin
or as an explicit list of regions that make up the grid hierarchy.
Traditionally, groups using \codename{CarpetRegrid} have employed
auxiliary thorns that are not part of the Einstein Toolkit to create
this list of boxes based on information obtained from apparent
horizon tracking or other means. \codename{CarpetRegrid2} provides a user-friendly
interface to define sets of nested boxes that follow BHs or
other tracked objects. 
Object coordinates are updated by 
\codename{CarpetTracker}, which provides a simple interface to
the object trackers
\codename{PunctureTracker} and \codename{NSTracker} (see
section~\ref{sec:object-tracking}) in order to have the refined region follow
the moving objects. 
\codename{CarpetRegrid2} contains code to handle the $\pi$-symmetry
provided by \codename{RotatingSymmetry180}, enforcing the symmetry on
the resulting grid layout (see Figure~\ref{fig:rot180-grid}).

\section{Examples}

To demonstrate the properties of the code and its capabilities, we have used it to simulate common astrophysical configurations of interest.  Given the community-oriented direction of the project, the parameter files required to launch these simulations and a host of others are included and documented in the code releases, along with the data files produced by a representative set of simulation parameters to allow for code validation and confirmation of correct code performance on new platforms and architectures.  As part of the internal validation process, 
nightly builds are checked against a set of benchmarks to ensure that consistent results are generated with the inclusion of all new commits to the code.

The performance of the Toolkit for vacuum configurations is demonstrated through evolutions of single, rotating BHs and the merger of binary black hole configurations (sections~\ref{sec:1bh-example} and \ref{sec:bbh-example}, respectively).   Linear oscillations about equilibrium for an isolated NS are discussed in section~\ref{sec:tov_oscillations}, and the collapse of a NS to a BH, including dynamical formation of a horizon, in section~\ref{sec:collapse_example}.  Finally, to show a less traditional application of the code, we show its ability to perform cosmological simulations by evolving a Kasner spacetime (see section~\ref{sec:cosmology}).

\subsection{Spinning BH}
\label{sec:1bh-example}
As a first example, we perform simulations of a single distorted rotating BH. 
We use \codename{TwoPunctures} to set up initial data for a single 
puncture of mass $M_{\mathrm{bh}}=1$ and dimensionless spin parameter 
$a = S_{\mathrm{bh}}/M_{\mathrm{bh}}^2 = 0.7$. Evolution of the data is 
performed by \codename{McLachlan}, apparent horizon finding by 
\codename{AHFinderDirect} and gravitational wave extraction by 
\codename{WeylScal4} and \codename{Multipole}. Additional analysis of the
horizons is done by \codename{QuasiLocalMeasures}. The runs were performed
with fixed mesh refinement provided by \codename{Carpet}, using 8 levels
of refinement on a quadrant grid (symmetries provided by 
\codename{ReflectionSymmetry} and \codename{RotatingSymmetry180}). The outer
boundaries were placed at $R=256M$. We performed runs at 3 different
resolutions: the low resolution was $0.024M (3.072M)$, medium was 
$0.016M (2.048M)$ and high was $0.012M (1.536M)$, where the numbers refer to the 
resolution on the finest (coarsest) grid. The runs where performed using the tapering
evolution scheme in \codename{Carpet} to avoid interpolation in
time during prolongation. The initial data corresponds to a rotating BH
perturbed by a Brill wave and, as such, has a non-zero
gravitational wave content. We evolved the BH using 4th-order finite differencing from
$T=0M$ until it had settled down to a stationary state at $T=120M$.

\begin{figure}
 \includegraphics[width=0.9\textwidth]{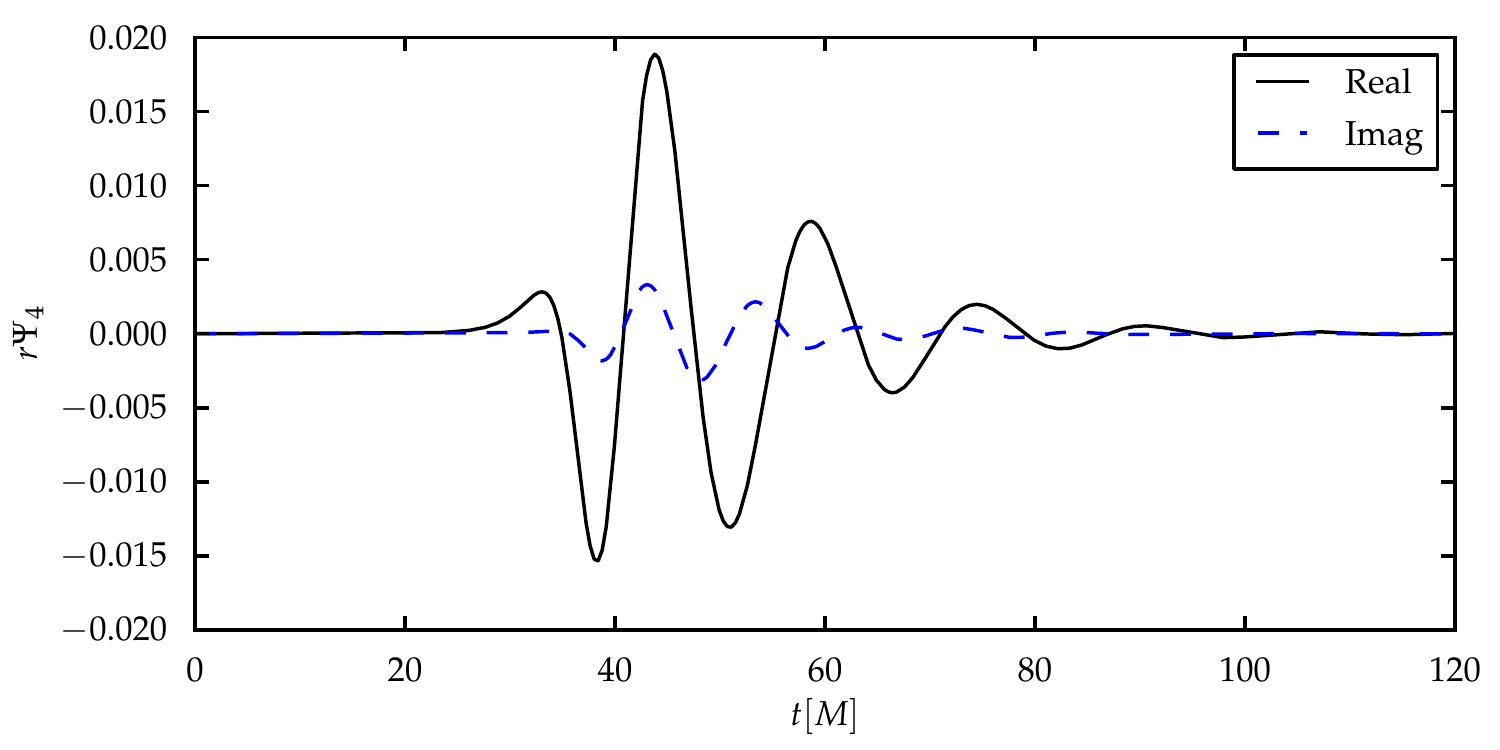}
 \includegraphics[width=0.9\textwidth]{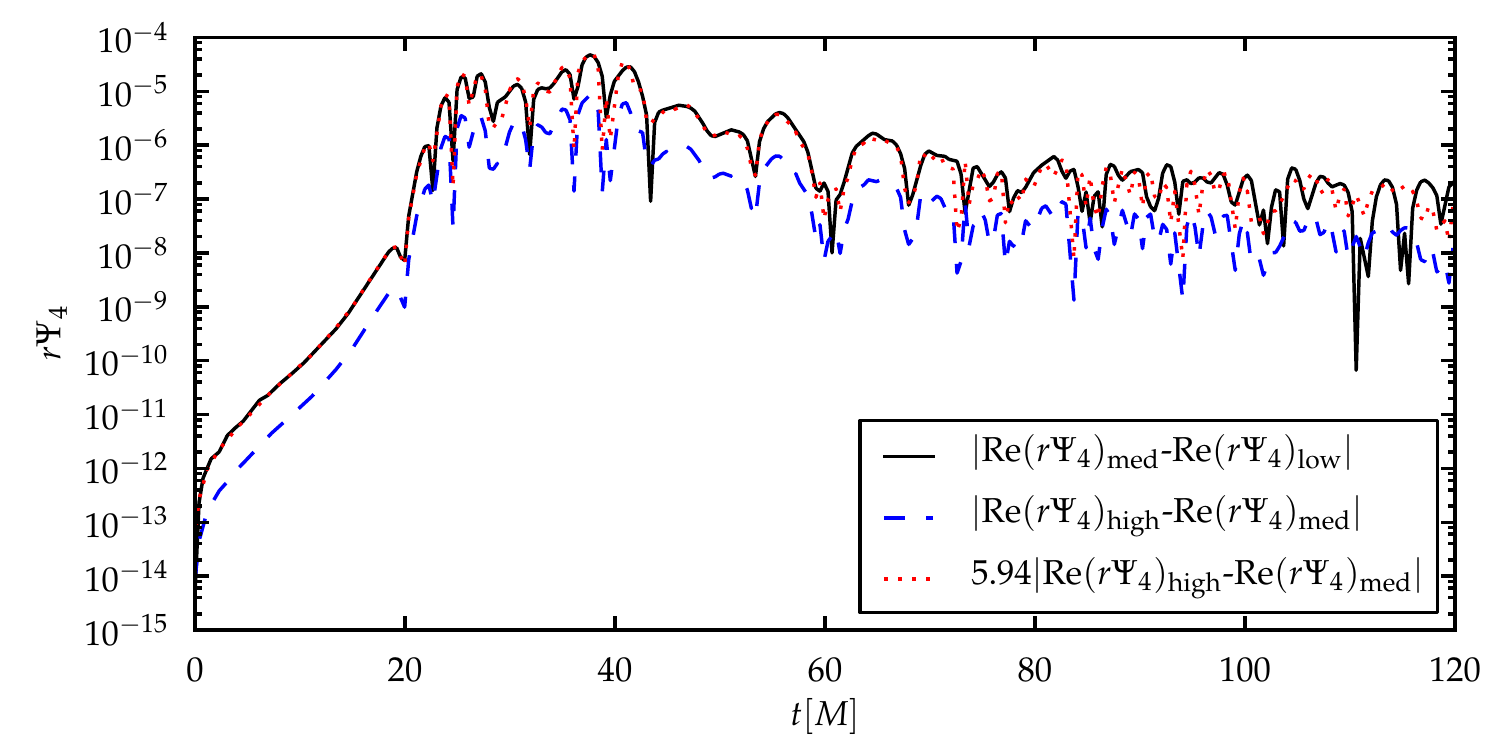}
 \caption{The extracted $\ell =2, m=0$ mode of $\Psi_4$
          as function of time from the high resolution run (top plot). The extraction was
          done at $R=30M$. Shown is both the real (solid black curve) and the
          imaginary (dashed blue curve) part of the waveform. At the bottom, we
          show the
          difference between the medium and low resolution runs (solid black
          curve), between the high and medium resolution runs
          (dashed blue curve), and the scaled difference (for 4th order
          convergence) between the medium and low resolution runs
          (dotted red curve) for the real part of the $\ell =2, m=0$ waveforms.}
 \label{fig:kerr_waves}
\end{figure}

Figure~\ref{fig:kerr_waves} shows the $\ell =2, m=0$ mode of $r\Psi_4$ 
extracted at $R=30M$, and its numerical convergence.
In the top plot the black (solid) curve is the real part and the blue (dashed)
curve is the imaginary part of $r \Psi_4$ for the high resolution run. Curves
from the lower resolution are indistinguishable from the high resolution curve
at this scale. In the bottom plot the black (solid) curve shows the absolute value
of the difference between the real part of the medium and low resolution
waveforms while the blue (dashed) curve shows the absolute value of the 
difference between the high and medium resolution waveforms in a log-plot.
The red (dotted) curve is the same as the blue (dashed) curve, except it is
scaled for 4th order convergence. With the resolutions used here this factor is
$\left (0.016^4-0.024^4\right )/\left ( 0.012^4-0.016^4\right) \approx 5.94$.

\begin{figure}
 \includegraphics[width=0.9\textwidth]{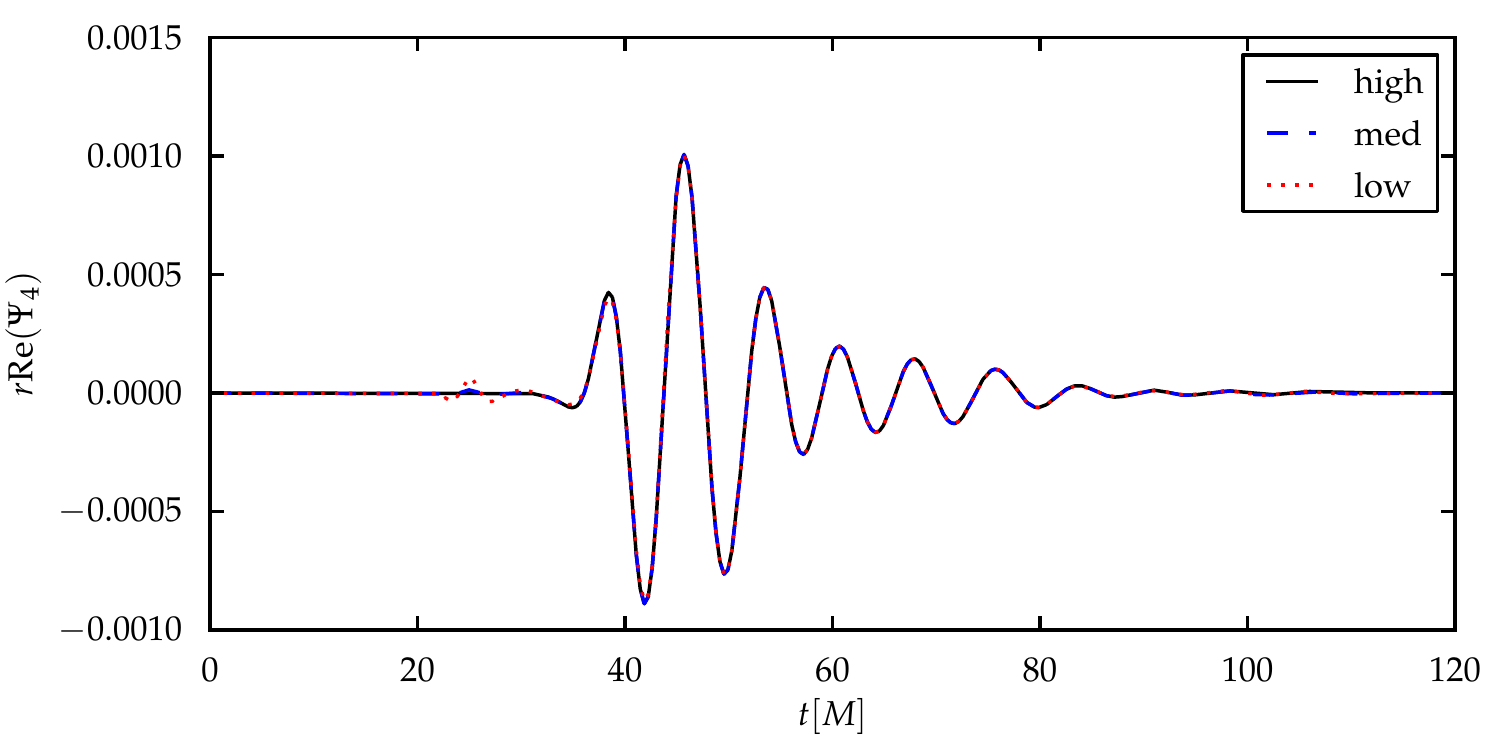}
 \includegraphics[width=0.9\textwidth]{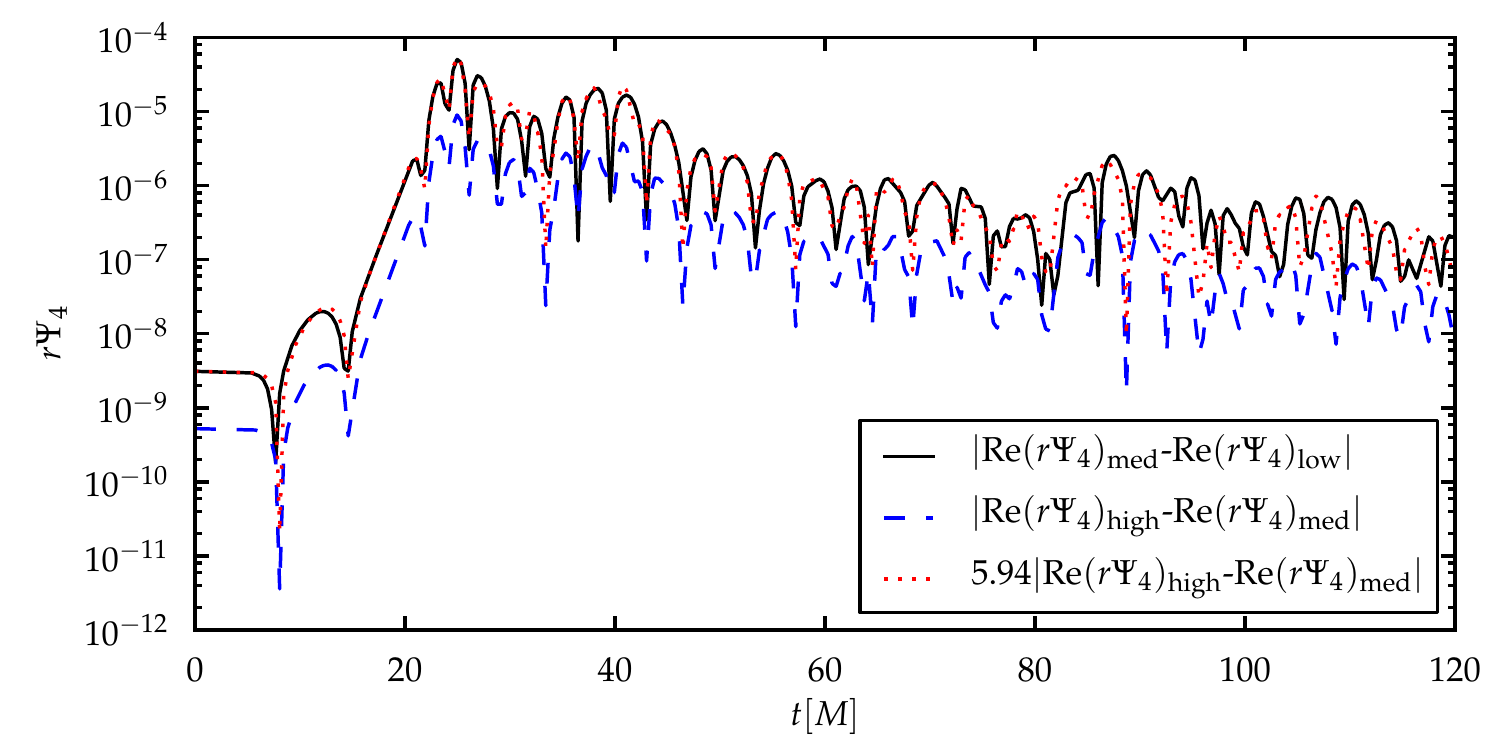}
 \caption{Real part of the extracted
          $\ell =4, m=0$ mode of $\Psi_4$ as function of time (top plot) for the high
          (solid black curve), medium (dashed blue curve) and low (dotted red
          curve) resolution runs. The extraction was done at $R=30M$.  The bottom
          plot shows for the real part of the $\ell =4, m=0$ waveforms the
          difference between the medium and low resolution runs (solid black
          curve), the difference between the high and medium resolution runs
          (dashed blue curve) as well as the scaled (for 4th order
          convergence) difference between the medium and low resolution runs
          (dotted red curve).}
 \label{fig:kerr_waves_l4}
\end{figure}

Figure~\ref{fig:kerr_waves_l4} shows similar plots for the $\ell =4, m=0$ mode
of $r\Psi_4$, again extracted at $R=30 M$.
The top plot in this case shows only the real part of the extracted waveform
but for all three resolutions (black solid curve is high, blue dashed curve is
medium and red dotted curve is low resolution). Since the amplitude of this
mode is almost a factor of 20 smaller than the $\ell =2, m=0$ mode there are
actually small differences visible between resolutions in the beginning of
the waveform. The bottom plot shows the convergence of the real part of the
$\ell =4, m=0$ mode (compare with the bottom plot in Figure~\ref{fig:kerr_waves})
and demonstrates that even though the amplitude is much smaller we still obtain close
to perfect fourth-order convergence. 

In addition to the modes shown in Figure~\ref{fig:kerr_waves} and 
\ref{fig:kerr_waves_l4} we note that the extracted $\ell =4, m=4$ mode
is non-zero due to truncation error, but shows fourth-order convergence to
zero with resolution (this mode is not present in the initial data and is not
excited during the evolution). Other modes are zero to round-off due to
symmetries at all resolutions. 

\begin{figure}
 \includegraphics[width=0.9\textwidth]{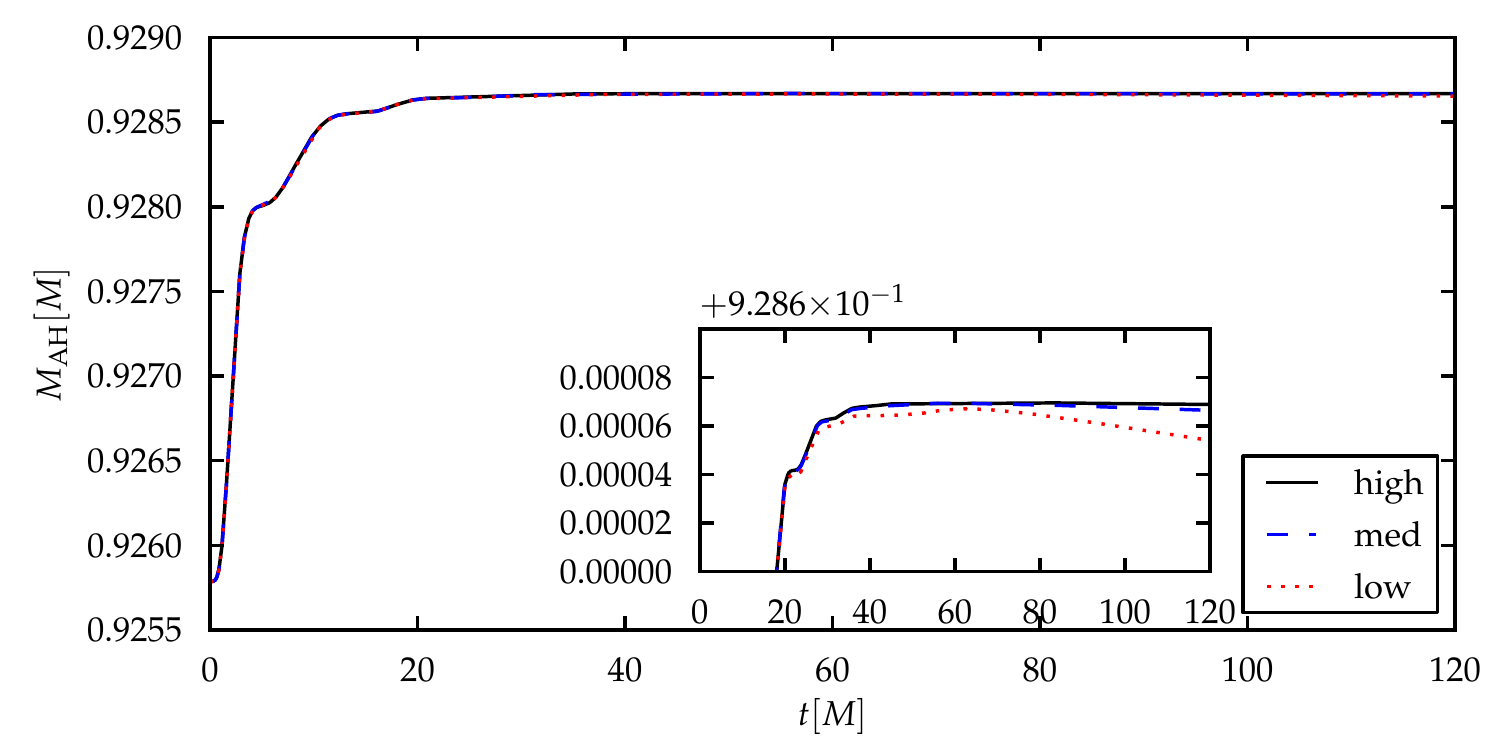}
 \includegraphics[width=0.9\textwidth]{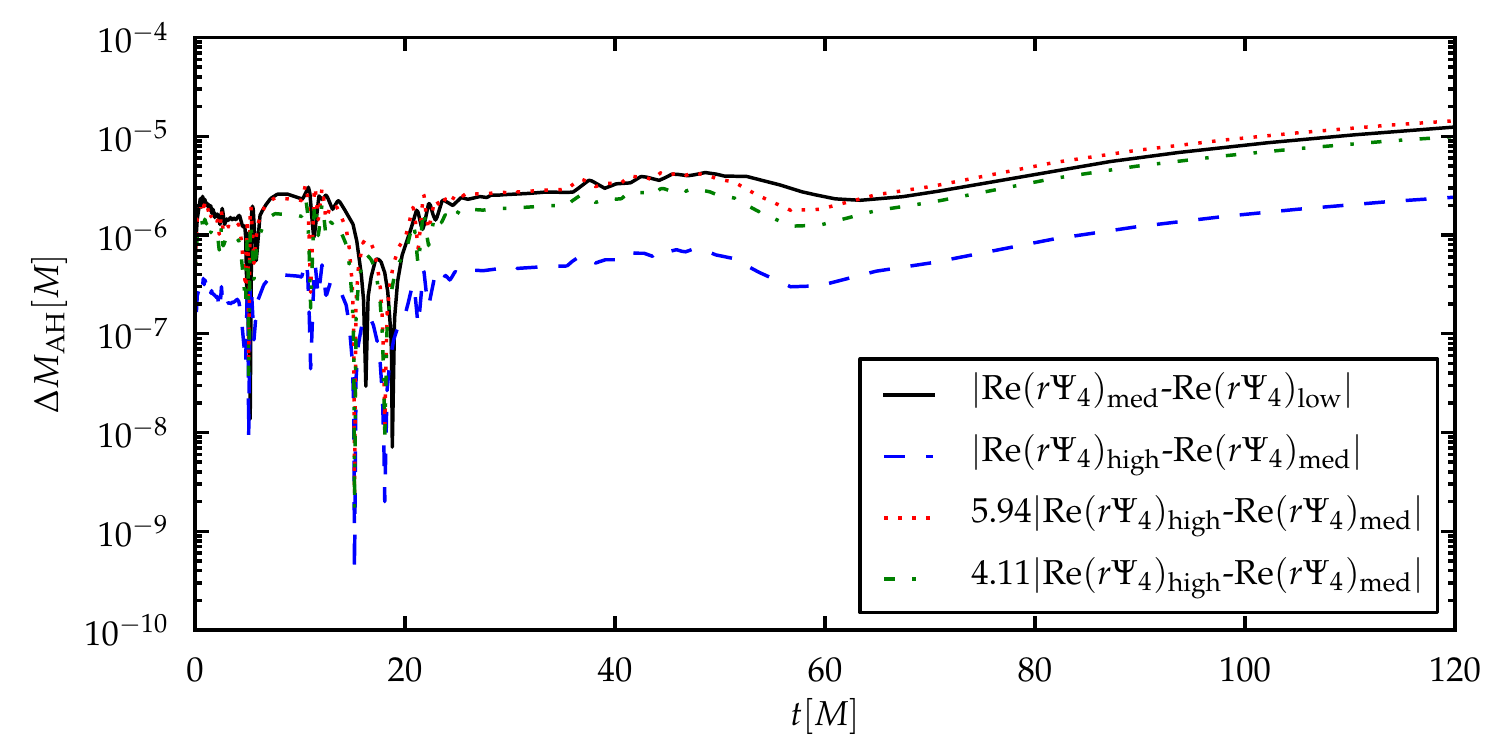}
 \caption{The top plot shows the irreducible mass of the apparent horizon
as a function of time at low (black solid curve), medium (blue dashed curve)
and high (red dotted curve) resolutions. The inset is a zoom in on the
$y$-axis to more clearly show the differences between the resolutions. The
bottom plot shows the convergence of the irreducible mass. The black (solid)
curve shows the difference between the medium and low resolution results,
the blue (dashed) curve shows the difference between the high and medium
resolution results. The red (dotted) and green (dash-dotted) show the 
difference between the high and medium resolutions scaled according to
fourth and third-order convergence respectively.} \label{fig:ah_mass}
\end{figure}

Since there is non-trivial gravitational wave content in the initial data,
the mass of the BH changes during its evolution. In figure~\ref{fig:ah_mass},
we show in the top plot the irreducible mass as calculated by
\codename{AHFinderDirect} as a function of time at the high (black solid curve),
medium (blue dashed curve) and low (red dotted curve) resolutions.
The inset shows in more detail the differences between the different 
resolutions. The irreducible mass increases by about 0.3\% during the first
$40M$ of evolution and then remains constant (within numerical error) for the
remainder of the evolution. The bottom plot shows the convergence of the
irreducible mass by the difference between the medium and low resolutions 
(black solid curve), the difference between the high and medium resolutions
(blue dashed curved) as well as the scaled difference between the
high and medium resolutions for fourth-order (red dotted curve) and
third-order (green dash-dotted curve). The convergence is almost perfectly
fourth-order until $T=50M$, then better than fourth-order until $T=60M$, and
finally between third-order and fourth-order for the remainder of the
evolution. The lack of perfect fourth-order convergence at late times may be attributed
 to non-convergent errors from the puncture propagating
to the horizon location at the lowest resolution.

\begin{figure}
 \includegraphics[width=0.9\textwidth]{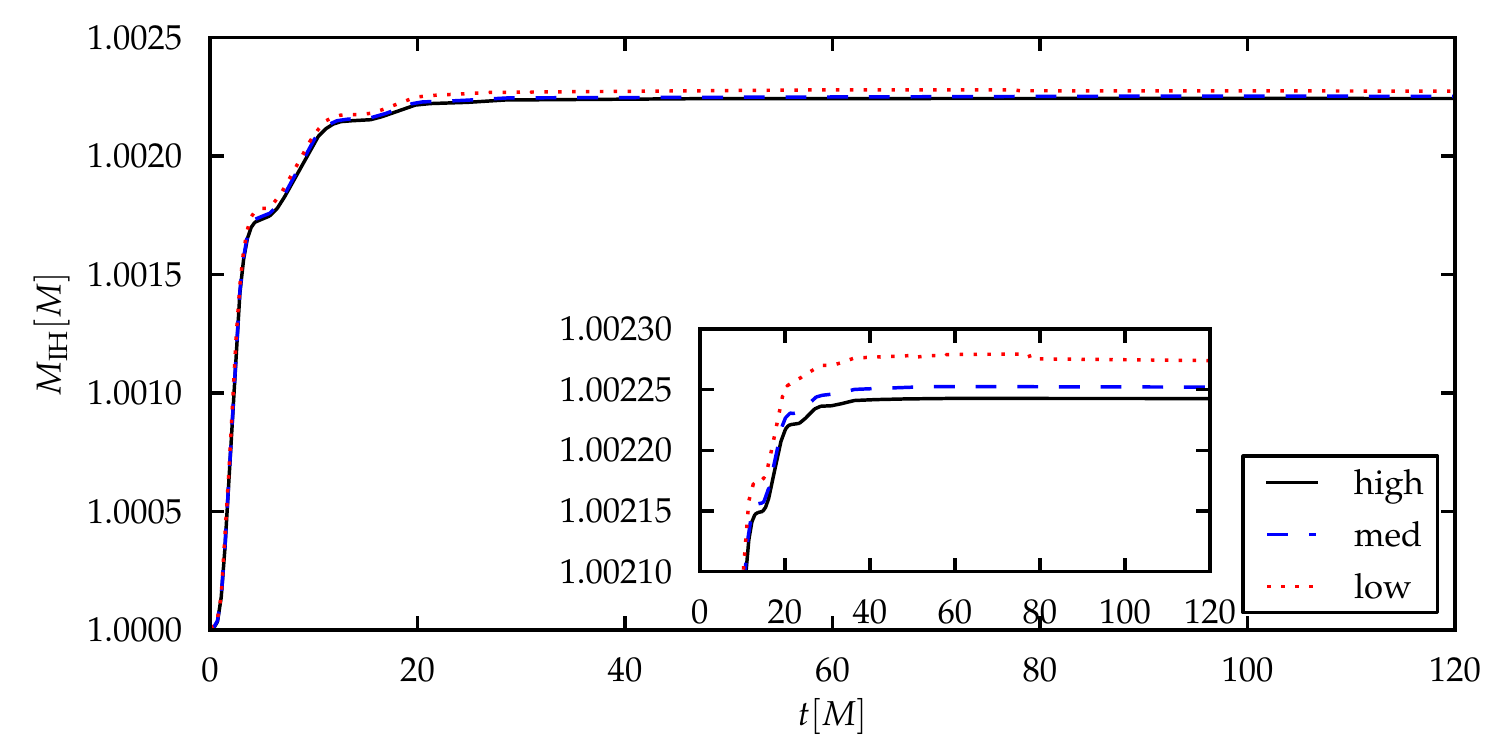}
 \includegraphics[width=0.9\textwidth]{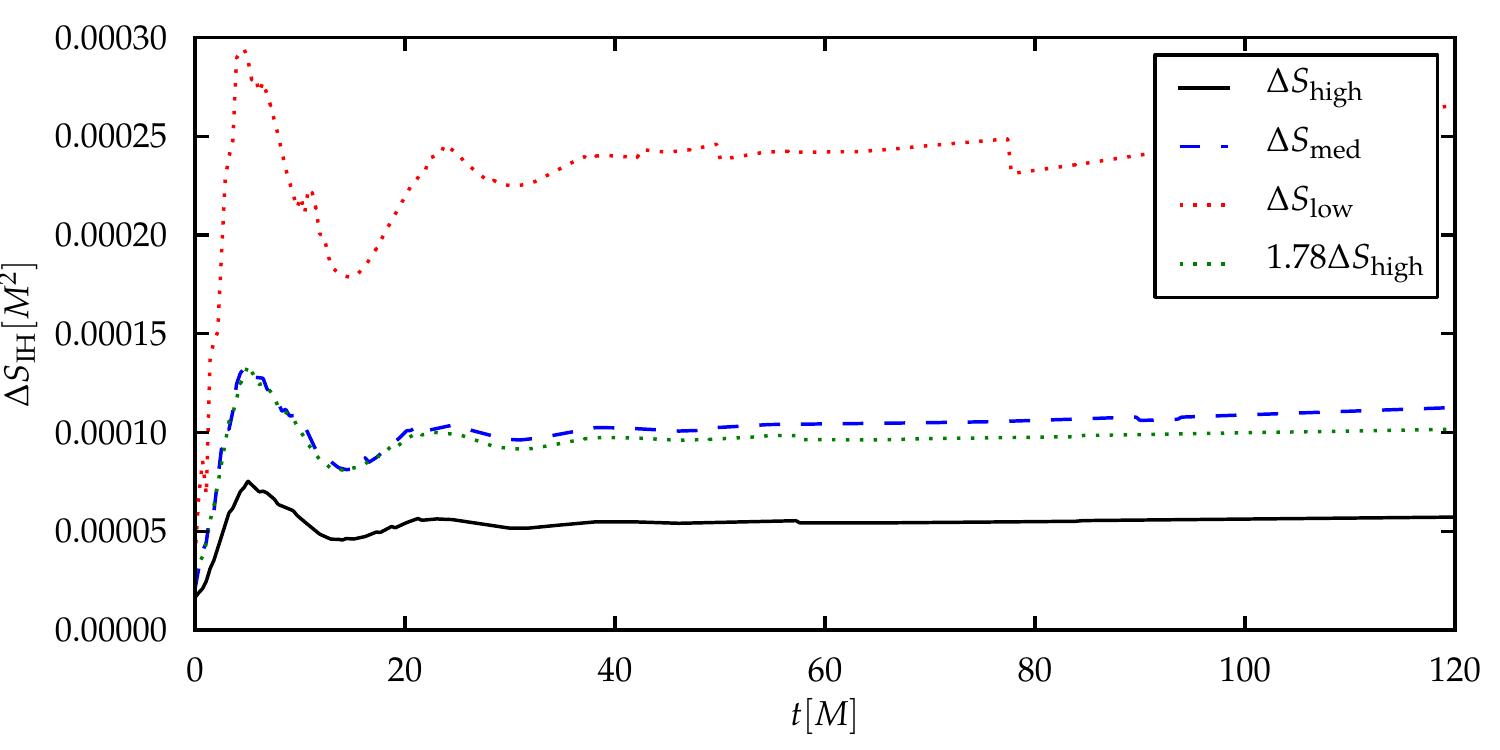}
 \caption{The top plot shows the total mass and the bottom plot shows the change in spin (i.e.\ $\Delta S=S(t)-S(t=0)$ of the BH as a function of time.
In both plots the black (solid) curve is for high, blue (dashed) for medium
and red (dotted) for low resolution. In the bottom plot the green (dash-dotted)
curve shows the high resolution result scaled for second-order convergence. The
agreement with the medium resolution curve shows that the change in spin
converges to zero as expected.}
\label{fig:ah_mass_spin}
\end{figure}

Finally, in Figure~\ref{fig:ah_mass_spin} we show the total
mass (top plot) and the change in the spin, $\Delta S = S(t) - S(t=0)$, as
calculated by \codename{QuasiLocalMeasures}.
In both cases the black (solid) curve is for high, blue (dashed) for medium and
red (dotted) for low resolution. Since the spacetime is 
axisymmetric the gravitational waves cannot radiate angular momentum. Thus
any change in the spin must be due to numerical error and $\Delta S$ should
converge to zero with increasing resolution. This is clearly shown in the
bottom plot of Figure~\ref{fig:ah_mass_spin}; the green
(dash-dotted) curve (the high resolution result scaled by a factor of $1.78$ for
second-order convergence to the resolution of the medium resolution) and the
blue (dashed) curve are on top of each other. Since the  
\codename{QuasiLocalMeasures} thorn uses an algorithm which is
only second-order accurate overall, this is the expected result. The increase of about 0.22\% in the mass of the
BH is caused solely by the increase in the irreducible mass.

\subsection{BH Binary}
\label{sec:bbh-example}
To demonstrate the performance in the code for a current problem of wide 
scientific interest, we have evolved a non-spinning equal-mass 
BH binary system.  
The initial data represent a binary system
in a quasi-circular orbit, with an initial separation chosen 
to be $r=6M$ so we may track the later inspiral, 
plunge, merger and ring down phases of the binary evolution.  
Table~\ref{table:BHB_ID} provides more details about 
the initial binary parameters used to generate the initial data. 
The \codename{TwoPunctures} module uses these initial parameters
to solve \eref{eq:twopunc_u}, the elliptic Hamiltonian constraint for 
the regular component of the conformal factor (see section~\ref{sec:twopunctures}). 
The spectral solution for this example was 
determined by using $[n_A,n_B,n_{\phi}]=[28,28,14]$ collocation 
points, and, along with the Bowen-York analytic solution for the 
momentum constraints, represents constrained GR
initial data $\{\gamma_{ij},K_{ij}\}$. The evolution is performed
by the \codename{McLachlan} module.

\begin{table}[!ht]
\label{table:BHB_ID}
{\centering
\begin{tabular}{l|llllll}
Configuration & $x_1$ & $x_2$ & $p_x$ & $p_y$ & $m$ & $M_{\rm ADM}$ \\
\hline
QC3 & 3.0 & -3.0 & 0.0 & 0.13808 & 0.47656 &  0.984618  \\
\end{tabular}\\}
\caption{Initial data parameters for a non-spinning equal mass 
BH binary. The punctures are located on the
$x$-axis at positions $x_1$ and $x_2$, with puncture bare 
mass parameters $m_1 = m_2 = m$, and momenta $\pm\vec p$.
}
\end{table}

The simulation domain spans the coordinate range 
$[[x_{\rm min},x_{\rm max}],[y_{\rm min},y_{\rm max}],[z_{\rm min},z_{\rm max}]]
= [[0,120],[-120,120],[0,120]]$, where we have taken advantage of
both the equatorial symmetry (implemented  by the 
\codename{ReflectionSymmetry} module) and the $180\degree$ rotational
symmetry around the $z$-axis, which we apply at the $x=0$ plane using the
\codename{RotatingSymmetry180} module. 
\codename{Carpet} provides a hierarchy of refined grids centered at 
each puncture.  Here, we used $7$ levels of refinement,
where the box edge coordinate lengths are given by 
$[128,32,16,8,4,2]$ in units of the total binary mass, which is set to unity. 
Note that overlapping boxes are automatically redefined by 
\codename{Carpet} into one unique region before the domain 
decomposition takes place.

\begin{figure}
        \includegraphics[width=0.45\textwidth]{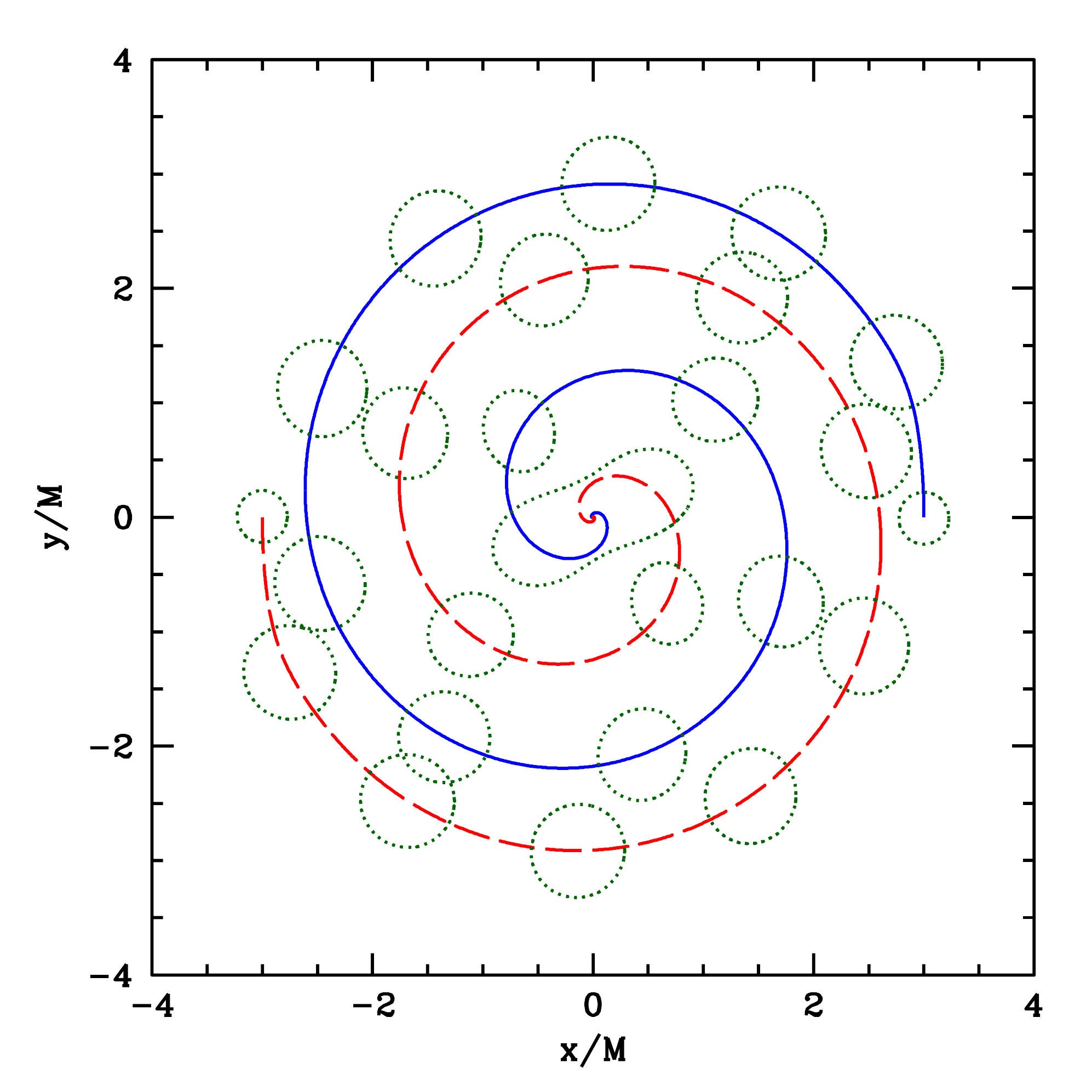}
        \includegraphics[width=0.45\textwidth]{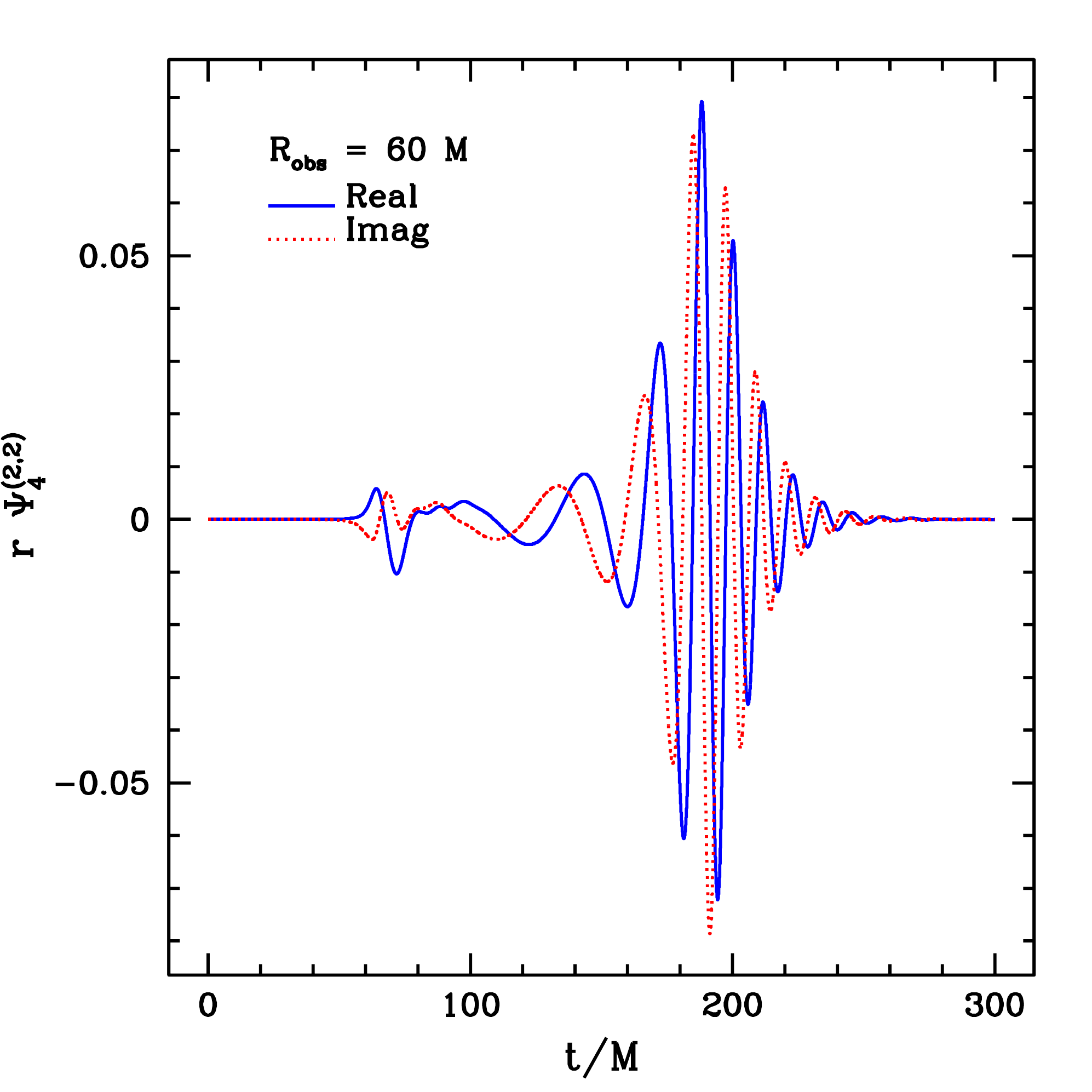}
    \caption{In the left panel, we plot the tracks corresponding to
the evolution of two punctures initially located on the $x$-axis at $x=\pm 3$.
The solid blue line represents puncture 1, and the dashed red line 
puncture 2. The circular dotted green lines are the intersections of the
apparent horizons with the $z=0$ plane plotted every $10M$ during the binary 
evolution. A common horizon appears at $t=116M$. In the right panel,
we plot the real (solid blue line) and imaginary (dotted red line) 
parts of the ($l=2$,$m=2$) mode of the Weyl scalar $\Psi_4$ as extracted 
at an observer radius of $R_{\rm obs}=60M$.}
    \label{fig:tracks_waveform}
\end{figure}

Figure~\ref{fig:tracks_waveform} shows the two puncture tracks 
throughout all phases of the binary evolution, 
provided by the \codename{PunctureTracker} module. In the same
plot we have recorded the intersection 
of the apparent horizon $2$-surface with the $z=0$ plane 
every time interval $t=10M$ during the evolution. 
A common horizon is first observed at $t=116M$. These apparent
horizons were found by the \codename{AHFinderDirect} module and their
radius and location information stored as a $2$-surface with
spherical topology by the \codename{SphericalSurface} module.
The irreducible mass and dimensionless spin of the merged BH were 
calculated by the \codename{QuasiLocalMeasures} module, 
and were found to be $0.647 M$ and $-0.243 M^{-2}$, respectively.

Two modules are necessary to perform the waveform extraction.
The first one, \codename{WeylScal4}, calculates the Weyl scalar
$\Psi_4$ in term of the metric components and its derivatives;
these were computed to be $4$-th order accurate in this example.
The second module, \codename{Multipole}, interpolates the 
Weyl scalars onto spheres with centers and radii specified by 
the user, and performs a spherical harmonic multipole 
mode decomposition. 
Figure~\ref{fig:tracks_waveform} shows the 
real and imaginary parts of the ($l=2$, $m=2$) mode for 
$\Psi_4$ extracted on a sphere centered at the origin  at $R_{\rm obs} = 60M$.
The number of grid points on the sphere was set to be 
$[n_{\theta},n_{\phi}]=[120,240]$, which yields an angular 
resolution of $2.6 \times 10^{-2}$ radians,
and an error of the same order, since the surface integrals were
calculated by midpoint rule -- a first order accurate method.  

In order to evaluate the convergence of the numerical 
solution, we ran five simulations with different
resolutions, and focus our analysis on the convergence
of the phase and amplitude of the Weyl scalar $\Psi_4$. 
The mesh spacings adopted for the coarser grid in the 
AMR hierarchy for these different runs were 
$\{h_{\rm low},h_{\rm med},h_{\rm medh},h_{\rm high},h_{\rm higher}\}
=\{2.0M,1.5M,1.25M,1.0M,0.75M\}$, respectively, while 
the finer grid spacings can be easily found by dividing 
them by $2^k$ for the $k$th level of mesh refinement.For this example, we set
$\{h^f_{\rm low},h^f_{\rm med},h^f_{\rm medh},h^f_{\rm high},h^f_{\rm higher}\}
=\{3.125M,2.344M,1.953M,1.563M,1.172M\}\times 10^{-2}$ for the
finest grid in the different AMR hierarchies, respectively. 
%

Here, we consider the phase $\phi(t)$ and 
the amplitude $A(t)$ of the Weyl scalar $\Psi_4$ at 
$R_{\rm obs}=60M$. In order to take differences between 
the numerical values at two different grid resolutions, we use
an $8$-th order accurate Lagrange operator to interpolate the higher-accuracy finite difference solution
into the immediately coarser grid.
We have experimented with $4$-th and $6$-th order as well,
to evaluate the level of noise these interpolations could
potentially introduce, but did not observe any noticeable
difference and we report here on results from  the higher-order option.

%

\begin{figure}
        \includegraphics[width=0.45\textwidth]{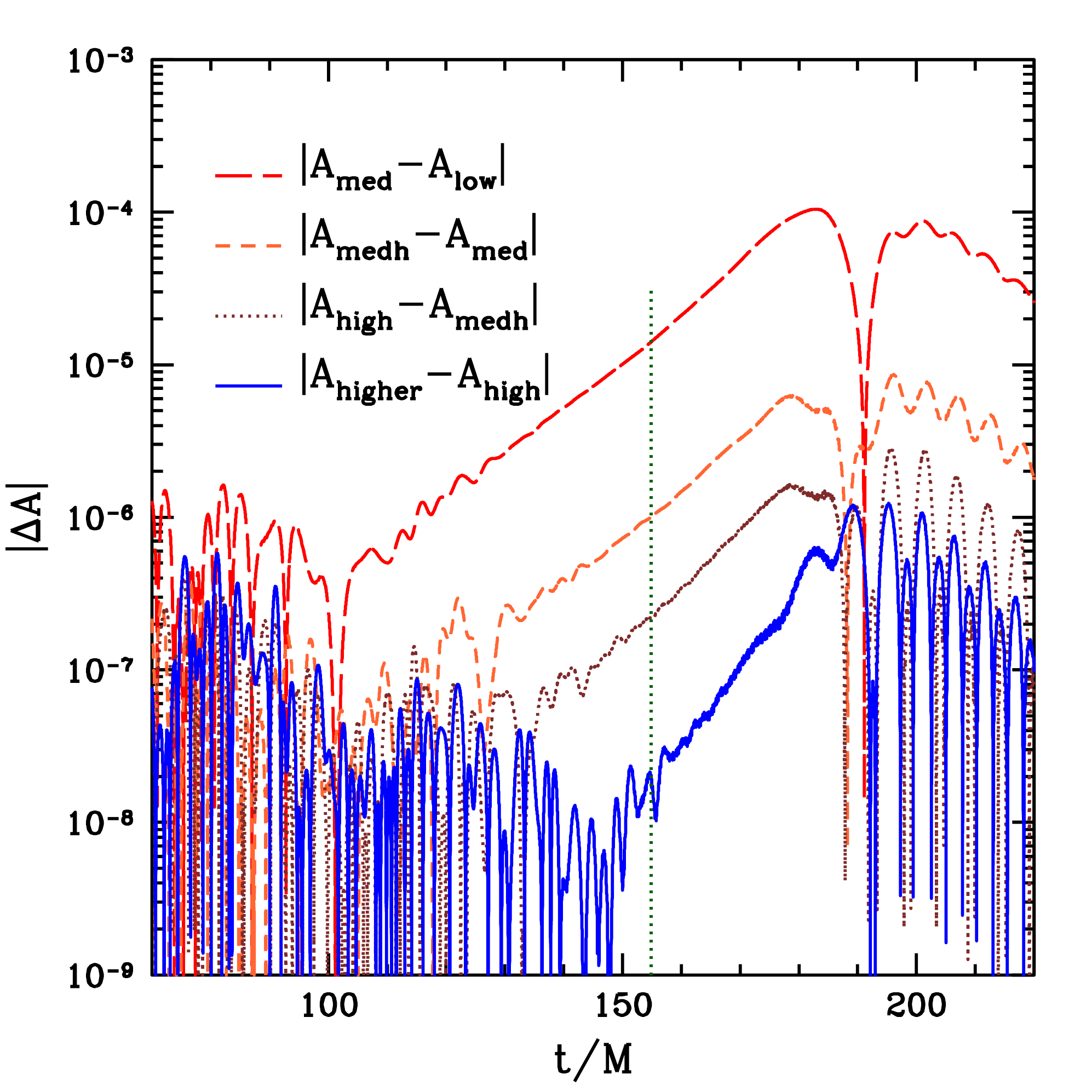}
        \includegraphics[width=0.45\textwidth]{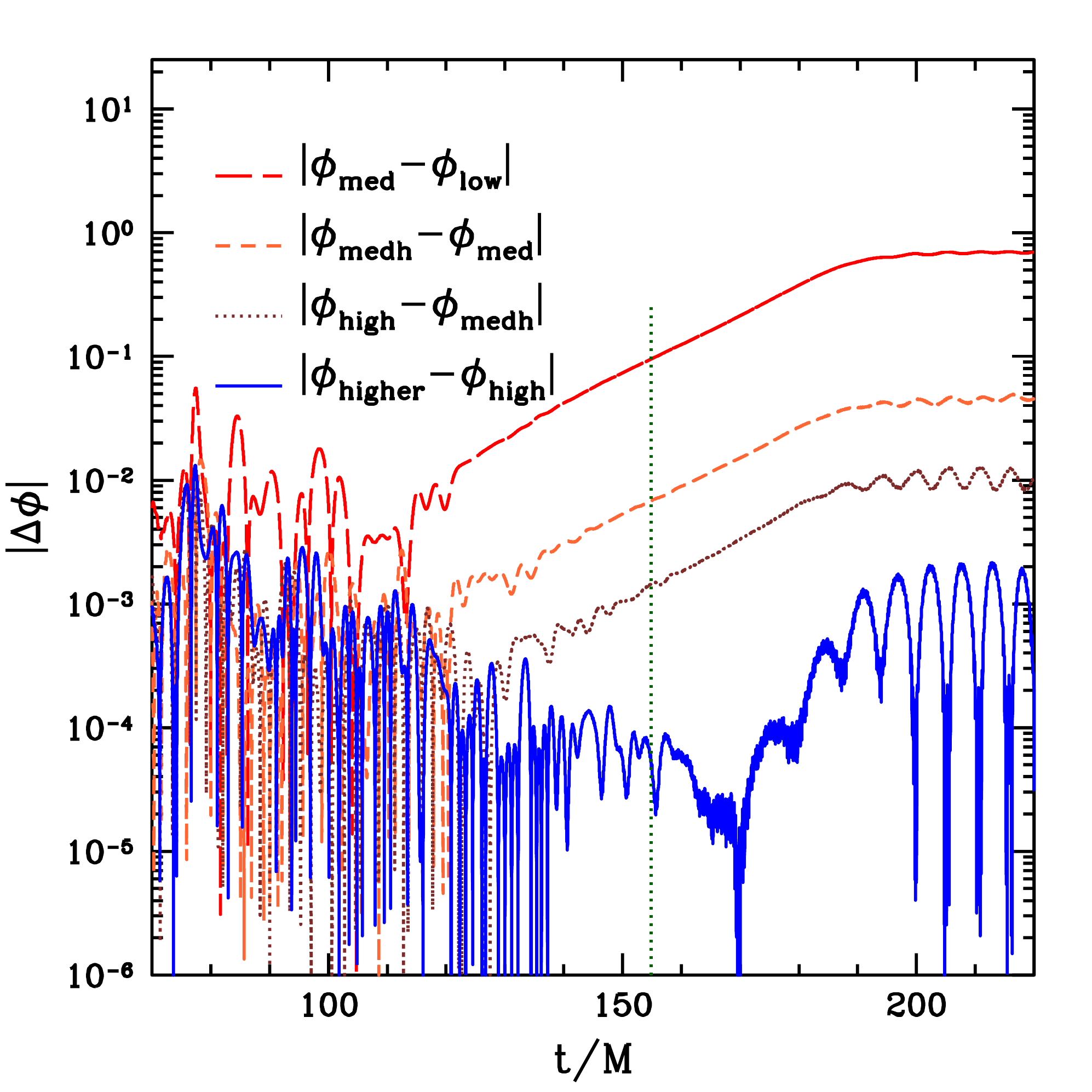}
    \caption{Weyl scalar amplitude (left panel) and phase (right panel) 
convergence. The long dashed red curves represent the difference between 
the medium and low-resolution runs. The short dashed orange curves show
the difference between the medium-high and medium resolution runs. The
dotted brown ones, the difference between high and medium-high resolutions,
while the solid blue curves represent the difference between the higher
and high resolution runs. The dotted vertical green line 
at $t=154M$ indicates the point during the evolution at which the Weyl 
scalar frequency reaches $\omega=0.2/M$. Observe that the three highest 
resolutions accumulate a phase error below the standard of $0.05$ radians 
required by the NRAR collaboration. }
    \label{fig:amp_phs_convergence}
\end{figure}

In Figure~\ref{fig:amp_phs_convergence}, we show the convergence
of the amplitude and phase of the Weyl scalar by plotting the 
logarithm of the absolute value of the differences between two levels 
of resolution. The differences clearly converge to zero as the resolution
is increased.  
We also indicate on both plots the time at which the gravitational
wave frequency reaches $\omega=0.2/M$. We follow a community standard, agreed 
to over the course of the NRAR\cite{NRAR:web} collaboration, that constrains
the numerical resolution so that the accumulated phase error is not
larger than $0.05$ radians at a gravitational wave frequency of
$\omega=0.2/M$. From the plot, we assert that the phase error between the 
higher and high resolutions and the one between high and medium-high
resolutions satisfies this criterion, while the phase error between 
the medium-high and medium resolutions barely satisfies the criterion; and the
one between medium and low resolutions does not. We conclude then
that the three highest resolution runs do have sufficient resolution
to extract waveforms for use in the construction of analytic waveform 
templates. 

\subsection{Linear oscillations of TOV stars}
\label{sec:tov_oscillations}
The examples in the previous subsections did not include the evolution of
matter within a relativistic spacetime. One interesting test of a coupled
matter-spacetime evolution is to measure the eigenfrequencies of a stable TOV
star~(see,
e.g.,~\cite{Gourgoulhon:1991aa, Romero:1996aa, Shibata:1998sg,
  Font:2001ew, Shibata:2003iy}).
These eigenfrequencies can be compared
to values known from linear perturbation theory.

We begin our simulations with  a self-gravitating fluid
sphere, described by a polytropic equation of state. This one-dimensional
solution is obtained by the code described in section~\ref{sec:TOVSolver}, and
is interpolated on the three-dimensional, computational evolution grid.
This system is then evolved using the BSSN evolution system implemented in
\codename{McLachlan} and the hydrodynamics evolution system implemented in
\codename{GRHydro}.

For the test described here, we set up a stable TOV star described by a
polytropic equation of state $p=K\rho^\Gamma$ with $K=100$ and $\Gamma=2$,
and an initial central density of $\rho_c=1.28\times10^{-3}$. This model can
be taken to represent a non-rotating NS with a mass of
$M=1.4\mathrm{M}_\odot$. The computational domain is a cube of length
$640\mathrm{M}$ with a base resolution of $2\mathrm{M}$ ($4\mathrm{M}$,
$8\mathrm{M}$) in each dimension. Four additional grids refine the region
around the star centered at the origin, each doubling the resolution, with sizes
of $120\mathrm{M}$, $60\mathrm{M}$, $30\mathrm{M}$ and $15\mathrm{M}$,
resulting in a resolution of $0.125\mathrm{M}$ ($0.25\mathrm{M}$,
$0.5\mathrm{M}$) across the entire star.

\begin{figure}
 \label{fig:tov_rho_max}
 \includegraphics[width=0.9\textwidth]{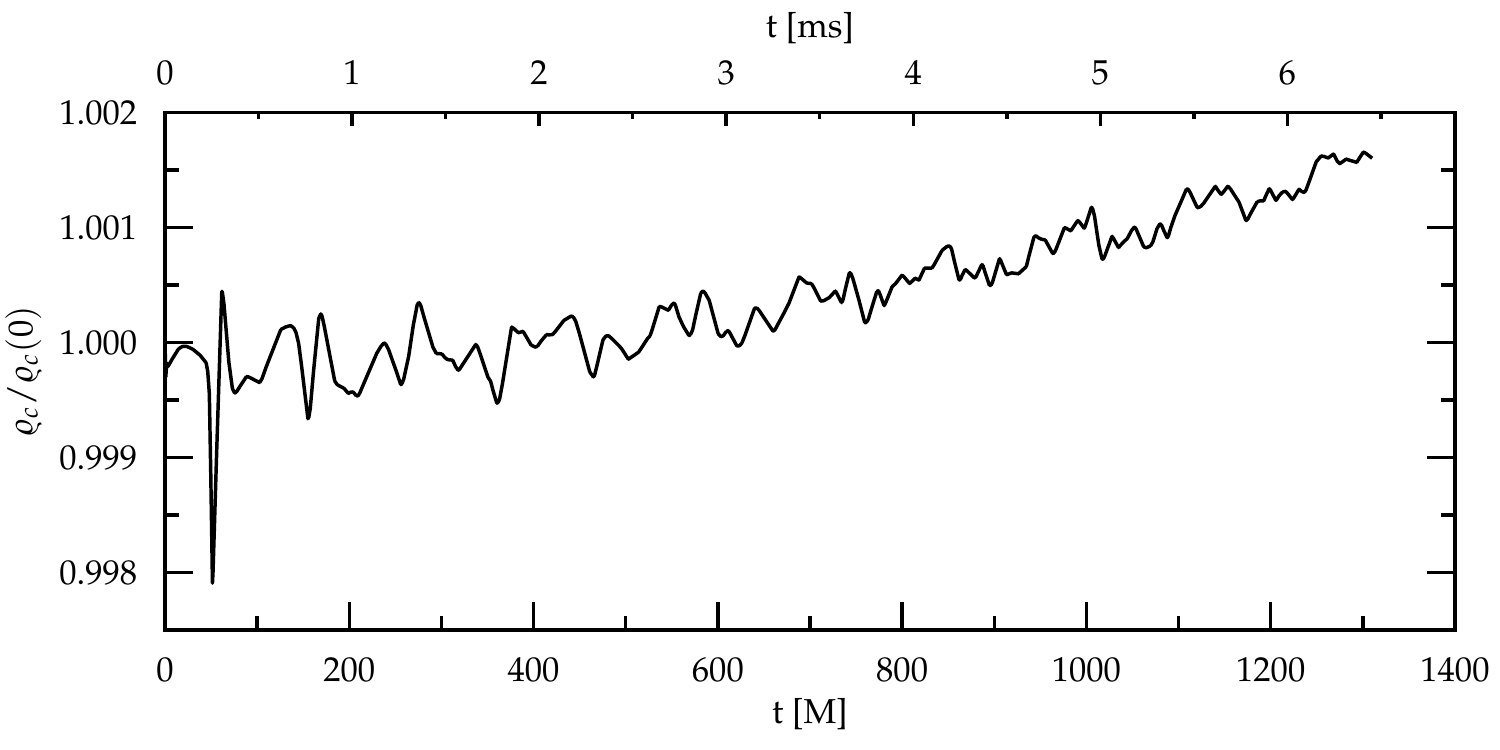}
 \caption{Evolution of the central density for the TOV star. Clearly visible is
 an initial spike, produced by the interpolation of the one-dimensional equilibrium
 solution onto the three-dimensional evolution grid. The remainder of the evolution
 however, the central density evolution is dominated by continuous excitations coming
 from the interaction of the stellar surface with the artificial atmosphere.}
\end{figure}

In Figure~\ref{fig:tov_rho_max} we show the evolution of the central density of
the star over an evolution time of $1300\mathrm{M}$ ($6.5\mathrm{ms}$). The
initial spike is due to the perturbation of the solution resulting from the
interpolation onto the evolution grid. The remaining oscillations are mainly
due to the interaction of the star and the artificial atmosphere and are
present during the whole evolution.  Given enough evolution time, the
frequencies of these oscillations can be measured with satisfactory accuracy.

\begin{figure}
 \label{fig:tov_mode_spectrum}
 \includegraphics[width=0.9\textwidth]{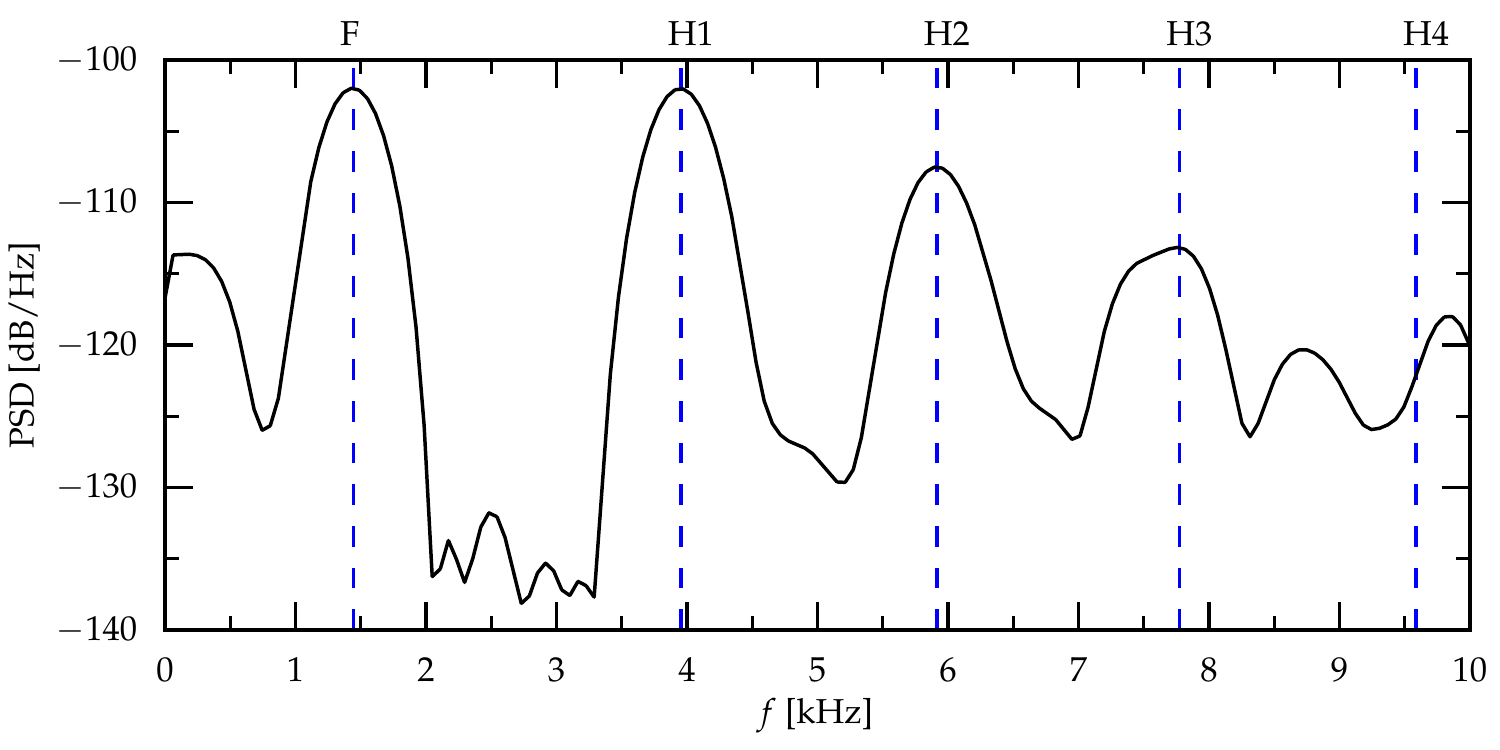}
 \caption{Eigenfrequency mode spectrum of a TOV star. Shown is the power
 spectral density of the central matter density, computed from a full 3D
 relativistic hydrodynamics simulation and compared to the values obtained by
 perturbation theory. The agreement of the frequencies of the fundamental mode
 and the first three overtones is clearly visible.}
\end{figure}

In Figure~\ref{fig:tov_mode_spectrum} we show the power spectral density (PSD)
of the central density oscillations computed from a full 3D relativistic
hydrodynamics simulation, compared to the corresponding frequencies as obtained
with perturbative techniques (kindly provided by Kentaro Takami and
computed using the method described in~\cite{Yoshida:1999vj}). The PSD was computed
using the entire time series of the high-resolution run, by removing the linear
trend and averaging over Hanning windows overlapping half the signal length after
padding the signal to five times its length. The agreement of the 
 fundamental mode and first three overtone frequencies is clearly visible, but
are limited beyond this by the finite numerical resolution.
Higher overtones should be measurable with higher
resolution, but at substantial computational
cost.

Within this test it is also interesting to study the convergence behavior of
the coupled curvature and matter evolution code. One of the variables often
used for this test is the Hamiltonian constraint violation. This violation
vanishes for the continuum problem, but is non-zero and resolution-dependent in
discrete simulations. The expected rate of convergence of the hydrodynamics
code lies between $1$ and $2$. It cannot be higher than $2$ due to the
directional flux-split algorithm which is of second order. Depending on
solution itself, the hydrodynamics code is only of first order in particular
regions, e.g., at extrema (like the center of the star), or at the stellar
surface.

\begin{figure}
 \label{fig:tov_ham_conv}
 \includegraphics[width=0.9\textwidth]{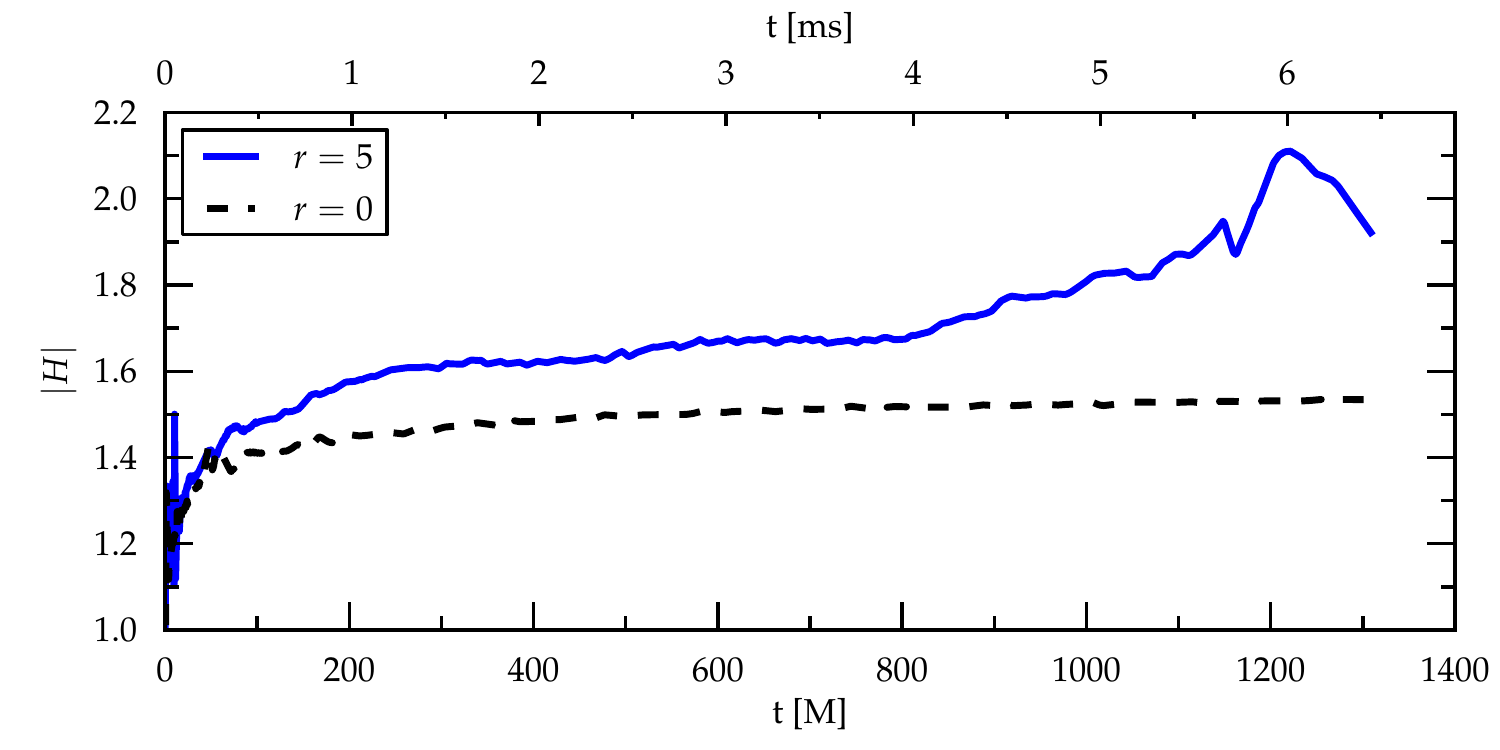}
 \caption{Convergence factor of Hamiltonian constraint violation at
   $r=0\mathrm{M}$ and $r=5\mathrm{M}$. The observed convergence order
   of about $1.5$ at the center of the star is lower then the general
   second order of the hydrodynamics evolution scheme. This is
   expected because the scheme's convergence rate drops to first order
   at extrema or shocks, like the stellar center or surface.
   Consequently, the observed convergence order about half way between
   the stellar center and surface is higher than $1.5$, but mostly below $2$.}
\end{figure}

Figure~\ref{fig:tov_ham_conv} shows the order of convergence of the Hamiltonian
constraint violation, using the three highest-resolution runs, at the stellar
center and a coordinate radius of $r=5\mathrm{M}$ which is about half way between the
center and the surface. The observed convergence rate for most of the
simulation time lies between $1.4$ and $1.5$ at the center, and between $1.6$ and
$2$ at $r=5\mathrm{M}$, consistent with the expected data-dependent convergence
order of the underlying hydrodynamics evolution scheme.

\subsection{Neutron star collapse}
\label{sec:collapse_example}
The previous examples dealt either with preexisting BHs, either single
or in a binary, or with a smooth singularity free spacetime, as in the
case of the TOV star.  The evolution codes in the toolkit are,
however, also able to handle the dynamic formation of a singularity,
that is follow a neutron star collapse into a BH. As a simple
example of this process, we study the collapse of a non-rotating TOV
star.  We create initial data as in section~\ref{sec:tov_oscillations}
using $\rho_c=3.154\times10^{-3}$ and $K_{\mathrm{ID}} = 100$, $\Gamma
= 2$, yielding a star model of gravitational mass $1.67\,M_\odot$, that
is at the onset of instability. As is common in such situations~(e.g.,
\cite{Baiotti:2005vi}), we trigger collapse by reducing the pressure
support after the initial data have been constructed by lowering the
polytropic constant $K_{\mathrm{ID}}$ from its initial value to
$K = 0.98 \, K_{\mathrm{ID}} = 98$.  To ensure that the pressure-depleted
configuration remains a solution of the Einstein constraint
equations~\eref{eqn:analysis_hamiltonian_constraint} in the presence
of matter, we rescale the rest mass density $\rho$ such that the total
energy density $T_{nn}$ does not change:
\begin{equation}
    \rho' + K (\rho')^2 = \rho + K_{\mathrm{ID}} \rho^2.
    \label{eqn:collapse_rho_rescaled}
\end{equation}
Compared to the initial configuration, this rescaled star possesses a
slightly higher central density and lower pressure.  This change in
$K$ accelerates the onset of collapse that would otherwise rely on
being triggered by numerical noise, which would not be guaranteed to
converge to a unique solution with increasing resolution. In order to
resolve the star as well as to push the outer boundary far enough away (so
that the star and the numerical outer boundary are not in causal
contact during the simulation) we employ a fixed mesh refinement
scheme.  The outermost box has a radius of $R_0 = 204.8\,M_\odot$ and
a resolution of $3.2\,M_\odot$ ($2.4\,M_\odot$, $1.6\,M_\odot$,
$0.6\,M_\odot$ for higher convergence levels).  Around the star,
centered about the origin, we stack $5$ extra boxes of approximate size
$8\times2^\ell\,M_\odot$ for $0 \le \ell \le 4$, where the resolution
on each
level is twice that of the surrounding level.  In order to resolve the large
density gradients developing during the collapse, two more levels with radii
$4\,M_\odot$ and $2\,M_\odot$ are placed inside the star.  We use the PPM
reconstruction method and the HLLE Riemann solver to obtain second-order
convergent results in smooth regions.  Due to the presence of the
density maximum at the center of the star and the non-smooth atmosphere at the
edge of the star, we expect the observed convergence rate to be somewhat lower
than second order, but higher than first order.  
\begin{figure}
 \label{fig:tov_collapse_radii}
 \includegraphics[width=0.9\textwidth]{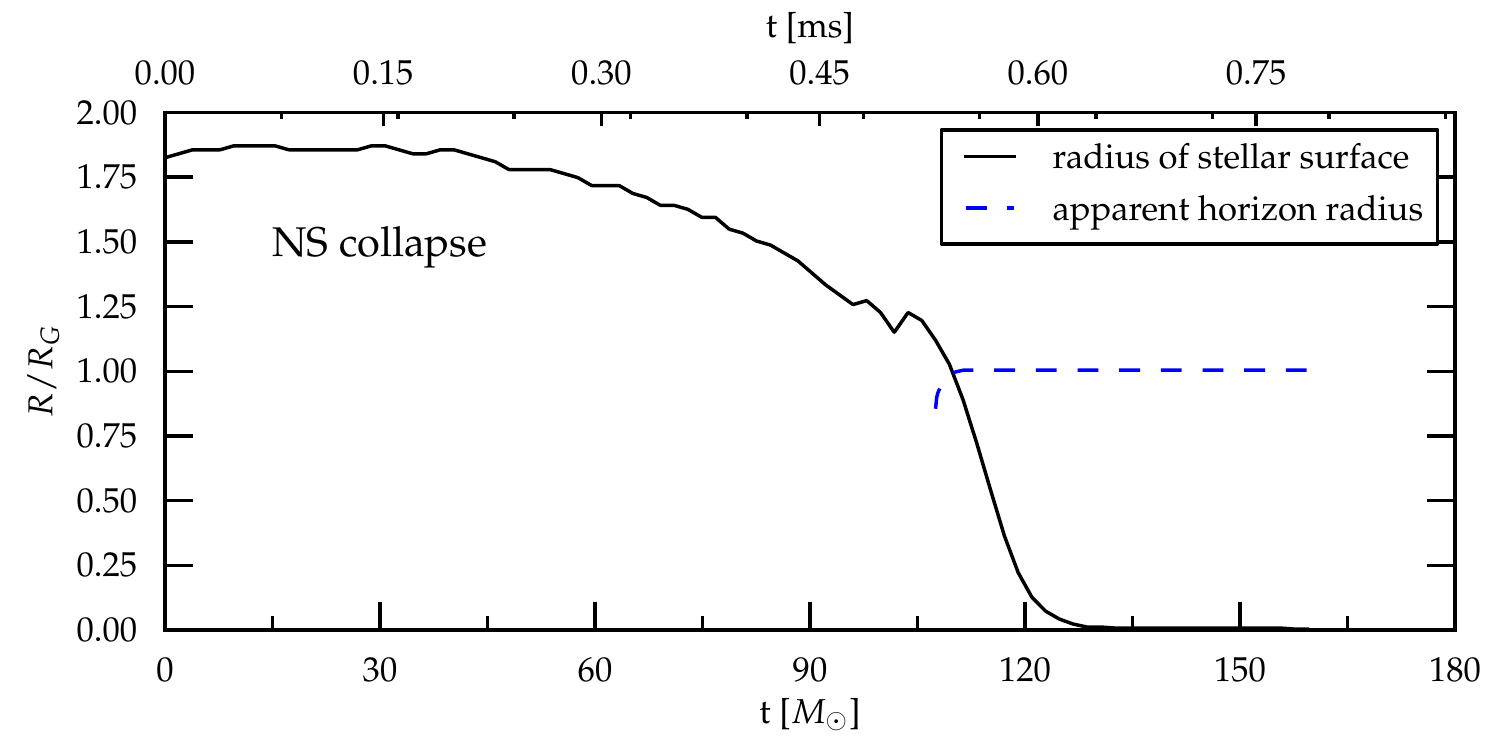}
 \caption{Coordinate radius of the surface of the collapsing star and radius
 of the forming
 apparent horizon. The stellar surface is defined as the point where $\rho$ is
 $100$ times the atmosphere density. $R$ is the circumferential radius of the
 apparent horizon and $R_g = 2\,M_\star = 2\times1.63\,M_{\mathord\odot}$. An
 apparent horizon forms at a time roughly equal to when the mass of the star
 is enclosed in its gravitational radius, forming a black hole and causally
 disconnecting the evolution in the interior from the outside spacetime. The
 lower $x$-axis displays time in code units where $M_\odot=G=c=1$, and the upper
 $x$-axis shows the corresponding physical time using $1\,M_\odot = 4.93\,\mu s$.}
\end{figure}
\begin{figure}
 \label{fig:tov_collapse_H_convergence_at0}
 \includegraphics[width=0.9\textwidth]{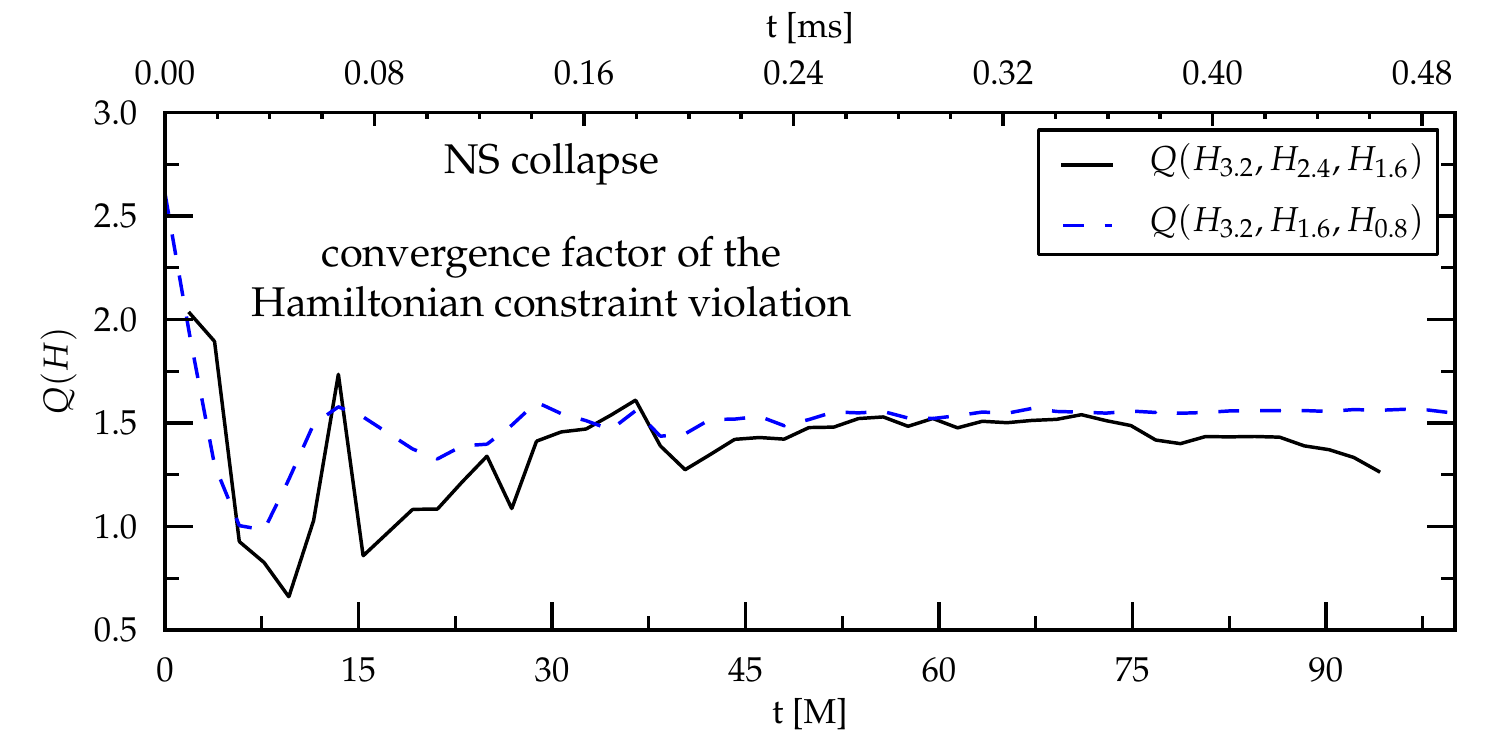}
 \caption{Convergence factor for the Hamiltonian constraint violation
   at the center of the collapsing star. We plot convergence factors
   computed using a set of 4 runs covering the diameter of the star
   with $\approx$ 60, 80, 120, and 240 grid points. The units of time
   on the upper and lower $x$-axes are identical to those of
   Figure~\ref{fig:tov_collapse_radii}.
 } 
\end{figure}
In Figure~\ref{fig:tov_collapse_radii}, we plot the approximate
coordinate size of the star as well as the circumferential radius of
the apparent horizon that eventually forms in the simulation.  The
apparent horizon is first found at approximately the time when the
star's coordinate radius approaches its Schwarzschild radius, though
one needs to keep in mind that the Schwarzschild radius is a
circumferential radius, whereas the meaning of the coordinate radius
in our BSSN calculation is likely somewhat different.
In Figure~\ref{fig:tov_collapse_H_convergence_at0}, we display the
convergence factor obtained from
\begin{equation}
    \frac{H_{h_1}-H_{h_2}}{H_{h_2}-H_{h_3}} = \frac{h_1^Q-h_2^Q}{h_2^Q-h_3^Q}\,,
    \label{eq:convergence-factor-definition}
\end{equation}
for the Hamiltonian constraint violation at the center of the collapsing star.
Here $H_{h_i}$ is the Hamiltonian constraint
violation~\eref{eqn:analysis_hamiltonian_constraint} at the center of the
star for a run with resolution $h_i$.
Up to the time when the apparent horizon forms, the convergence order is
an expected $\approx 1.5$. At later times, the singularity forming at the
center of the collapsing star renders a pointwise measurement of the
convergence factor at the center impossible.

\subsection{Cosmology}\label{sec:cosmology}
The Einstein Toolkit is not only designed to evolve compact-object
spacetimes, but also to solve the initial-value
problem for spacetimes with radically different topologies and global
properties. In this section, we illustrate the evolution of an
initial-data set representing a constant-$t$ section of a
spacetime from the Gowdy $T^3$ class~\cite{Gowdy:1971jh,New:1997me}. Models in
this class have the line element:
\begin{equation}
\label{eq:gowdyT3}
ds^2=\tau^{-1/2}e^{\lambda/2}(-d\tau^2+dz^2)+\tau[e^P(dx+Qdy)^2+e^{-P}dy^2]
\end{equation}
defined on a 3-torus $-x_0 \leq x \leq x_0$, $-y_0 \leq y \leq y_0$,
$-z_0 \leq z \leq z_0$, with the functions $P$, $Q$ and $\lambda$ to be 
determined by the Einstein equations. For $P=Q=\lambda=0$, a coordinate
transformation $t=4/3 \, \tau^{3/4}$ (plus a rescaling of the spatial
coordinates) casts the line element into the form:
\begin{equation}
\label{eq:kasner}
ds^2=-dt^2+t^{4/3}(dx^2+dy^2)+t^{-2/3}dz^2
\end{equation}
which represents the familiar Kasner spacetime for a homogeneous but 
anisotropically expanding universe. In the 3+1 decomposition described
above, this reads:
\begin{widetext}
\begin{eqnarray}
\alpha(t) &=& 1 \\
\beta^i(t) &=& 0 \\
\gamma_{ij}(t) &=& {\rm diag}(t^{4/3},t^{4/3},t^{-2/3}) \\
K_{ij}(t) &=& - {\rm diag}(\frac{2}{3} \, t^{4/3},\frac{2}{3} \, t^{4/3},\frac{1}{3} \, t^{-2/3})
\end{eqnarray}
\end{widetext}

\begin{figure}
 \includegraphics[width=0.9\textwidth]{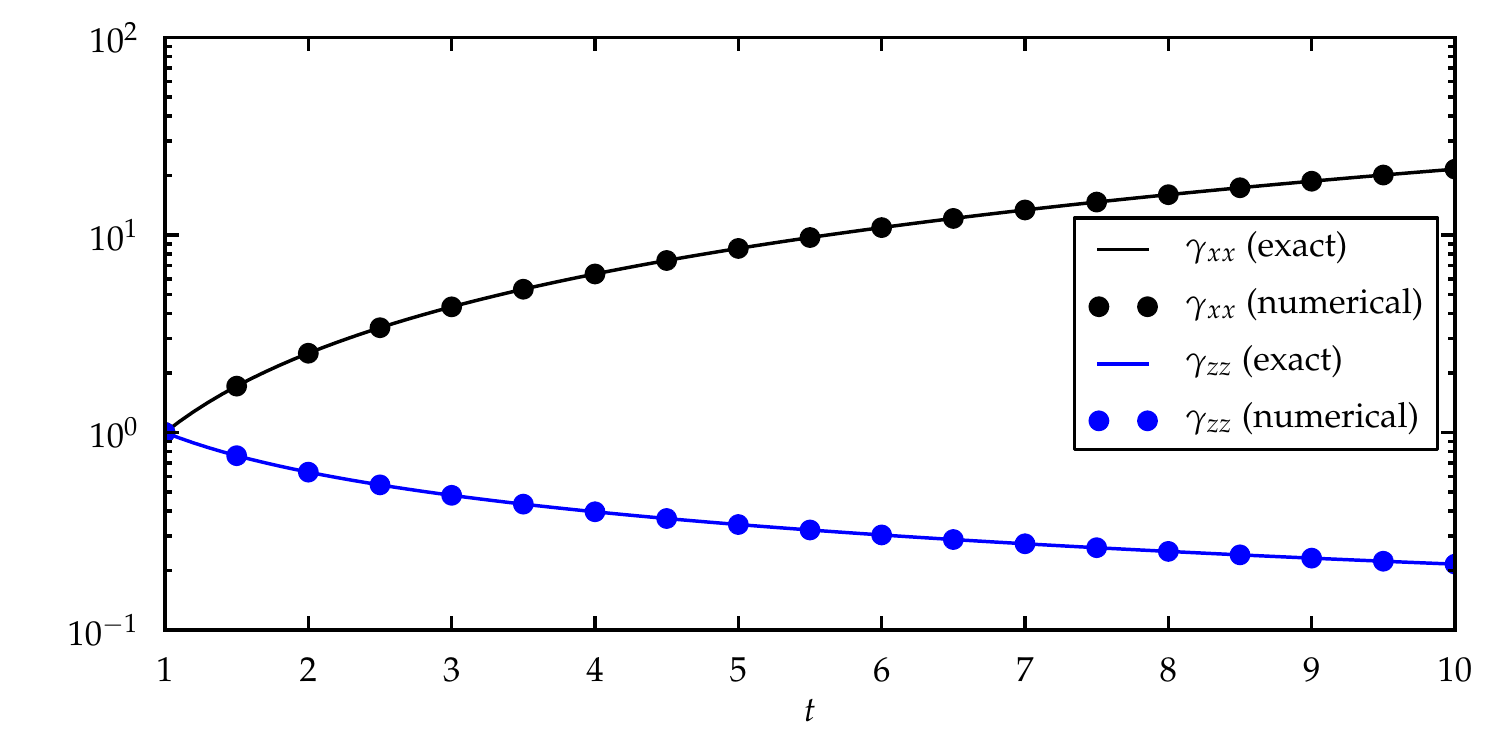}
 \includegraphics[width=0.9\textwidth]{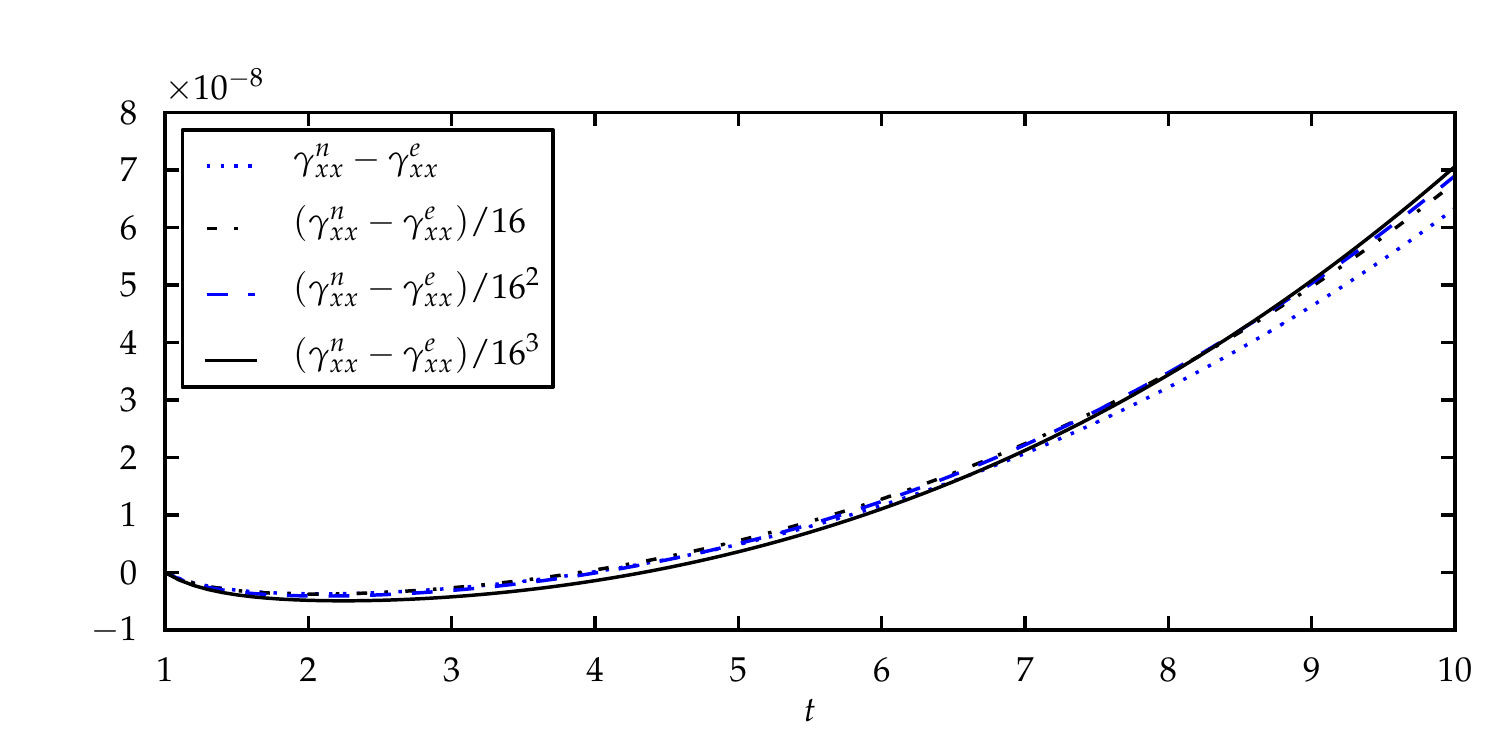}
 \caption{Top: the evolution of a vacuum spacetime of the type~\eref{eq:gowdyT3},
 with $P=Q=\lambda=0$; the initial data are chosen as
 $\gamma_{ij}=\delta_{ij}$ and $K_{ij}={\rm diag}(-2/3,-2/3,1/3)$.
 Bottom: the numerical error for a sequence of four time resolutions $dt=[0.0125,0.025,0.05,0.1]$;
 the errors are scaled according to the expectation for fourth-order convergence.
 \label{fig:kasner}}
\end{figure}

In Figure~\ref{fig:kasner}, we show the full evolution of the $t=1$ slice 
of spacetime~\eref{eq:kasner}, along with the associated error for a sequence of 
time resolutions.

\section{Conclusion and Future Work}

In this article, we described the Einstein Toolkit, a collection
of freely available and easy-to-use computational codes for numerical
relativity and relativistic astrophysics. The code details and example
results present in this article represent the state of the Einstein
Toolkit in its release ET\_2011\_05 ``Curie,'' released on April 21,
2011. 

The work presented here is but a snapshot of the Einstein Toolkit's
ongoing development, whose ultimate goal it is to provide an
open-source set of robust baseline codes to realistically and
reproducibly model the whole spectrum of relativistic astrophysical
phenomena including, but not limited to, isolated black holes and
neutron stars, binary black hole coalescence in vacuum and gaseous
environments, double neutron star and neutron star -- black hole mergers,
core-collapse supernovae, and gamma-ray bursts.

For this, much future work towards including proper treatments of
magnetic fields, more complex equations of state, nuclear reactions,
neutrinos, and photons will be necessary and will need to be matched
by improvements in infrastructure (e.g., more flexible AMR on general
grids) and computing hardware for the required fully coupled 3-D,
multi-scale, multi-physics simulations to become reality.  These tasks, 
as well as the others mentioned below, are likely to occupy a great deal of the 
effort spent developing future versions of the Einstein Toolkit over the next few years.

  Without a doubt, collapsing stars and merging BH-NS and NS-NS binaries must be simulated with GRMHD to capture
  the effects of magnetic fields that in many cases will
  alter the simulation outcome on a qualitative level and may be 
  the driving mechanisms behind much of the observable EM signature
  from GRBs (e.g.,~\cite{Woosley:2006fn}) 
  and magneto-rotationally exploding core-collapse supernovae
  (e.g.,~\cite{Burrows:2007yx}). To date, all simulations that have
  taken magnetic fields into account are still limited to the
  ideal MHD approximation, which assumes perfect conductivity. 
  Non-ideal GRMHD schemes are just becoming 
  available~(see, e.g.,~\cite{Palenzuela:2008sf,DelZanna:2007pk}), 
  but have yet to be implemented widely in many branches of numerical relativity.
  
  Most presently published 3D GR(M)HD simulations, with the
  exception of recent work on massive star collapse 
  (see, e.g.,~\cite{Ott:2006eu}) and binary mergers 
  (see, e.g.,~\cite{Sekiguchi:2011zd}),
  relied on simple zero-temperature descriptions of
  NS stellar structure, with many assuming simple polytropic forms. 
  Such EOSs are computationally
  efficient, but are not necessarily a good description for matter in
  relativistic astrophysical systems. The inclusion of 
  finite-temperature EOSs, derived from the microphysical descriptions of
  high-density matter, will lead to qualitatively different and much
  more astrophysically reliable results (see, e.g.,~\cite{Ott:2006eu}).
  In addition, most GR(M)HD studies
  neglect transport of neutrinos and photons
  and their interactions with matter. Neutrinos in
  particular play a crucial role in core-collapse supernovae and in
  the cooling of NS-NS merger remnants, thus they must not be left out when
  attempting to accurately model such events.  Few studies have
  incorporated neutrino and/or photon transport and interactions in
  approximate ways (see, e.g.,~\cite{Ott:2006eu,Farris:2008fe,Sekiguchi:2011zd,Sekiguchi:2011mc}).



Besides new additions of physics modules, existing techniques require
improvement. One example is the need for the gauge invariant
extraction of gravitational waves from simulation spacetimes as realized
by the Cauchy Characteristic Extraction (CCE) technique recently studied
in~\cite{Babiuc:11,Reisswig:2010cd,Reisswig:2011a}.  The authors of one such
CCE code~\cite{Babiuc:11} have agreed to make their work available to the
whole community by integrating their CCE routines into the Einstein Toolkit
release 2011\_11 ``Maxwell,'' which will be described elsewhere.

A second  much needed improvement of our existing methods
is a transition to cell-centered AMR for GR hydrodynamic simulations,
which would allow for exact flux conservation across AMR interfaces
via a refluxing step that adjusts coarse and/or fine grid fluxes for
consistency (e.g., \cite{Berger:1984zza}). This is also a prerequisite
for the constrained transport method \cite{Toth:00} for ensuring the
divergence-free condition for the magnetic field in a future
implementation of GRMHD within the Einstein Toolkit.  Work towards
cell-centered AMR, refluxing, and GRMHD is underway and will be
reported in a future publication.

  While AMR can increase resolution near regions of interest within
  the computational domain, it does not increase the convergence
  order of the underlying numerical methods. Simulations of BHs
  can easily make use of high-order numerical methods, with eighth-order
  convergence common at present. However,
  most GRMHD schemes, though they implement high-resolution 
  shock-capturing methods,
  are limited to 2nd-order numerical accuracy in the 
  hydrodynamic/MHD sector while performing
  curvature evolution with 4th-order accuracy or more. Higher order
  GRMHD schemes are used in fixed-background simulations
  (e.g.,~\cite{Tchekhovskoy:2007zn}), but still await implementation in 
  fully dynamical simulations.


Yet another important goal is to increase the scalability of the {\tt
  Carpet} AMR infrastructure. As we have shown, good scaling is
limited to only a few thousand processes for some of the most widely used
simulation scenarios.  Work is in progress to eliminate this
bottleneck~\cite{Zebrowski:2011bl}.  On the other hand, a production simulation is typically
composed of a large number of components, and even analysis and I/O
routines have to scale well to achieve overall good performance. This
is a highly non-trivial problem, since most Einstein Toolkit physics
module authors are neither computer scientists nor have they had
extensive training in parallel development and profiling
techniques. Close collaboration with experts in these topics has been
fruitful in the past and will be absolutely necessary for the
optimization of Einstein Toolkit codes for execution on the upcoming
generation of true petascale supercomputers on which typical compute jobs
are expected to be running on 100,000 and more compute cores.

\section*{Acknowledgments}

The authors wish to thank Ed Seidel whose
inspiration and vision has driven work towards the Einstein Toolkit
over the past 15 years.
We are also grateful to the large number of people who contributed to
the Einstein Toolkit via ideas, code, documentation, and testing;
without these contributions, this toolkit would not exist today.

The Einstein Toolkit is directly supported by
the National Science Foundation in the USA under the grant numbers
0903973/0903782/0904015 (CIGR\@).  Related grants contribute directly
and indirectly to the success of CIGR, including NSF OCI-0721915, NSF
OCI-0725070, NSF OCI-0832606, NSF OCI-0905046, NSF OCI-0941653, NSF
AST-0855535, NSF DMS-0820923, NASA 08-ATFP08-0093, EC-FP7
PIRG05-GA-2009-249290 and Deutsche Forschungsgemeinschaft grant
SFB/Transregio~7 ``Gravitational-Wave Astronomy''.  Results presented
in this article were obtained through computations on the Louisiana
Optical Network Initiative under allocation loni\_cactus05 and
loni\_numrel07, as well as on NSF XSEDE under allocations
TG-MCA02N014, TG-PHY060027N, TG-PHY100033, at the National Energy
Research Scientific Computing Center (NERSC), which is supported by
the Office of Science of the US Department of Energy under contract
DE-AC03-76SF00098, at the Leibniz Rechenzentrum of the Max Planck
Society, and on Compute Canada resources via project cfz-411-aa.

G. Allen acknowledges that this material is based upon work supported while
serving at the National Science Foundation. Any opinion, findings, and
conclusions or recommendations expressed in this material are those of the
authors and do not necessarily reflect the views of the National Science
Foundation.


\bibliographystyle{iopart-num-edit}
\bibliography{manifest/einsteintoolkit}

\end{document}